\pdfoutput=1
\documentclass[journal]{IEEEtran}
\usepackage{amsmath,amsfonts}
\usepackage{algorithmic}
\usepackage{algorithm}
\usepackage{array}
\usepackage[caption=false,font=normalsize,labelfont=sf,textfont=sf]{subfig}
\usepackage{textcomp}
\usepackage{stfloats}
\usepackage{url}
\usepackage{verbatim}
\usepackage{graphicx}
\usepackage{cite}
\usepackage{xcolor}
\usepackage{multirow}
\usepackage{makecell}
\usepackage{booktabs} 
\usepackage{ragged2e}
\usepackage{epsfig}
\usepackage[caption=false]{subfig} 
\usepackage{newunicodechar}
\usepackage{caption}

\begin{document}

\title{Physical-layer Adversarial Robustness for \\ Deep Learning-based Semantic Communications}

\author{Guoshun Nan,~\IEEEmembership{~Member,~IEEE,} Zhichun Li, Jinli Zhai, Qimei Cui,~\IEEEmembership{~Senior Member,~IEEE,} Gong Chen, Xin Du, Xuefei Zhang,~\IEEEmembership{~Member,~IEEE,} Xiaofeng Tao,~\IEEEmembership{~Senior Member,~IEEE,} \\ Zhu Han,~\IEEEmembership{~Fellow,~IEEE,} Tony Q.S. Quek,~\IEEEmembership{~Fellow,~IEEE}
\thanks{G. Nan, Z. Li, J. Zhai, Q. Cui, G. Chen, X. Du, X. Zhang, X. Tao are with Beijing University of Posts and Telecommunications, China and National Engineering Research Center for Mobile Network Technologies. (e-mail:nanguo2021@bupt.edu.cn; monstry@bupt.edu.cn; chenest@bupt.edu.cn; cuiqimei@bupt.edu.cn; cgvxpr@bupt.edu.cn; juliusxin@bupt.edu.cn; zhangxuefei@bupt.edu.cn; taoxf@bupt.edu.cn).}
\thanks{Z, Han is with University of Houston, USA (e-mail:zhan2@uh.edu).}
\thanks{T. Q. S. Quek is with the Singapore University of Technology and Design, Singapore 487372, and also with the Yonsei Frontier Lab, Yonsei University, South Korea (e-mail: tonyquek@sutd.edu.sg).}
\thanks{Corresponding author: Q. Cui (e-mail: cuiqimei@bupt.edu.cn).}}
\markboth{Journal of \LaTeX\ Class Files,~Vol.~14, No.~8, August~2021}%
{Shell \MakeLowercase{\textit{et al.}}: A Sample Article Using IEEEtran.cls for IEEE Journals}


\maketitle

\begin{abstract}
End-to-end semantic communications (ESC) rely on deep neural networks (DNN) to boost communication efficiency by only transmitting the semantics of data, showing great potential for high-demand mobile applications. We argue that central to the success of ESC is the robust interpretation of conveyed semantics at the receiver side, especially for security-critical applications such as automatic driving and smart healthcare. However, robustifying semantic interpretation is challenging as ESC is extremely vulnerable to physical-layer adversarial attacks due to the openness of wireless channels and the fragileness of neural models. Toward ESC robustness in practice, we ask the following two questions: Q1: For attacks, is it possible to generate semantic-oriented physical-layer adversarial attacks that are imperceptible, input-agnostic and controllable? Q2: Can we develop a defense strategy against such semantic distortions and previously proposed adversaries? To this end, we first present \texttt{MobileSC}, a novel semantic communication framework that considers the computation and memory efficiency in wireless environments. Equipped with this framework, we propose \texttt{SemAdv}, a physical-layer adversarial perturbation generator that aims to craft semantic adversaries over the air with the abovementioned criteria, thus answering the Q1. To better characterize the real-world effects for robust training and evaluation, we further introduce a novel adversarial training method $\texttt{SemMixed}$ to harden the ESC against \texttt{SemAdv} attacks and existing strong threats, thus answering the Q2.
Extensive experiments on three public benchmarks verify the effectiveness of our proposed methods against various physical adversarial attacks. We also show some interesting findings, e.g., our \texttt{MobileSC} can even be more robust than classical block-wise communication systems in the low SNR regime. 
\end{abstract}


\section{Introduction}
\subsection{Background}
ITU reported that the mobile traffic would grow at an annual rate of around 55\% in $2020$-$2030$ \cite{ITU} due to a huge amount of data generated by newly emerging mobile applications such as Internet of Things (IoT), VR/AR, and machine-to-machine (M2M) networks. On this basis, many studies on wireless communication\cite{Lin2023TransmissiveMA,Yang2020ReconfigurableIS,Li2010ANF} and privacy protection\cite{Xu2016EnhancingSC,Liu2014JointBA,Li2021PrivacyPreservedFL} have been conducted. The remarkable trend \cite{ITU,you2021towards,xiao2020toward,strinati20216g} indicates that these applications require interpretations of the information more intelligently on the receiver side, and the increasing popularity of such applications puts unprecedented pressure on both network capacities and intelligence capabilities of conventional wireless communication systems, and motivates the community to develop a more efficient and smart way of information transmission in future wireless networks. 

Recently proposed end-to-end semantic communications (ESC) \cite{xie2020deep,luo2022semantic}, as a revolutionary wireless communication paradigm, holds great potential to meet the requirements of high-demand applications. Specifically, ESC merges all physical
layer blocks in traditional communication systems and replaces the block-wise structures as end-to-end neural networks, facilitating joint transceiver optimization. Thereby, ESC can inherently rely on deep neural networks (DNN) \cite{lecun2015deep} to boost communication efficiency by learning to transmit only the semantics of the data rather than the whole. The concept of such semantic communications \cite{weaver1953recent,bao2011towards} has been discussed for a long history and can be dated from the open question raised by Weaver \cite{weaver1953recent}: ``How precisely do the transmitted symbols convey the desired meaning'' in $1949$. Towards this direction, a line of DNN-based ESC systems \cite{o2017introduction,farsad2018deep,xie2020deep} have been proposed recently, which employ an autoencoder-like neural architecture \cite{kramer1991nonlinear} to extract semantic representations with a semantic encoder at the sender side and then reconstruct the information with a decoder at the receiver side.    
\subsection{Motivation}
As highlighted in prior efforts \cite{xie2021deep,luo2022semantic}, the goal of ESC is not often to fully recover the underlying message delivered, but to empower the receiver to make the correct understanding or to take the proper actions in the right context. Thereby, \textit{we argue that central to the success of semantic communication systems is the robust interpretation of semantics conveyed at the receiver side}. However, ensuring robust semantic interpretation is challenging as ESC is extremely susceptible to physical adversarial attacks \cite{sadeghi2019physical} due to the openness of wireless channels and the fragileness of DNNs. 

For DNN, small perturbations  added to the input can mislead the model to make arbitrarily incorrect results, causing significant security concerns for safety-critical tasks such as automatic driving, unmanned aerial vehicle, and smart healthcare. In computer vision and machine learning, there are plenty of works for adversarial training \cite{moosavi2017universal,wong2020learning,maini2020adversarial,madaan2021learning,leino2021globally}. The success of these approaches provides insightful ideas for the robustness of DNN-based semantic communication. However, directly adapting these methods to ESC is challenging as a communication system needs to take the time-varying wireless channels into consideration. Another difference is that these prior effects focus more on producing adversaries based on input data. While in ESC, the threats are more likely to be added to signals in wireless channels rather than the input at the transmitter side. 

Early adversarial robustness methods \cite{bahramali2021robust} proposed for the end-to-end communication system generate universal perturbations \cite{moosavi2017universal} for physical adversarial training. As the bits/symbols are treated equally on these systems, such content-oriented attacks are not applicable to semantic communications. We will also empirically show that these adversarial robustness approaches can be easily broken with crafted semantic attacks (see details in Section \ref{sec:perforance_under_attacks}). A very recent work \cite{hu2022robust} considers adversarial examples for semantic communications. However, this work is limited to the semantic noise injected in input data at the transmitter side, and such data-specific attacks may not be practical in real-world communication scenarios. 
\textit{To the best of our knowledge, the physical-layer adversarial robustness of DNN-based semantic communications, although non-trivial to robust semantic interpretation for safety-critical applications in practice, is still largely underexplored.} 

\subsection{Our Method}
To fill the aforementioned gap, this paper studies semantic adversarial robustness that augments the training procedure with carefully-crafted perturbation signals over the wireless channel. Central to this approach is the generation of physical adversarial distortions, aiming to mislead semantic interpretations at the receiver side. In order to characterize the real-world effect for robust training and evaluation, we consider perturbation signals that attack an ESC system to meet the following criteria: 
\begin{itemize}
    \item \textbf{Semantic-oriented:} The perturbation signals can be tailored to focus on attacking the targeting semantic objects, such as cars and signposts for automatic driving while securing the interpretation of others' semantics. Such semantic-specific adversarial attacks are much more destructive than existing attacks crafted for conventional end-to-end communication systems. Our framework has two optional settings, i.e., reconstructing images based on semantics for the applications, and directly interpreting semantic information for goal-oriented communications.
    \item \textbf{Imperceptibility:} The perturbations added to the signals will be considered as the natural noise in the wireless channel and cannot be detected by the receiver, fooling a semantic communication system into making incorrect decisions. This is practical in real-world cases.
    \item \textbf{Input-agnostic:} In reality, an attacker doesn't have any knowledge of the input data when generating the perturbation signals in the test stage.
    \item \textbf{Controllability:} An attacker may manipulate the communication system by misleading the model to interpret the semantics as expected results. Such attacks may cause severe accidents in safe-critical applications. Defense against such attacks will be non-trivial for semantic communication systems.
\end{itemize}

Keeping the above goals in mind, we first introduce a deep learning-based semantic communication framework \texttt{MobileSC} that considers the memory and computation constraints in a wireless environment. The framework consists of four modules including a semantic encoder, an Orthogonal Frequency Division Multiplexing (OFDM)-transmitter, an OFDM receiver, and a semantic decoder. We additionally develop a classifier to interpret the semantics from the reconstructed images by the semantic decoder. Inspired by MobileNet V2 \cite{sandler2018mobilenetv2}, we introduce SemBlock, which is a more lightweight structure for semantic encoder and decoder. Our SemBlock uses depth-wise separable convolutions to build lightweight DNN \cite{sandler2018mobilenetv2}. Based on this framework, we then present a novel perturbation generator \texttt{SemAdv}, which aims to learn to craft physical adversarial distortions based on a semantic loss. In the following parts, we will outline how our perturbation generation procedure meets the four criteria mentioned above. 

For the semantic-oriented goal, our proposed perturbation generator crafts distortions that aim to attack certain semantics over the air by introducing awards and penalties in an objective function. For imperceptibility, we propose a regularizer to encourage the variation of the generated perturbations, as well as a normalization operation with a power constraint to adjust the distribution of the distortions. For the input-agnostic goal, our generator only relies on a random value without any knowledge of signals. For the last goal, we simply replace the labels in the objective function to mislead the model to make expected decisions with a manipulated semantic interpretation.   

 Based on $\texttt{SemAdv}$, we then introduce an adversarial training method $\texttt{SemMixed}$, a novel AT method that is capable of defending against both $\texttt{SemAdv}$ and PGM attacks, to improve the robustness of our $\texttt{MobileSC}$ against physical-layer adversaries. Our $\texttt{SemMixed}$ is able to mislead classifiers by distorting target semantics. Experiments on three popular benchmarks show the effectiveness of our proposed approaches. Our $\texttt{MobileSC}$ is also compatible with other downstream applications that require an understanding of semantics to take the right actions on the receiver side. 

\subsection{Main Contributions}
The main contributions of this paper are four-fold: 

\begin{itemize}
    \item We present \texttt{MobileSC}, a novel ESC framework for image transmission. The proposed \texttt{MobileSC} takes the computation and memory efficiency in wireless environments into account. Our semantic encoder/decoder SemBlock, the key component of \texttt{MobileSC}, needs nearly 40\% and 64\% fewer parameters respectively than the popular MobileNet 
 V2\cite{sandler2018mobilenetv2} and JSCC \cite{yang2022ofdm}. 
    \item Equipped with \texttt{MobileSC}, we introduce a novel approach \texttt{SemAdv} that aims to learn to generate perturbations for physical-layer adversarial attacks to the semantic communication system. To be practical in real-world scenarios, our attacks meet four criteria including semantic-oriented, imperceptibility, input-agnostic, and controllability. This is the key contribution of this paper.
    \item We propose an adversarial training method $\texttt{SemMixed}$ that is able to harden the interpretation of the semantic communication system against multiple physical adversarial perturbations, including \texttt{SemAdv} attacks, as well as some other destructive attacks such as PGM \cite{bahramali2021robust}. 
    \item We conduct extensive experiments to show the effectiveness of our proposed \texttt{SemAdv} attacks and defense strategy $\texttt{SemMixed}$ for \texttt{MobileSC}. To evaluate the ``controllability'' and ``imperceptibility'' of our semantic attacks, we also introduce a novel performance metric besides traditional accuracy, Peak signal-to-noise ratio (PSNR), and Structural similarity index measure (SSIM). Some insights are also given in discussions and case studies. 
\end{itemize}

\subsection{Related Work}

\subsubsection{Semantic Communication System}
Deep learning has been widely used in wireless communication to
refine the traditional block-structure systems with end-to-end ones \cite{dorner2017deep,aoudia2018end,ye2018channel,ye2018channel,bourtsoulatze2019deep,aoudia2019model,kurka2020deepjscc,ye2020deep,yang2022ofdm}, showing impressive improvement by jointly optimizing the processing blocks. Towards this direction, many deep learning-based semantic communication systems have been proposed, including DeepSC \cite{xie2020deep} \cite{xie2021deep} for textual data,  UDSem\cite{nan2023udsem} for texts and images, DVST for video transmission\cite{wang2022wireless}, SCS\cite{weng2021semantic-jsac}\cite{Han2022SemanticPreservedCS} for speech signals, MU-DeepSC \cite{xie2021task} for question answering applications, and L-DeepSC \cite{xie2020lite} for text and speech over 
IoT. Different from the above previous works, our proposed \texttt{MobileSC} considers the memory and computation constraints of mobile devices for image-based semantic communications. Compared with the most relevant end-to-end communication systems that are based on JSCC \cite{yang2022ofdm} and MobileNet V2 \cite{sandler2018mobilenetv2}, our \texttt{MobileSC} is much more lightweight, with much fewer parameters and processing time. 

\subsubsection{Robustness of Semantic Communication System}
Deep learning models are susceptible to attacks \cite{singh2019beyond}\cite{Li2023BoostingPL}. The idea of attacking deep neural networks was first introduced in \cite{szegedy2013intriguing}. In recent years there have been plenty of works along this line \cite{tramer2019adversarial,sadeghi2019physical,wong2020learning,maini2020adversarial,madaan2021learning,leino2021globally}\cite{Qin2023SecuringSC}, which aim to develop various defense strategies for model robustness.  
Among these works, the most relevant ones to this paper are \cite{hu2022robust} and \cite{bahramali2021robust}. The very recent work \cite{hu2022robust} is limited to injecting semantic noise into input data, which may not be practical over wireless communication systems. PGM \cite{hu2022robust}, which was proposed for content-level communications, generates universal perturbations \cite{moosavi2017universal} for end-to-end systems. The two key differences between our work and the previous ones are: 1) Our proposed attacks consider the cases of semantic communications in real-world scenarios, and the physical adversarial attacks generated by our proposed \texttt{SemAdv} are semantic-oriented, imperceptible, input-agnostic, and controllable. Such attacks are specifically tailored for semantic communication systems to distort the receiver to make incorrect decisions, while the PGM is proposed for content-oriented communications. 2) Our \texttt{SemMixed} is capable of defending against both \texttt{SemAdv} attacks and PGM attacks, while the existing efforts proposed for end-to-end communications only consider a single attack, such as the fast gradient method (FGM) attack \cite{sadeghi2019physical} and the PGM \cite{bahramali2021robust} attack.

\subsubsection{Other Research of Semantic Communications} There is a line of works that apply semantic communications to various scenarios. 
 Semantic communication systems may suffer some privacy issues as it is learned from a large volume of data, and hence privacy-preserving is discussed in the previous work \cite{luo2022encrypted}\cite{tung2022deep}. The authors also rely on semantic communication to facilitate wireless cognition\cite{chen2022Neuromorphic}, energy saving\cite{yang2023energy}, mobile edge computing\cite{huang2023semantic},  image transmission \cite{Zhang2022SemanticCA}\cite{Tang2023ContrastiveLB} and MIMO communication\cite{shi2023excess}. However, these systems don't consider the robustness of semantic communications and hence are fragile to various attacks.

\subsection{Paper Organization and Notations}
The remainder of the paper is organized as follows. Section \ref{sec:system_model} describes our proposed framework for deep learning-based semantic communication systems. Section \ref{sec:semadv_semmixed} presents the proposed perturbation method and the adversarial training approach. Section \ref{sec:experiments} shows the experiments and discusses attacks and defense strategies. Section \ref{sec:discussion} gives some insightful discussions, as well as a case study. Finally, conclusions are drawn in Section \ref{sec:conclusion}. Table \ref{tab:notation} lists the notations used in this paper.




\begin{figure*}[!t]
\centering
\captionsetup{justification=centering}
\includegraphics[width=6.5in]{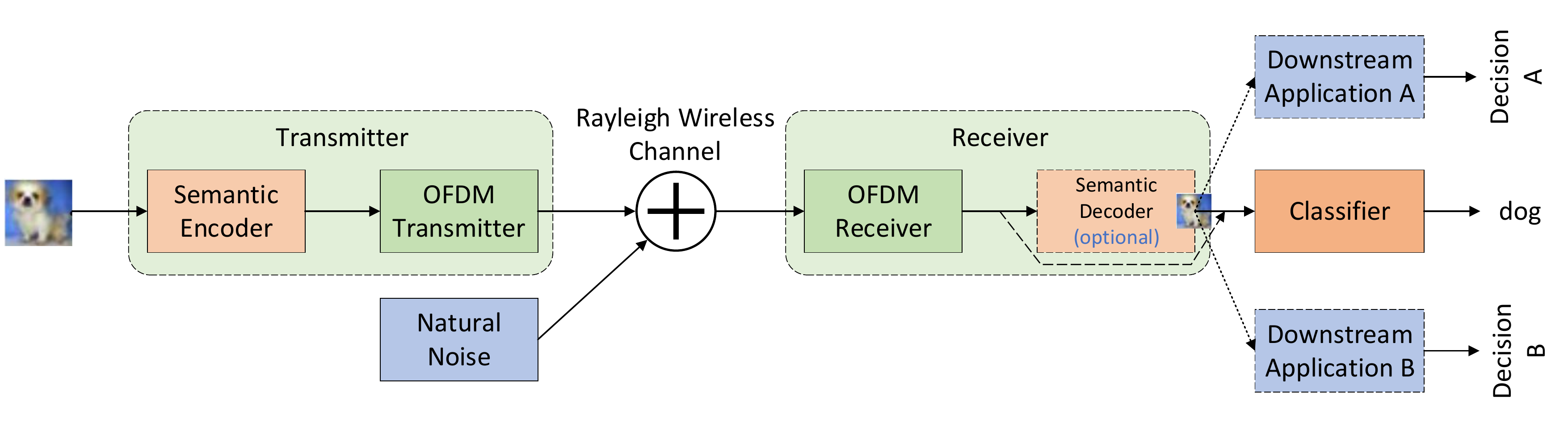}
\caption{System model of our proposed \texttt{MobileSC}.}
\label{fig:arc}
\end{figure*}

\begin{table}
\caption{Definitions of Notations\label{tab:notation}}
\centering
\scalebox{0.9}{
\begin{tabular}{ll}
\toprule
Notation & Definition \\
\midrule
$\mathbf{x}, \mathbf{x}' \in \mathbb{R}^{c \times w \times h}$ & input image, recovered image \\
$c$, $h$, $w$                                & num of channel, height and weight of image $\mathbf{x}$ and $\mathbf{x}'$ \\
$\mathcal{E}_\psi$, $\mathcal{D}_{\pi}$       & semantic encoder, semantic decoder \\
$\mathcal{S}_\theta$, $\mathcal{M}_\omega$                         & MobileSC, image classifier \\
$\mathcal{G}_\eta$, $\mathcal{G}_\varrho^{pgm}$                           & perturbation generator, PGM attacks generator \\
$\mathcal{R}$                                & regularizer of perturbations\\
$\mathcal{L}_{sys}$, $\mathcal{L}_{rec}$, $\mathcal{L}_{cls}$ & system loss, reconstruction loss, classification loss \\
$\lambda_r$, $\lambda_c$                     & weights of reconstruction and classification loss \\
$N_{cp}$, $N_c$ &length of cyclic prefix, num of OFDM subcarriers \\
$N_p$, $N_d$ &num of pilot symbols, num of information symbols \\
$\mathbf{X} \in \mathbb{R}^d$                                 & contextualized representation of the image $\mathbf{x}$ \\
$\mathbf{\widetilde{X}} \in \mathbb{R}^{(N_p + N_d) \times N_c \times 2}$                                 & complex OFDM symbols over the air \\
$\mathbf{H}$, $\mathbf{H}_{es}$ & channel gain, channel estimation \\
$\mathbf{X}_r \in \mathcal{R}^d$ & received latent representations of the image $\mathbf{x}$\\
$\Delta_{trg}$, $\Delta_{ino}$  & target set and innocent 
set in a mini-batch \\
\bottomrule  
\end{tabular}}
\end{table}

\section{System Model}
\label{sec:system_model}
Figure \ref{fig:arc} shows the architecture of our proposed semantic communication framework \texttt{MobileSC}, which consists of five modules including a semantic encoder, an OFDM transmitter, an OFDM receiver, an optional semantic decoder, and a classifier. Next, we detail how each module works.

\subsection{Semantic Encoder} 
Recently proposed semantic neural encoders \cite{yang2022ofdm,xie2020deep}, which are based on Transformer \cite{vaswani2017attention} and ResNet \cite{he2016identity}, require high computational resources beyond the capabilities of mobile devices, which may not be practical in real-world scenarios. To bridge this gap, the goal of our semantic encoder is to generate symbols for input data in a timely and memory-efficient fashion. We propose a novel semantic encoder that is specifically tailored for resource-constrained wireless environments. The two key ingredients are listed as follows. 
\begin{itemize}
    \item  Our encoder adapts DWConvolution block in MobileNet V2\cite{sandler2018mobilenetv2} and relies on SemBlock for non-linear transformation and lightweight depthwise convolution. The proposed structure simplifies inverted residual structure \cite{sandler2018mobilenetv2} and uses shortcuts connections between thin bottleneck layers to improve the expressiveness of the feature representations, where the thin bottleneck layers indicate the ones with a much less number of channels. By doing so, our encoder is able to generate semantic representations for the input data with much fewer parameters.
    \item Our semantic encoder jointly considers the source encoding and channel encoding in the above single neural networks to reduce the network complexity and facilitate the end-to-end optimization during the training procedure.
\end{itemize}

We show the formulations as follows. We say $\mathbf{x} \in \mathbb{R}^{c \times h \times w}$ is an input image, where $c$, $h$ and $w$ are the numbers of channels, height, and width of the image, respectively. We feed image $\mathbf{x}$ to our semantic encoder $\mathcal{E}_\phi$ to obtain the contextualized semantic representations $\mathbf{X} \in \mathbb{R}^{d}$, where $\phi$ and $d$ are the trainable parameters of the encoder and the dimension of the representation, respectively
Our semantic encoder consists of multiple convolution layers and SemBlocks. The number of these sub-modules can be flexibly configured based on the customized requirements of various wireless applications, as well as the capacities of different mobile devices. We give the detailed implementations in Section \ref{implementation_details}.  

\subsection{OFDM Transmitter}
\label{ofdm_trans}
For the input image $\mathbf{x}$, so far we have  symbols $\mathbf{X}$ generated by the semantic encoder. To make efficient use of the spectrum and reduce the computational overhead, we employ OFDM \cite{weinstein2009history} as our wireless transmission scheme, which is able to encode $\mathbf{X}$ into multiple carrier frequencies without inter-symbol interference (ISI). Noted the transmitter also supports the single-carrier OFDM mode with configurable parameters. We reshape the symbols $\mathbf{X} \in \mathbb{R}^{d}$ to $\mathbf{X} \in \mathbb{R}^{d/2 \times 2}$ to facilitate the following computations that are based on the complex symbol stream. 

\subsubsection{Performing IFFT and Adding CP} Specifically, we first perform Inverse Fast Fourier Transform (IFFT) on $\mathbf{X}$.
Then we add a cyclic prefix (CP) $\mathbf{X}_{cp} \in \mathbb{R}^{N_{cp} \times 2}$ to $\mathbf{X}$ to secure the reliability of OFDM signal and overcome the negative impact of ISI, i.e.,
\begin{equation}
    \mathbf{\hat{X}} = [{\rm IFFT}(\mathbf{X});\mathbf{X}_{cp}],
\end{equation}
where the dimension of the updated $\mathbf{\hat{X}}$ is $(d/2 + N_{cp}) \times 2$ and $N_{cp}$ indicates the number of CP that is truncated from $\mathbf{X}$. Generally, we select the last $N_{cp}$ symbols out of $d/2$ ones in $\mathbf{X} \in \mathbb{R}^{d/2 \times 2}$. 

\subsubsection{Clipping Signal} One of the disadvantages of OFDM is the high peak-to-average power ratio (PAPR), which is caused by the linear combination of Quadrature Amplitude Modulation (QAM) symbols in the IFFT operation, leading to excessive power assumption at the power amplifier \cite{ochiai2002performance}. To mitigate this issue, we refer to previous signal clipping techniques \cite{ochiai2002performance, yang2022ofdm} to reduce PAPR by introducing an additional non-linear activation function to the time domain OFDM signal $\mathbf{\hat{X}}$. 

\subsubsection{Adding Pilot}
Pilot signals in OFDM can be used for measurement of the channel conditions, describing how a signal propagates from the transmitter to the receiver such as  fading and power decay. Such a way makes it possible to adapt transmissions to current channel conditions. We say $N_c$ is the number of subcarriers, and then data symbols can be derived as $N_d = \lceil \frac{(d/2 + N_{cp})}{N_c} \rceil$. Hence output symbols $\mathbf{\hat{X}} \in \mathbb{R}^{(d/2 + N_{cp}) \times 2}$ can be reshaped as  $\mathbf{\hat{X}} \in \mathbb{R}^{N_d \times N_c \times 2}$. We denote the number of pilot symbols as $N_p$. The pilot symbols can be transmitted on a part of the OFDM subcarriers of all subcarriers. We choose the latter for simplicity and denote the pilot symbols as $\mathbf{X_{pt}} \in \mathbb{R}^{N_c \times 2}$. The output complex stream of the OFDM transmitter can be expressed as $\mathbf{\widetilde{X}} \in \mathbb{R}^{(N_p + N_d) \times N_c \times 2}$. Although pilot-assisted channel estimation benefits reliable wireless transmission, the main drawback lies in the reduction of the transmission rate when inserting a large number of pilot symbols. We need to carefully select $N_p$ for a better trade-off between the cost and benefit. 

In briefly, for the input symbols $\mathbf{X} \in \mathbb{R}^{d}$, the output $\mathbf{\widetilde{X}} \in \mathbb{R}^{(N_p + N_d) \times N_c \times 2}$ of OFDM transmitter ${\rm OFDM_T}$ can be expressed as

\begin{equation}
    \mathbf{\widetilde{X}} = {\rm OFDM_T}(\mathbf{X}).
\end{equation}

\subsection{Wireless Channel} 
Without loss of generality, for $\mathbf{{\widetilde{X}}}$, we denote the received signal at the receiver as $\mathbf{Y} \in \mathbb{R}^{(N_p + N_d) \times N_c \times 2}$, which can be expressed as
\begin{equation}
    \mathbf{Y} = \mathbf{H} \mathbf{\widetilde{X}} + \mathbf{N},
\end{equation}
where $\mathbf{H}$ refers to channel gain and $\mathbf{N} \sim \mathcal{CN}(0, \sigma^2_n)$ represents the additive Gaussian noise. As discussed in previous works \cite{ye2018channel,bourtsoulatze2019deep,kurka2020deepjscc}, deep neural networks are able to model the wireless physical channels, including additive white Gaussian noise (AWGN), the erasure channel, and the Rayleigh fading channel. In this paper, we mainly consider the Rayleigh fading channel as it better models the effect of a propagation environment for semantic communication systems. We use Rayleigh fading channel $\mathbf{H} \sim \mathcal{CN}(0, \sigma^2)$. 

\subsection{OFDM Receiver}
The OFDM receiver takes received symbols $\mathbf{Y}$ as the input and performs inverse operations that have been done in the OFDM transmitter. As we have already detailed the intuition of each module in the transmitter, here we only briefly outline the procedure. We first separate pilot symbols $\mathbf{X}'_{pt} \in \mathbb{R}^{N_c \times 2}$ and information symbols $\mathbf{\hat{X}}' \in \mathbb{R}^{N_d \times N_c \times 2}$ from $\mathbf{Y}$. Then we remove the cyclic prefix for both symbols and perform Fast Fourier Transform (FFT) to obtain $\mathbf{X}{''}_{pt} \in \mathbb{R}^{N_c \times 2}$ and $\mathbf{\hat{X}}{''} \in \mathbb{R}^{d/2 \times 2}$. We follow the previous work \cite{yang2022ofdm} to perform the channel estimation and equalization by exploring channel state information based on $\mathbf{X}{''}_{pt}$ and $\mathbf{\hat{X}}{''}$, i.e., 
\begin{equation}
    \mathbf{X}_r = {\rm EQ}( {\rm CE}(\mathbf{X}{''}_{pt}), \mathbf{\hat{X}}{''}),
\end{equation}
where ${\rm CE}$ and ${\rm EQ}$ represent channel estimation and equalization, respectively, and $\mathbf{X}_r \in \mathbb{R}^{d}$ indicates the received symbols. In briefly, our OFDM receiver $\rm OFDM_r$ can be formulated as

\begin{equation}
    \mathbf{X}_{r} = {\rm OFDM_r} (\mathbf{Y}).
\end{equation}

\subsection{Semantic Decoder}
Here we detail how the receiver reconstructs the image with the decoder. Noted that such a decoder is optional in our framework. Now we have received symbols $\mathbf{X}_r$, which can be considered as the latent representations of image $\mathbf{x}$ after transmission. To build small and efficient neural networks at the receiver side, we mainly use SemBlocks and the convolution layer in our semantic decoder $\mathcal{D}_{\pi}$, where $\pi$ denotes the trainable parameters of the decoder. The decoder jointly considers the channel and source decoding. We feed semantic representation $\mathbf{X}_r$ to $\mathcal{D}_{\pi}$ to reconstruct image $\mathbf{x}' \in \mathbb{R}^{c \times h \times w}$, which can be expressed as
\begin{equation}
    \mathbf{x}' = \mathcal{D}_{\pi}(\mathbf{X}_r).
\end{equation}

Our semantic decoder can also be flexibly configured with different numbers of the SemBlock and convolution layers to meet mobile and resource-constrained environments. Detailed implementations are available in Section \ref{implementation_details}.

Based on the recovered $\mathbf{x}'$, we are able to measure the semantic loss with widely used image quality metrics including PSNR and SSIM. For some popular downstream applications in the real world, such as image recognition and classification, we need to further evaluate semantic interpretations based on a classification loss with trainable neural networks. In the following, we will show such a classification module. 
\subsection{Classifier}
So far we have got the reconstructed image $\mathbf{x}'$ from the semantic decoder. We consider real cases that need to interpret the semantics by categorizing the image into a class type, such as dog, car, and airplane. For example, the remote operating system that supports automatic driving cars\cite{Li2021PrivacyPreservedFL} uses the recovered image to make decisions. Compared with PSNR and SSIM which focus more on visual quality and image quality, measuring the semantic loss by recognition/classification loss for the above applications is non-trivial for the practical deployment of the communication system. We introduce a simple classifier to predict the category of $\mathbf{x}'$ based on the semantic interpretation. Here we use ResNet-preact \cite{he2016identity} to compute the probability for each class type $t$, i.e.,
\begin{equation}
    p(t|\mathbf{x}') = \mathcal{M}_{\omega}(\mathbf{x}'),
\end{equation}
where $\omega$ is the trainable parameters of the classifier. The prediction class $t^{*}$ of the image $x'$ can be expressed as 
\begin{equation}
    t^{*} = \underset{t}{\mathrm{argmax}} \quad p(t|\mathbf{x}').
\end{equation}

While the model is specifically designed for recognition and classification applications, our conceptual 
ideas underlying the design are general for semantic communications. As illustrated in Figure \ref{fig:arc}, our \texttt{MobileSC} is compatible with other applications, such as downstream applications A and B, that require accurate semantic interpretations to make the right inference or to take the right actions at the receiver side.

\subsection{Objective Function}
To train our semantic communication system, we consider a multi-task learning procedure composed of a reconstruction loss and a semantic loss. The reconstruction loss $\mathcal{L}_{rec}$ is expressed by the cross-entropy between the recovered image $x'$ and the original one $x$, and the semantic loss is described as the classification loss  $\mathcal{L}_{cls}$. The total loss $\mathcal{L}_{sys}$ can be formulated as the weighted sum of the above two losses, which can be expressed as
\begin{equation}
    \mathcal{L}_{sys} = \lambda_r \mathcal{L}_{rec} + \lambda_c \mathcal{L}_{cls},
\end{equation}
where $\lambda_r$ and $ \lambda_c$ are two weights to configure the importance of each loss. 
\begin{figure}[!t]
\centering
\includegraphics[width=3.5in]{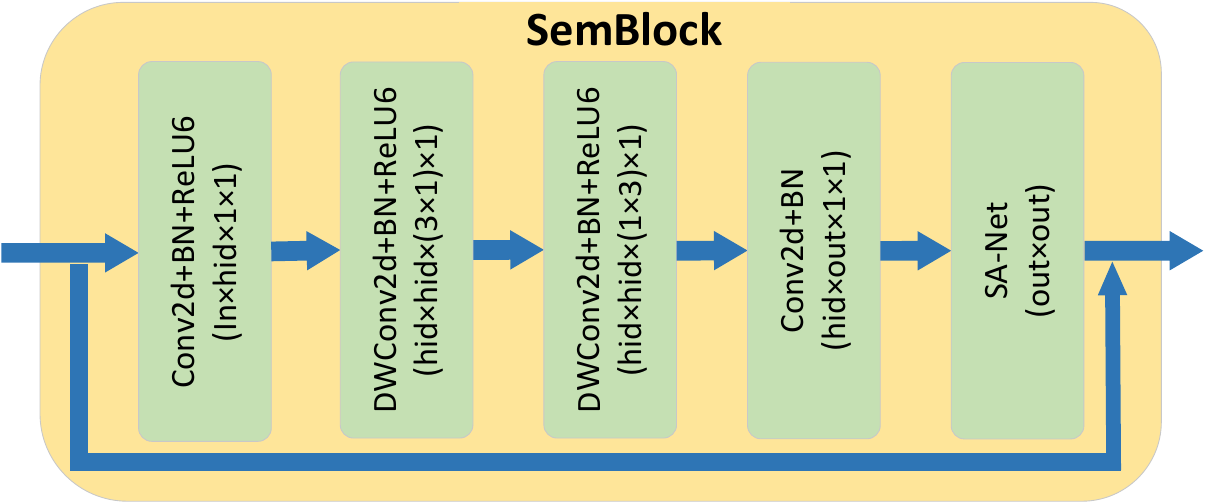}
\caption{Structure of our proposed \texttt{SemBlock}.}
\label{fig:semblock}
\end{figure}
\subsection{\texttt{SemBlock}}
Our semantic encoder/decoder is comprised of convolution layers and SemBlocks. As shown in figure \ref{fig:semblock}, to reduce the neural parameters, our SemBlock splits the $3$x$3$ convolution kernel to $1$x$3$ and $3$x$1$ and then adds an additional attention layer and an SANET structure \cite{fan2017sanet} to improve the model's capability in capturing the meaningful semantics. Each block contains a much less number of parameters than popular ResNet \cite{he2016identity}, taking account of the computation and energy efficiency in wireless environments, and thus can be practically deployed on mobile devices.
\section{Semantic Adversarial Robustness}
\label{sec:semadv_semmixed}
This Section shows how we generate physical-layer adversarial attacks for the ESC and how we harden the system against the attacks. Specifically, we first outline the high-level design of our semantic adversarial attack in Section \ref{sec:overview_attacks}, and then Section \ref{sec:attack_gen} delves into the details of the proposed perturbation generator \texttt{SemAdv} that aims to craft semantic adversarial distortions over the air. Our \texttt{SemAdv} attacks are semantic-oriented, imperceptible, input-agnostic, and controllable. Section \ref{sec:overview_at} shows the general idea of adversarial training, and finally Section \ref{sec:overview_semmixed} presents our adversarial training method \texttt{SemMixed} 
that is able to harden the semantic communication system against multiple physical adversarial perturbations. 

\begin{figure*}[!t]
\centering
\includegraphics[width=6.5in]{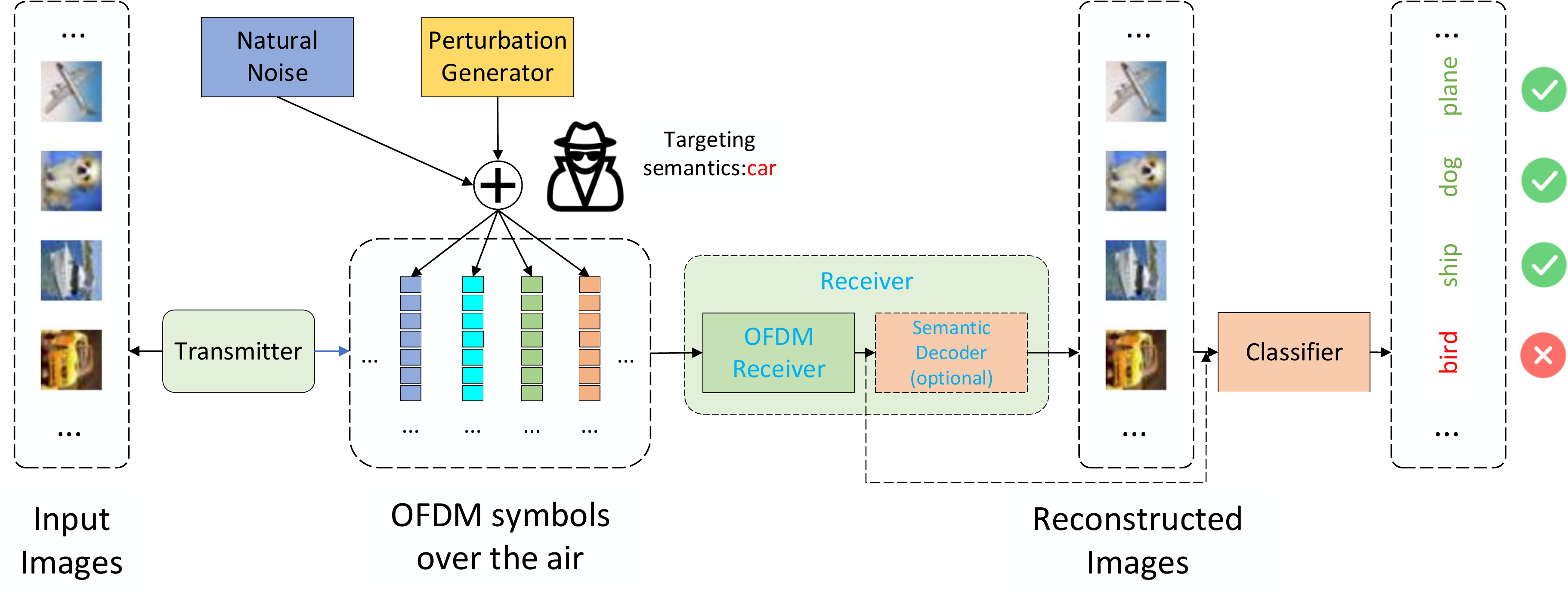}
\caption{Physical adversarial attacks to \texttt{MobileSC}. The attacker aims to deceive the receiver to recognize a ``car'' as an incorrect type of ``bird'' by generating adversarial perturbations that can distort the semantics of ``car'' in signals over the air.}
\label{fig:semattack}
\end{figure*}

\subsection{Overview of Physical-layer Adversarial Attacks}
\label{sec:overview_attacks}
Figure \ref{fig:semattack} demonstrates how we train a perturbation generator to produce adversarial attacks that meet the four criteria mentioned at the beginning. Our goal here is to attack \texttt{MobileSC} over a wireless channel to fool the classifier to make incorrect decisions. Such a procedure is described as follows. 
\begin{itemize}
    \item The transmitter sends OFDM symbols that contain semantics of the images to a wireless channel, where each image contains one class such as airplane, dog, ship, and car, as shown in Figure \ref{fig:semattack}.
    \item Then the additive white Gaussian noise  and adversarial perturbations are added to the signal, where the former can be considered as natural noise and the latter perturbations aim to attack the semantics that is related to cars. To be more practical, the perturbations produced by the generator are input-agnostic.
    \item The receiver directly feeds signals to the classifier or reconstructs all images and then feeds them to the classifier to output the predictions based on semantic information. Since we have injected semantic-oriented (car) adversarial perturbations into the signals, the recovered image with targeting semantics will deceive the system to make incorrect predictions. As demonstrated in the bottom right of Figure \ref{fig:semattack}, the image of a ``car'' may be misclassified as the ``bird'' category under the adversarial attacks. Such prediction results may be pipelined to systems such as automobiles and smart health, which may cause serious accidents for these security-crucial applications.
\end{itemize}
Next, we will go into the detailed procedure for the generation of the above adversarial perturbations. 

\begin{figure}[!t]
\centering
\includegraphics[width=3.5in]{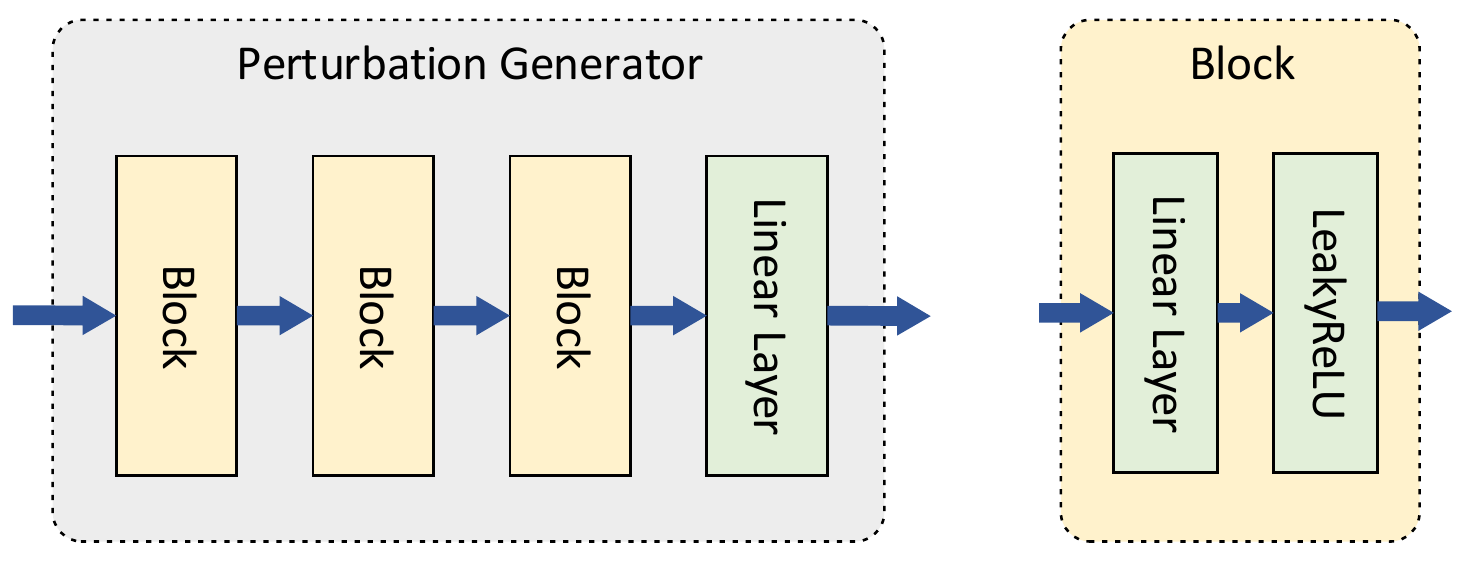}
\caption{Our proposed perturbation generator \texttt{SemAdv}.}
\label{fig:generator}
\end{figure}

Generally, the physical adversarial perturbations $\delta$ for the input can be derived by the following optimization problem: 
\begin{equation}
\begin{aligned}
    & \underset{\delta}{\mathrm{argmin}}\, \Vert \delta \Vert_2 \\
    &s.t. \quad \forall \mathbf{x}: \mathcal{S}_\theta(\mathbf{x}, \delta) \neq \mathcal{S}_\theta(\mathbf{x}),
    \label{eq:general_opt}
\end{aligned}
\end{equation}

The goal of the above formulation is to generate  destructive adversarial perturbations that can mislead the semantic communication systems to make incorrect interpretations, and the constraint indicates that such perturbations should be imperceptible so that they can not be easily detected, meeting the criteria that we discussed in the Section I.C.

However, directly solving (\ref{eq:general_opt}) is very challenging as the structure of \texttt{MobileSC} is not convex. A common approach is to employ Gradient-based approximations to obtain $\delta$, such as the fast gradient method (FGM) \cite{sadeghi2019physical} and projected gradient descent (PGD) \cite{tramer2019adversarial}. One drawback of such approaches is the high computational cost for training and inference. Another main issue is that the formulation in  (\ref{eq:general_opt}) aims to attack a content-oriented wireless network and does not take the semantic attack into consideration, where the semantics rather than content are used by the receiver to make the right inference or to take the right actions. Therefore, we introduce our proposed attack generation method for semantic communications as follows.

\subsection{Attacks Generation}
\label{sec:attack_gen}
Figure \ref{fig:generator} presents the architecture of our adversarial perturbation generator, which is an MLP neural network with multiple computation blocks. Each block consists of a linear layer and LeakyReLU \cite{xu2015empirical} for nonlinear transformation. Algorithm \ref{alg:attack} shows the process of perturbation generation in two steps. 
\begin{enumerate}
    \item We train our deep learning-based \texttt{MobileSC} system $\mathcal{S}_\theta$ on the dataset $\Omega$ in an end-to-end manner, where $\theta$ indicates the learnable parameters of the system. Then we save the model to facilitate the training of our perturbation generator in the next step.
    \item Our perturbation generator learns to produce  adversarial distortions for symbols over the air by attacking the pre-trained \texttt{MobileSC}.
\end{enumerate}

As we discussed at the beginning, the physical adversarial attacks crafted by our \texttt{SemAdv} are expected to meet four criteria including  semantic-oriented, imperceptibility, input-agnostic, and controllability. Next, we show how our approach achieves the above goals.  

\begin{algorithm}
	\caption{The generation process of physical adversarial attacks.}\label{alg:attack}
	\renewcommand{\algorithmicrequire}{\textbf{Input:}}
	\renewcommand{\algorithmicensure}{\textbf{Output:}}
	\begin{algorithmic}
		\STATE \textbf{Input:} Pre-trained \texttt{MobileSC} ${\mathcal{S}}_{\theta}$ \\
		\STATE \hspace{1.1cm}Training Dataset $\Omega$ \\
		\STATE \hspace{1.1cm}Distance Regularization Function $\mathcal{R}(\cdot)$
		\STATE \textbf{Output:} Semantic Perturbation Generator $\mathcal{G}_\eta(\cdot)$
		\STATE \textbf{for} $t \leftarrow 1$ to $T$ \textbf{do}
		\STATE \hspace{0.5cm}\textbf{for each} mini-batch $\Omega_i$ in $\Omega$ \textbf{do}
		\STATE \hspace{1cm}\textbf{for each} $sample_{ij}$ in $\Omega_i$ \textbf{do}
		\STATE \hspace{1.5cm}\textbf{if} $sample_{ij}.label$ == $targetlabel$ \textbf{then}
		\STATE \hspace{2cm}$sample_{ij}.label$ $\leftarrow$ random other label
		\STATE \hspace{1.5cm}\textbf{end}
		\STATE \hspace{1cm}\textbf{end}
		\STATE \hspace{1cm}$\mathcal{L} = \sum_{\mathbf{x}_j \in \Omega_i} \mathcal{L}_{cls}(\mathcal{S}_\theta(\mathbf{x}_j,\mathcal{G}_\eta(z)))-\mathcal{R}(\mathcal{G}_\eta(\mathbf{z}))$
		\STATE \hspace{1cm}Update $\mathcal{G}_\eta$ to minimize $\mathcal{L}$
		\STATE \hspace{0.5cm}\textbf{end}
		\STATE \textbf{end}
		\STATE \textbf{return} $\mathcal{G}_\eta$
	\end{algorithmic}
\end{algorithm}

We quantify the complexity of the Algorithm. \ref{alg:attack} in terms of complexity and space complexity. The time complexity of Algorithm \ref{alg:attack} can be expressed as $O(n^2)$. We denote the number of samples as $N$ and the number of mini-batches as $i$. Hence, the number of images in each mini-batch can be expressed as $N/i$. The complexity of replacing labels can be expressed as $T*(N/i)*i$, i.e., $T*N$. Hence, the time complexity of the algorithm can be expressed as:
\begin{equation}
    T(n)=O(n^2),
\end{equation}

For space complexity, we use arrays to store labels of all $N$ images in the dataset. It needs $T*N$ storage units for $T$ epochs. Hence, the space complexity can be expressed as:

\begin{equation}
    S(n)=O(n^2),
\end{equation}

\subsubsection{Semantic-oriented} We categorize training examples in a mini-batch into two sets based on their semantics, i.e., the targeting set and the innocent set. We denote the indices of all $b$ examples in two sets as $\Delta_{trg}$ and $\Delta_{ino}$, respectively. Our aim is to attack the examples in $\Delta_{trg}$ to mislead the system to make incorrect semantic interpretations, while at the same time securing the interpretations of the innocent ones in $\Delta_{ino}$ under the same attacks. We denote our training examples in a mini-batch as $\{\mathbf{x}_i, y_i\}_{i \in \Delta_{trg}}$ and $\{\mathbf{x}_j, y_j\}_{j \in \Delta_{ino}}$. We say $\mathcal{G}_\eta(z_i)$ is our proposed perturbation generator, where $\eta$ is trainable neural parameters and $z_i, z_j \sim \mathcal{U} (0,1)$ refers to two random seeds that follow uniform distributions. The optimization problem for generating the adversarial perturbations for the \texttt{MobileSC} can be expressed as
\begin{equation}
    \begin{aligned}
    \underset{\mathcal{G}_\eta(\mathbf{z}) \subset \Lambda_{p,\epsilon}} {\mathrm{maximize}} 
     \{ & \sum_{i \in \Delta_{trg},j \in \Delta_{ino}} \{ \mathcal{L}_{trg}(\mathcal{S}_\theta(\mathbf{x}_i,\mathcal{G}_\eta(z_i)),y_i)
    \\& - \mathcal{L}_{ino}(\mathcal{S}_\theta(\mathbf{x}_j,\mathcal{G}_\eta(z_j)), y_j) 
     - \mathcal{R}(\mathcal{G}_\eta(\mathbf{z}))\}\}, 
    \end{aligned}
    \label{eq:perturabtion_gen}
\end{equation}
where $\mathbf{z} = \{z_i: i \in [1, b]\}$, $\mathcal{L}_{trg}$ and $\mathcal{L}_{ino}$ refer to the losses for the examples with target semantics and the ones for the innocent examples, and $\Lambda_{p,\epsilon}$ is an $\ell_p$ ball around the unperturbed example, defined as $\Lambda_{p,\epsilon} = \{\mathcal{G}_\eta(\mathbf{x}_i): \forall i, \Vert \mathcal{G}_\eta(\mathbf{x}_i) \Vert \leq \epsilon \}$ for certain norm $p$ and radius $\epsilon$. For simplicity, we replace the label in $\Delta_{ino}$ as a random category $y'_i \in \mathbf{C} \backslash y_i$, where $\mathbf{C}$ refers to the set of pre-defined class labels. Then (\ref{eq:perturabtion_gen}) can be reformulated as
\begin{equation}
    \begin{aligned}
    \underset{\mathcal{G}_\eta(\mathbf{z}) \subset \Lambda_{p,\epsilon}} {\mathrm{maximize}} 
     \{ & \sum_{i \in \Delta_{trg},j \in \Delta_{ino}} \{ \mathcal{L}_{trg}(\mathcal{S}_\theta(\mathbf{x}_i,\mathcal{G}_\eta(z_i)),y_i)
    \\& - \mathcal{L}_{ino}(\mathcal{S}_\theta(\mathbf{x}_j,\mathcal{G}_\eta(z_j)), y'_j) 
     - \mathcal{R}(\mathcal{G}_\eta(\mathbf{z}))\}\}.
    \end{aligned}
    \label{eq:perturabtion_gen_new}
\end{equation}
Such an objective function awards the generator to craft perturbations that can only distort certain semantics in signals.    
\subsubsection{Imperceptibility} For imperceptibility, we introduce a regularizer $\mathcal{R}(\cdot)$ to encourage the variation of generated perturbations, making it hard for detection at the receiver side. The regularizer $\mathcal{R}(\cdot)$ can be formulated as follows.

\begin{equation}
\begin{aligned}
    \mathcal{R}(\mathcal{G}_\eta(\mathbf{z})) = \frac{\sum_{i \in [1,k],j \in [i+1,k]} \Vert \mathcal{G}_\eta(z_i)-\mathcal{G}_\eta(z_j) \Vert}{\frac{1}{2}k(k-1)}. 
\end{aligned}
\end{equation}
The above formulation is able to calculate the average distance of generated perturbations. Our generation process also preserves the power constraints with an upper bound $p_{max}$ to further improve the imperceptibility of the perturbations. We adjust the perturbations by normalizing every value in the vector to a Gaussian distribution if the power is larger than $p_{max}$, such that the mean of all of the values is $0$ and the standard deviation is $1$. We show the formulations as follows.
\begin{equation}
\mathcal{G}_\eta(z_i)=\left\{
\begin{aligned}
& \mathcal{G}_\eta(z_i) & \Vert \mathcal{G}_\eta(z_i) \Vert^2_2 \leq p_{max},  \\
& \sqrt{p_{max}}\frac{\mathcal{G}_\eta(z_i)-\overline{ \mathcal{G}_\eta(\mathbf{z})}}{\sigma(p)} &  \Vert \mathcal{G}_\eta(z_i) \Vert^2_2 > p_{max},
\end{aligned}
\right.
\end{equation}
where $z_i \in \mathbf{z}$. Equipped with the above operations, our perturbation can be more imperceptible and can be disguised as natural noise. 

\subsubsection{Input-agnostic} During the training phase, our generator $\mathcal{G}_\eta$ learns to craft the adversarial distortions by attacking pre-trained \texttt{MobileSC} model. The input of the generator is a random value $z_i \in \mathbf{z}$ that follows a uniform distribution. During the testing stage, the attacker generates perturbations to distort targeting semantics without any knowledge of the input data and signals over the air. 

\subsubsection{Controllability} Our generator can also manipulate the receiver to take some actions by controlling the semantic interpretation. To achieve this goal, we can simply replace the labels in $\Delta_{trg}$ in  (\ref{eq:perturabtion_gen_new}) as the given one, i.e., $y_0$, so that the generator can learn to craft the adversarial examples that can mislead target semantics to $y_0$. In other words, the attacker can control the semantic communication system to make expected decisions, where the triggers are the targeting semantics.   

We have described how we attack \texttt{MobileSC} by learning to generate adversarial examples. Next, we show how our proposed adversarial training approach defends against such physical-layer semantic attacks, as well as strong adversarial attacks that are crafted for end-to-end communication systems.

\begin{algorithm}[H]
	\caption{Our proposed adversarial training method \texttt{SemMixed}.}\label{alg:defense}
	\renewcommand{\algorithmicrequire}{\textbf{Input:}}
	\renewcommand{\algorithmicensure}{\textbf{Output:}}
	\begin{algorithmic}
		\STATE \textbf{Input:} Training Dataset $\Omega$
		\STATE \hspace{1.0cm} Two attackers our $\mathcal{G}_\eta$ and PGM $\mathcal{G}^{pgm}_\varrho$
		\STATE \hspace{1.0cm} Perturbation set $\mathcal{P}$ 
		\STATE \textbf{Output:} Robust semantic communication system $\mathcal{S}^*_\theta$
		\STATE \textbf{Initialize:} mini-batch size $\gets $ $d$
		\STATE \hspace{1.5cm}$\mathcal{P} \gets $ $b$ zero vectors 
		\STATE \hspace{1.5cm}$\mathcal{P}_0 \gets $ $b$ zero vectors 
		\STATE \textbf{for} $epoch \gets 1$ to $T$ \textbf{do}
		\STATE \hspace{1cm}\textbf{for each} mini-batch $\Omega_i$ in $\Omega$ \textbf{do}
		\STATE \hspace{1.5cm}\textbf{if} i\%2 == 0 \textbf{then}
		\STATE \hspace{2cm}$\mathcal{P}$ $ \gets $ $\mathcal{P}_0$
		\STATE \hspace{1.5cm}\textbf{else}
		\STATE \hspace{2cm}$\mathcal{P}$ $ \gets $ $\mathcal{G}_\eta(\Omega_i)$ or $\mathcal{G}^{pgm}_\varrho(\Omega_i)$
		\STATE \hspace{1.5cm}\textbf{end}
		\STATE \hspace{1.5cm}$\mathcal{L} = 
		\sum_{\mathbf{x}_j \in \Omega_i} \mathcal{L}_{sys}(\mathcal{S}_{\theta}(\mathbf{x}_j,\mathcal{P}),y_j)$
		\STATE \hspace{1.5cm}Update $\theta$ to minimize $\mathcal{L}$
		\STATE \hspace{1cm}\textbf{end}
		\STATE \textbf{end}
	\end{algorithmic}
\end{algorithm}


\subsection{Overview of Adversarial Training}
\label{sec:overview_at}
Adversarial training (AT) is a popular defense strategy against attacks by augmenting the training data with adversarial examples. Without loss of generality, the adversarial training for our mobile system $\mathcal{S}_\theta$ can be formulated as minimizing the worst-case loss with $\ell_p$ norm-bounded perturbations with radius $\epsilon$. Such a formulation can be expressed as
\begin{equation}
    \min \limits_{\theta} \sum \limits_{i}^n \max \limits_{\delta \in \Lambda_{p,\epsilon}} \mathcal{L}_{sys}(\mathcal{S}_\theta(\mathbf{x}_i, \delta), y_i),
\end{equation}
where $\Lambda_{p,\epsilon} = \{\delta: \Vert \delta \Vert_{p}\}$ is the $\ell_p$ norm ball with the radius $\epsilon$ centered around the symbols $\mathbf{\widetilde{X}}$, representing the perturbation magnitude. The above training procedure is capable of assuring the system's worse-case performance against possible adversaries. 
The inner optimization hopes the model can be spoofed by the optimal perturbation $\delta$. The typical solutions such as Fast Gradient Method (FSM) \cite{sadeghi2019physical} and PGD are computationally expensive and we have shown our solution in the (\ref{eq:perturabtion_gen}) of Section \ref{sec:attack_gen}. The outer minimization step learns to update the neural parameters $\theta$ under the adversarial loss constructed at the maximization step. We will discuss the details of our proposed AT method in the following parts. 

\subsection{Mixed Adversarial Training}
\label{sec:overview_semmixed}
The goal of our proposed AT is to harden \texttt{MobileSC} against the aforementioned physical-layer attacks that aim to mislead the classifier by distorting the targeting semantics. Additionally, our method is also expected to be robust under the threats of the state-of-the-art attacker. Towards this goal, we introduce \texttt{SemMixed}, a novel AT method that is able to defend against both SemAdv and PGM attacks simultaneously. 

Algorithm \ref{alg:defense} shows the procedure of our proposed AT approach. Each step is explained as follows.
\begin{enumerate}
    \item We first train our \texttt{MobileSC} $\mathcal{S}_\theta$ without adversarial examples, as we empirically observed that AT from scratch with strong adversarial perturbations may not benefit the system robustness. Furthermore, learning from scratch is also time-consuming. To generate PGM attacks for AT in the following steps, we also pre-trained a PGM generator $\mathcal{G}^{pgm}_\varrho$ \cite{bahramali2021robust}.
    \item Equipped with pre-trained $\mathcal{S}_\theta$ and $\mathcal{G}^{pgm}_\varrho$, we tune $\mathcal{S}_\theta$ with \texttt{SemAdv} attacks and PGM attacks for each min-batch. The two attacks are randomly selected during the training procedure. More specifically, our \texttt{SemAdv} attempts to craft physical adversarial perturbations for targeting semantics in each mini-batch $\Omega_i \in \Omega$. During AT, the targeting semantics are randomly selected in different mini-batch, such that semantics for each class type can be properly protected during the evaluation stage when the system is under different attacks. The PGM attacks, which are originally proposed for content-level communications, are added to signal symbols $\widetilde{\mathbf{X}}$ to fool the system without taking semantics into consideration. 
    \item To assure the model's robustness as well as accuracy, we also use half of the training instances for clean training to calibrate our $\mathcal{S}_\theta$ during AT, as previous studies reveal that there exist trade-offs between robustness and accuracy \cite{tsipras2018robustness}. Here we use $\mathcal{P}_0$ to denote the zero vector, and hence there is no distortion if we add $\mathcal{P}_0$ to the signal. By doing so, we are able to carry out cleaning training equipped with $\mathcal{P}_0$.
\end{enumerate}

\section{Experiments and Numerical Results}
\label{sec:experiments}
\subsection{Datasets}
We conduct experiments on MNIST, CIFAR10 \cite{hendrycks2019benchmarking} and ImageNet, three popular datasets that are widely used for training various image processing systems. The MNIST consists of handwritten digits and contains $60,000$ $28$x$28$ pixel grayscale training images and $10,000$ testing images in $10$ classes. CIFAR10 includes $60,000$ $32$x$32$-pixel color images in $10$ classes, with $6,000$ images per class. The training and testing image of CIFAR10 are $50,000$ and $10,000$ respectively. ImageNet contains $27,0000$ $256$x$256$ pixel color images in $10$ types, with $1,000$ images per class. We have $26,000$ and $10,000$ instances for training and testing, respectively.

\begin{figure}[!t]
\centering
\includegraphics[width=3.5in]{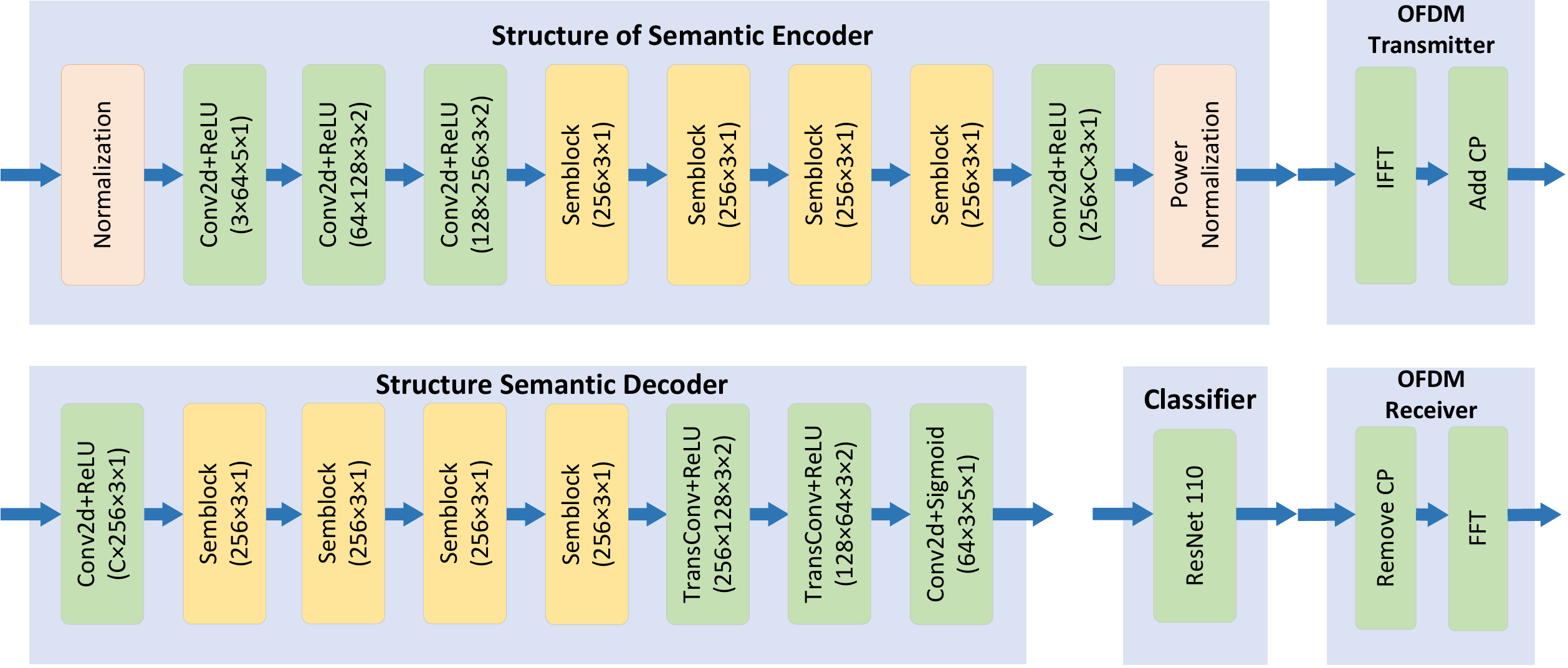}
\caption{Detailed modules of our \texttt{MobileSC}.}
\label{fig:implementation}
\end{figure}

\subsection{Implementation Details}
\label{implementation_details}
\subsubsection{\texttt{MobileSC} System} We implement our \texttt{MobileSC} based on PyTorch \cite{pytorch}, one of the most popular machine learning library. Figure \ref{fig:implementation} demonstrates the architecture of our semantic encoder, semantic decoder, and classifier.
Each image is firstly normalized within the range of [-1, 1]. Then, it is passed through a series of down-scaling convolutional layers and SemBlocks. The structure of the semantic decoder is almost symmetric (in the reverse order) to the encoder network, except that the final activation function is set to the Sigmoid function to enforce a valid dynamic range of image output pixels. For the two OFDM modules, we adapt the OFDM-transmitter and OFDM-receiver from the prior work OFDM-guided JSCC \cite{yang2022ofdm}. The OFDM transmitter performs IFFT transformation on signal, and then adds CP and pilot frequency. The OFDM receiver removes CP and pilot from the signal and then performs FFT transformation on the signal. The receiver optionally feeds latent representations to the classifier or to the semantic decoder. Exhaustive experiments are conducted in both two settings. For the classifier, we refer to the official implementation of ResNet-preact \cite{he2016identity} and adapt it in our case. 


\begin{table}
\caption{Parameters\label{tab:param}}
\centering
\scalebox{0.90}{
\begin{tabular}{lll}
\toprule
\textbf{Category}             & \textbf{Parameter}                          & \textbf{Value} \\ \midrule
\multirow{9}{*}{Input Image}
                              & Image shape (MNIST)  & (1, 28, 28) \\
                              & Image shape (CIFAR10) & (3, 32, 32) \\
                              & Image shape (ImageNet) & (3, 256, 256) \\
                          &Num of class type (MNIST)& 10 \\
                          &Num of class type (CIFAR10)& 10 \\
                          &Num of class type (ImageNet)& 10 \\
                          &Num of train/test instances (MNIST) &60,000/10,000\\
                          &Num of train/test instances (CIFAR10) &50,000/10,000\\
                          &Num of train/test instances (ImageNet) &26,000/10,000\\                          
\midrule
\multirow{5}{*}{Neural model} 
                              & Weight of the classification loss $\lambda_{cls}$            & 0.5           \\
                              & Weight of  the reconstruction loss $\lambda_{rec}$        & 0.5            \\
                              & Num of layers in the discriminator       & 3              \\
                              & Num of SemBlocks           & 4              \\                              
                              & Dimension of the representation $d$ & 1134 \\

\midrule
\multirow{2}{*}{Attacks}                       & PSR(dB)               & -10, -8, -6, -4, -2             \\
& $p$ normal               & 2             \\
\midrule
\multirow{5}{*}{Training}   
                              & Maximum epochs                                      & 200            \\
                              & Batch size                                  & 256            \\
                              & Learning rate                               & 5.00E-04       \\
                              & Optimizer                          & Adam  \\ &Parameter for Adam                          & 0.5            \\
\midrule
\multirow{5}{*}{OFDM module}  & Num of pilot symbols $N_p$                    & 1              \\
                              & Num of subcarriers per symbol $N_c$            & 64             \\
                              & Length of Cyclic Prefix $N_{cp}$                               & 16             \\
                              & The parameter of Rayleigh channel $\sigma^2$ & 1.5              \\
                              & SNR(dB)                                         & 3,5,7,9,10,11,13,15     \\ 
                              \bottomrule
\end{tabular}}
\end{table}

\subsubsection{Computation platform and Settings} 

We run our model on a DELL server with Ubuntu $18.04$ operating system, equipped with $64$GB RAM and an RTX3090 GPU card. The PyTorch and Python versions are $1.7.1$ and $3.7$, respectively. For drivers, the CUDA and cudnn versions installed in the server are $11.0$ and $8,004$. It takes about $3$ hours to train our semantic communication system, $3$ hours for our proposed perturbation generator, and $4$ more hours for the adversarial training on the three datasets. Our model takes up $3$GB of GPU memory during the training procedure. The parameters of neural networks are randomly initialized and then iteratively updated by the Adam optimizer with an initial learning rate of $0.00005$. We set the batch size as $256$ for all datasets. The parameter $\sigma^2$ in the Rayleigh channel is set as $1.5$. We refer to previous works \cite{sadeghi2019physical,bahramali2021robust} to calculate $p_{max}$ as
\begin{equation}
p_{max} = p_{sig} \times 10^{PSR/10},
\end{equation}
where the Perturbation-to-signal ratio (PSR) indicates the ratio of the received perturbation power to received signal power, $p_{sig}$ refers to the power of a signal. For simplicity, we also set the radius $\epsilon$ in $\ell_p$ ball of (\ref{eq:perturabtion_gen}) as $p_{max}$.
Table \ref{tab:param} shows the detailed parameters for neural models, attacks, training procedures, and the OFDM modules.  


\subsection{Performance Metrics}

\subsubsection{Commonly used metrics}
We evaluate the performance of our \texttt{MobileSC} with three widely used metrics including PSNR, SSIM and classification accuracy. 
\begin{itemize}
    \item \textbf{PSNR}: PSNR is the most commonly used metric to quantify the reconstruction quality of the lossy image and video compression. Generally, a higher PSNR indicates the reconstruction is of higher quality.
    \item \textbf{SSIM}: SSIM can be used for predicting the perceived quality of images and videos. It is a full reference metric that leverages the initial uncompressed or distortion-free image as a reference. 
    \item \textbf{Classification accuracy (CA)}: In our ESC system, CA quantifies the performance of a neural classification model as the number of correct predictions divided by the total number of predictions. We have some pre-defined class categories. 
\end{itemize}

\subsubsection{Misleading Rate} We introduce a new metric ``misleading rate'' to evaluate the capabilities of our model in misleading the predictions to a certain category. Such a metric measures the capability of our model conducts ``controlled attacks" on the semantic communication system. 


\begin{figure*}[htp]
\centering
    \subfloat[]{
        \includegraphics[width=.33\textwidth]{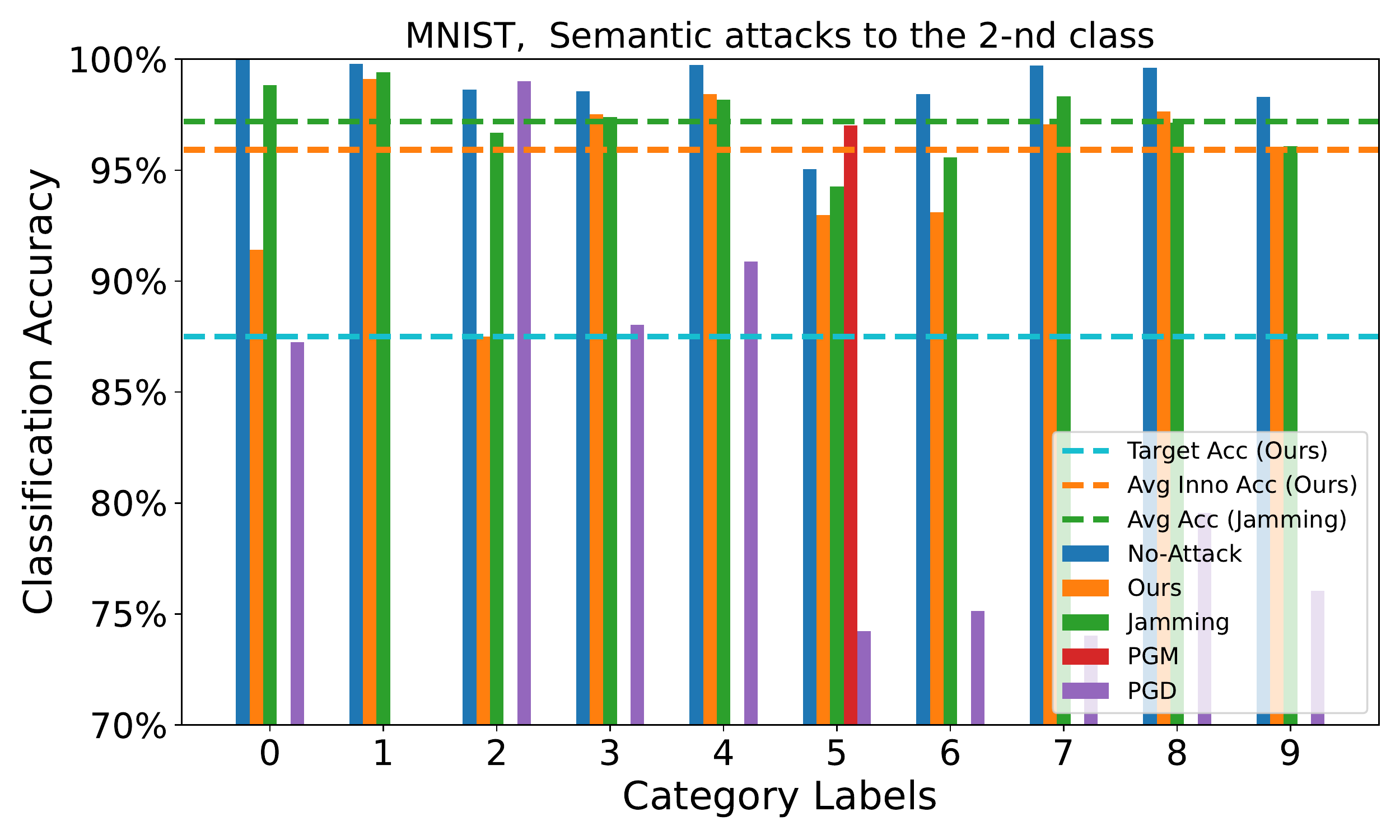}\hfill
    }
    \subfloat[]{
        \includegraphics[width=.33\textwidth]{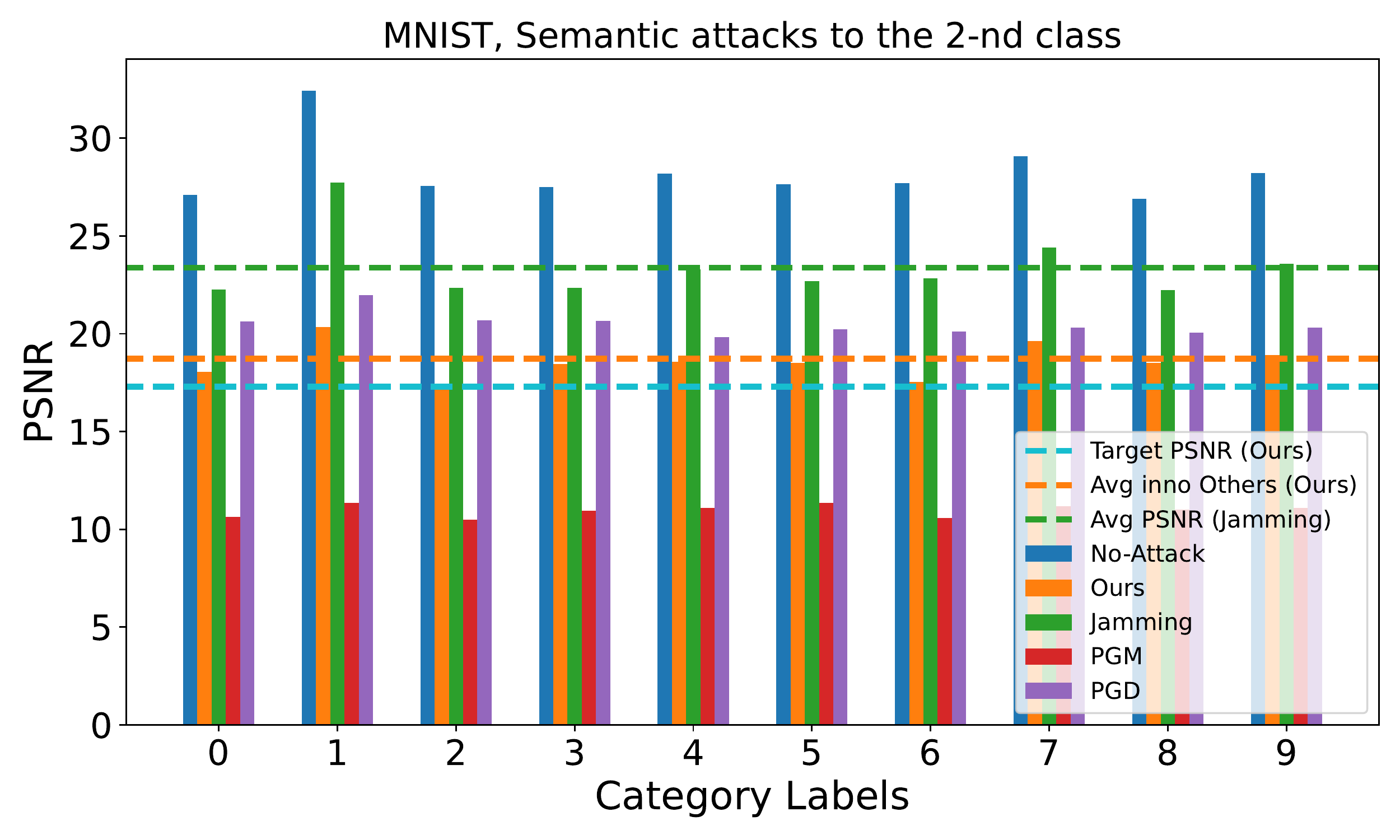}\hfill
    }
    \subfloat[]{
\includegraphics[width= .33\textwidth]{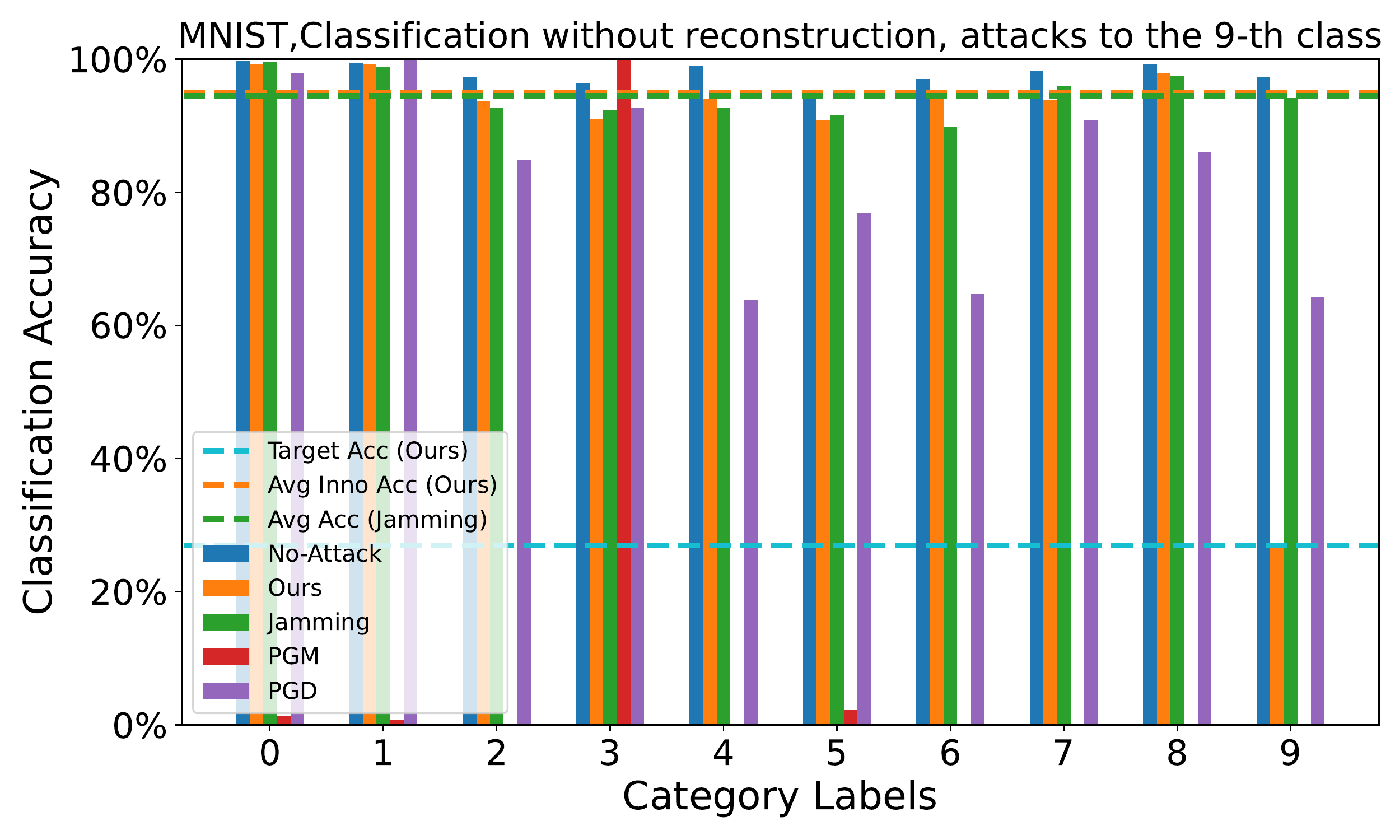}\
\hfill }
    
    \subfloat[]{
        \includegraphics[width=.33\textwidth]{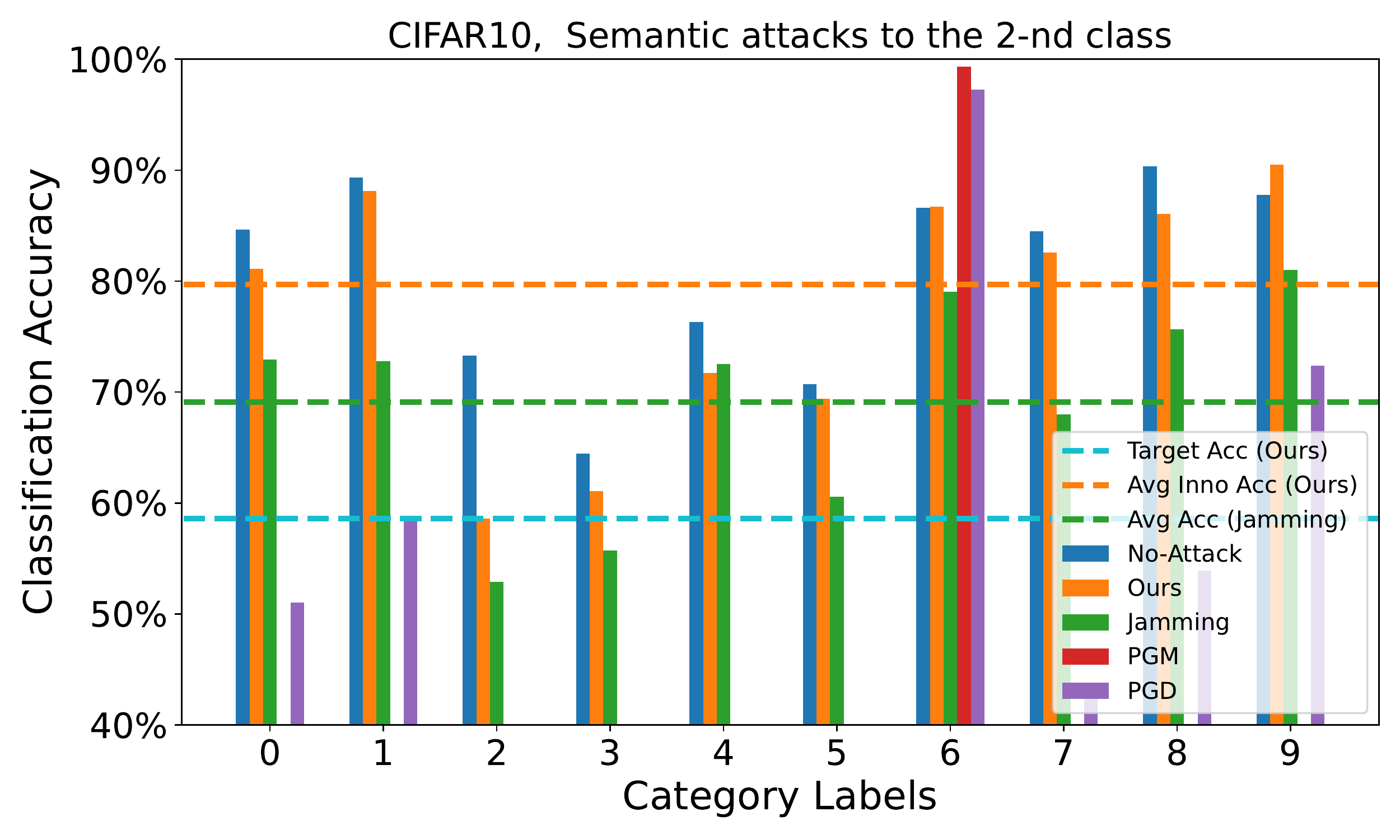}\hfill
    }
    \subfloat[]{
    \includegraphics[width=.33\textwidth]{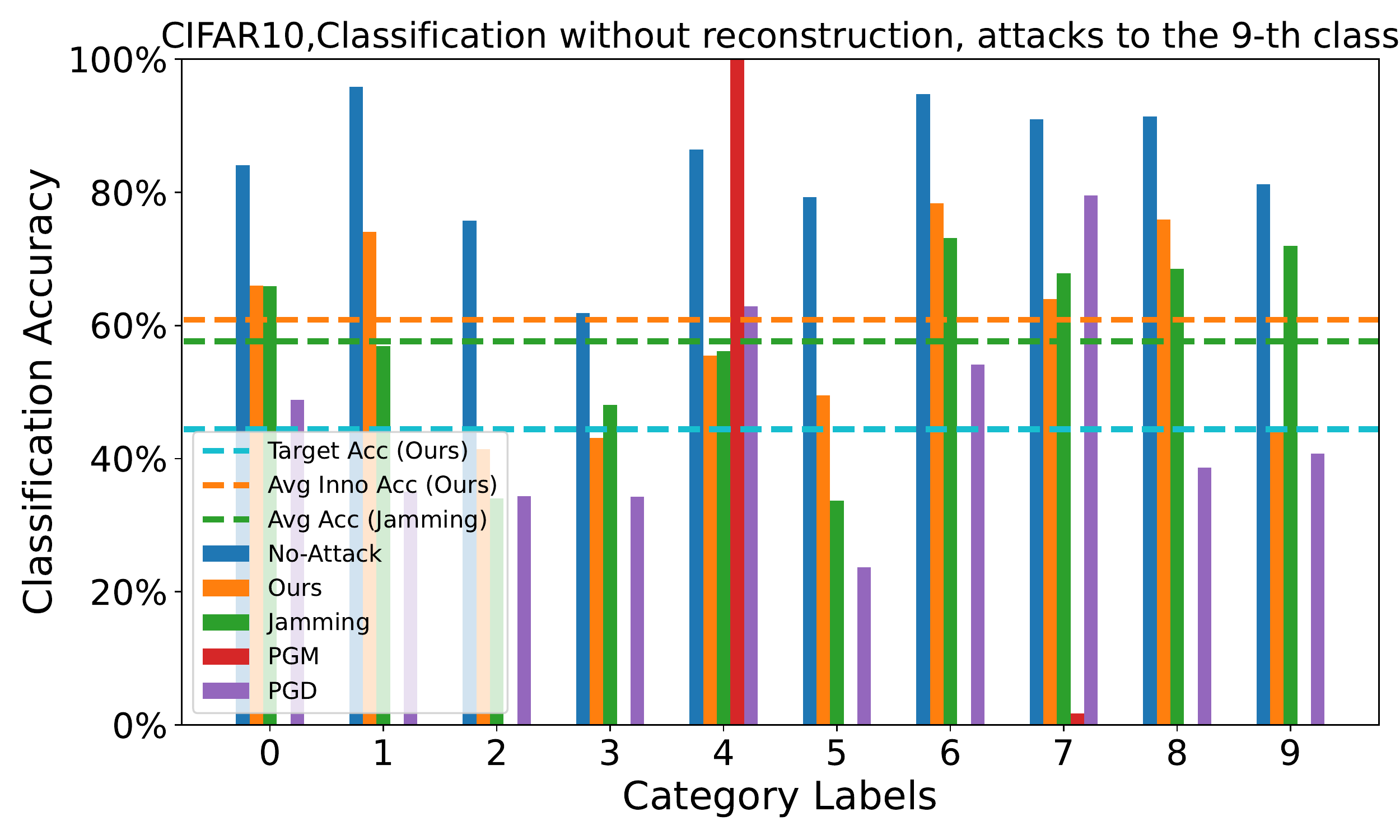}\
}
    \subfloat[]{
        \includegraphics[width=.33\textwidth]{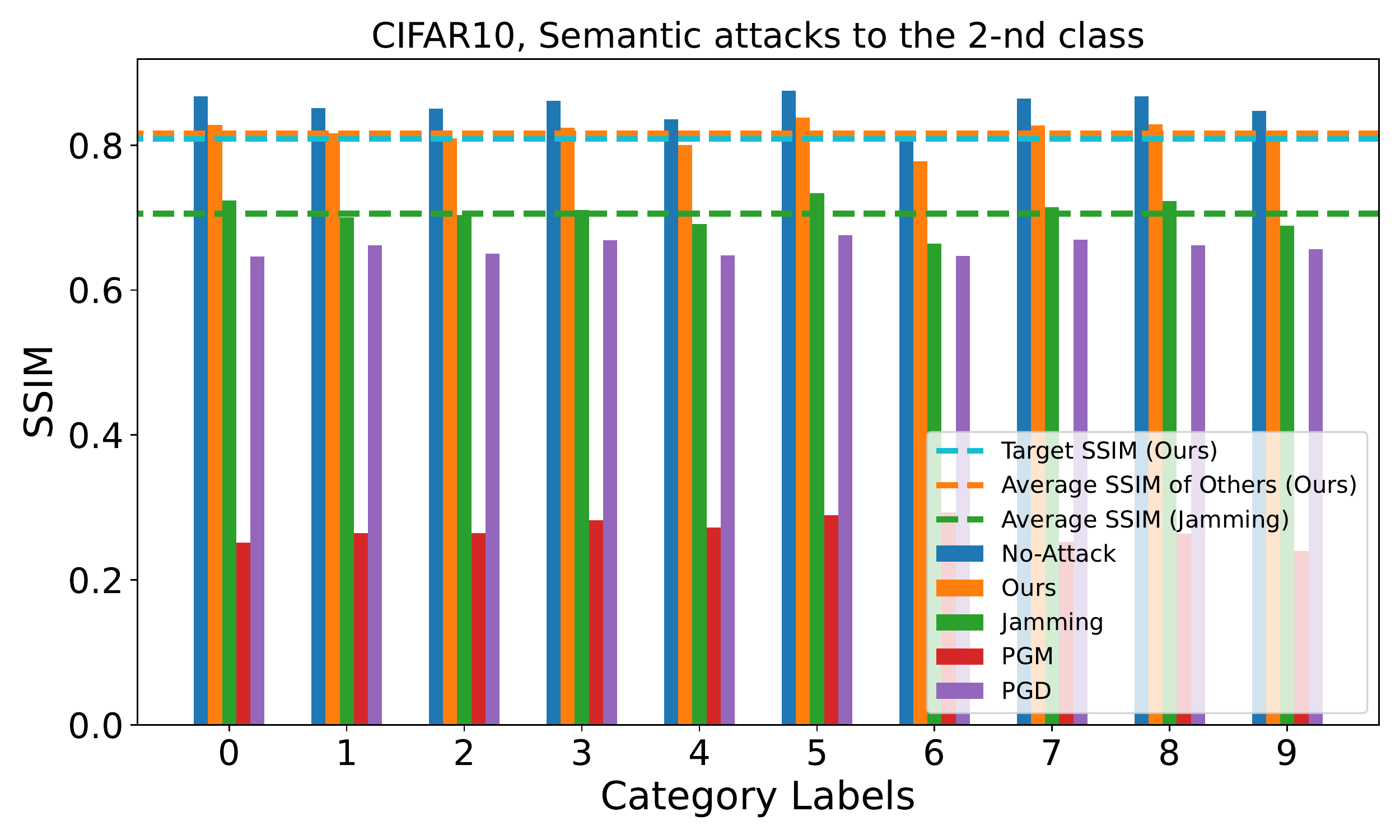}\hfill
    }
    
    \subfloat[]{
         \includegraphics[width=.33\textwidth]{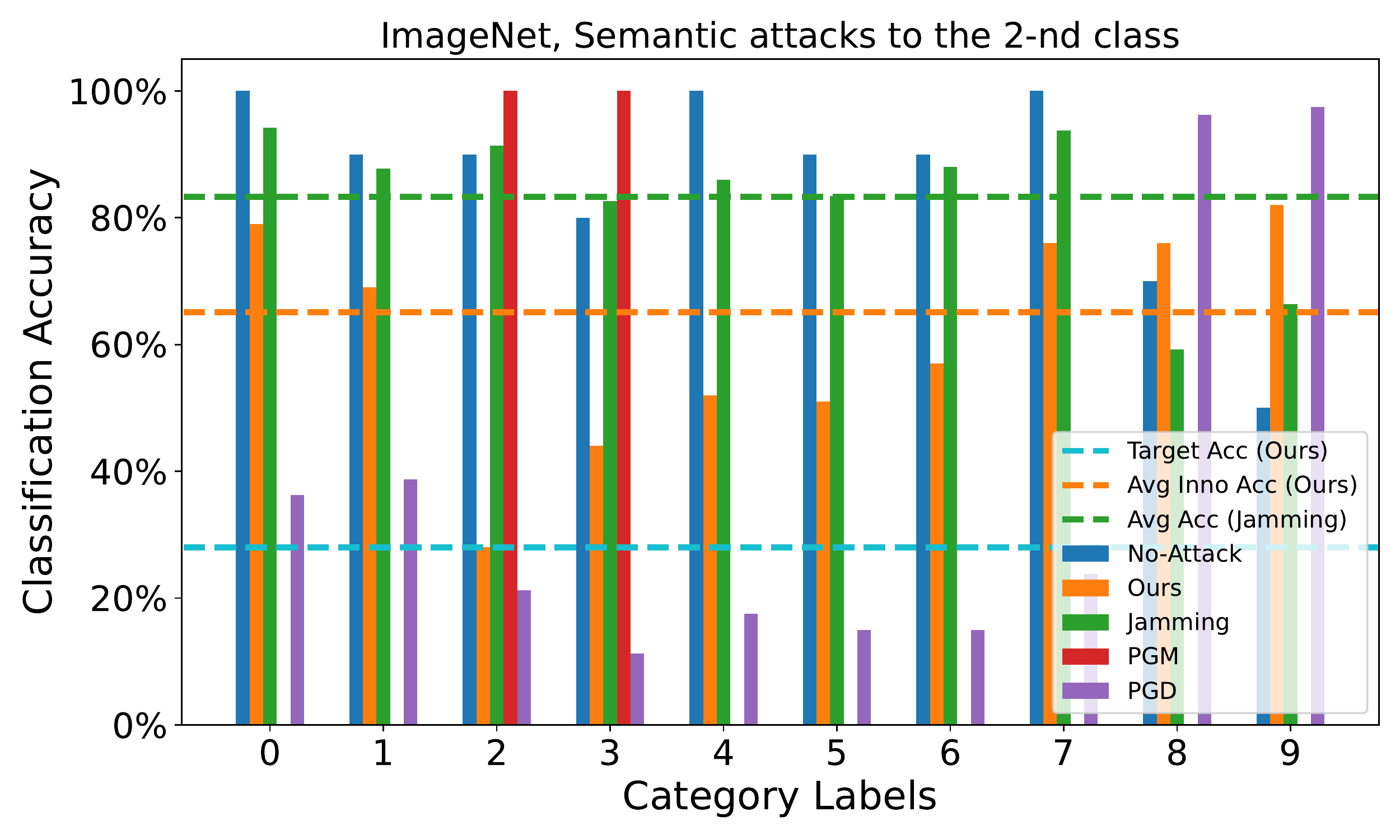}
     }
    \subfloat[]{
        \includegraphics[width=.33\textwidth]{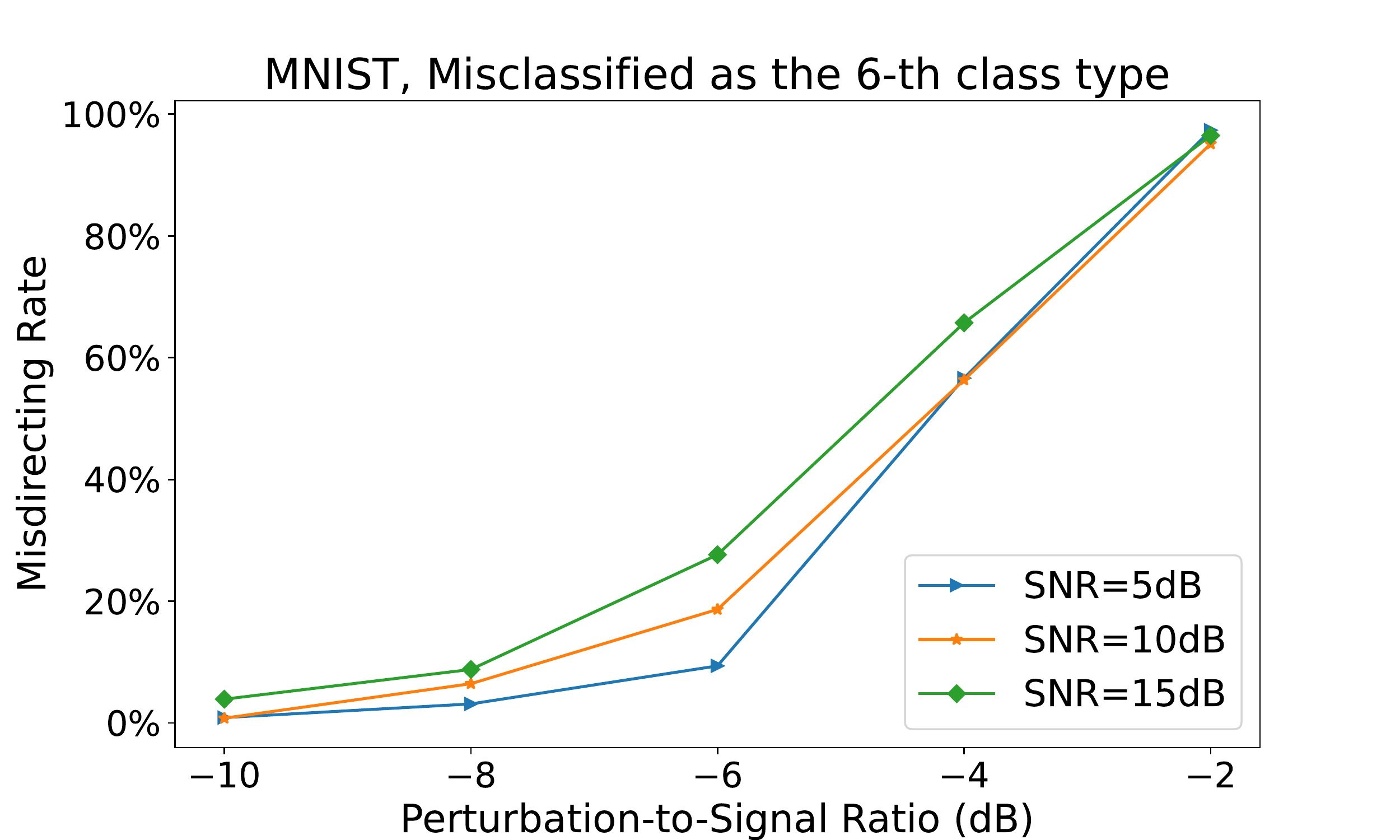}\hfill
    }
    \subfloat[]{
        \includegraphics[width=.33\textwidth]{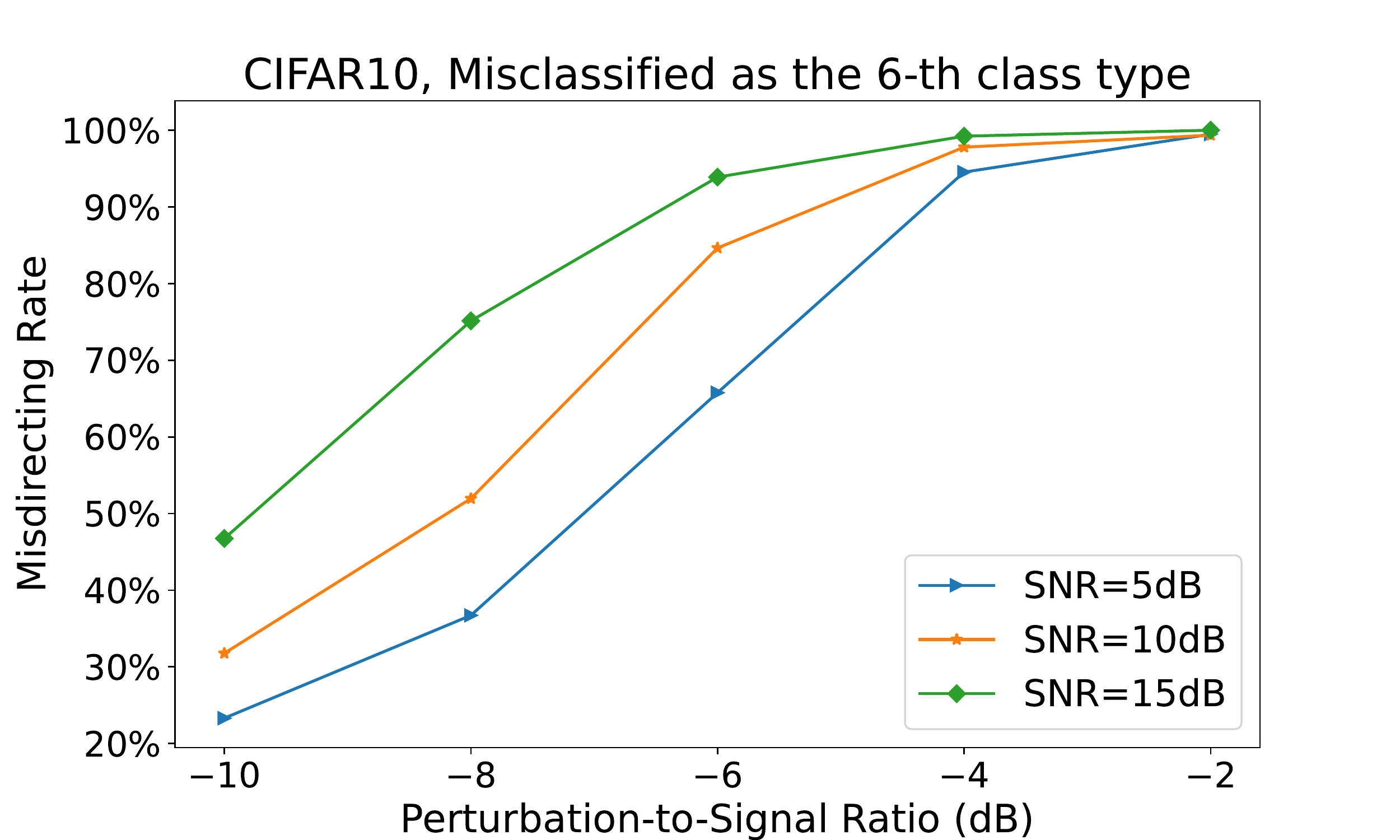}\hfill
    }
    \caption{ Performance comparisons under various attacks. (a), (b), (d), (f) and (g)  are comparisons  after reconstructing images in terms of three metrics including classification accuracy, PSNR and SSIM under four attacks including jamming attack, PGD, PGM and our proposed $\texttt{SemAdv}$ with SNR=10dB and PSR=-4dB. (c) and (e) are comparisons without reconstructing images.} (a)-(c), (d)-(f), (g) are based on MNIST, CIFAR10 and ImageNet dataset respectively. (h) and (i) measure the controllability of our proposed attacks. 
    \label{fig:attacks}
\end{figure*}

\subsection{Baselines}
We compare our model with three baseline models including Jamming attack, PGD and PGM.

\begin{itemize}
    \item Jamming attack: This attack follows the same distribution as the Gaussian noise. It is widely used in conventional communication systems. 
    \item PGD: PGD is a standard method for adversarial training and it generates adversarial examples based on gradients. We integrate the publicly released PGD implementation into our system for generating the perturbations. 
    \item PGM: PGM is proposed for the generation of adversarial attacks for content-level wireless communication systems. We re-implement this baseline in PyTorch as it is not publicly available.
\end{itemize}

\subsection{Performance Under Attacks}
\label{sec:perforance_under_attacks}
Figure \ref{fig:attacks} shows the comparison results in terms of classification accuracy, PSNR, SSIM, and Misleading rate. Different from previous works that quantify the performance of all categories with a single score, our comparisons are conducted in a much more fine-grained manner, which is based on each category of the images. Each class type indicates certain semantics. We also give the average values of these performance scores (the dotted line in the same color). Such comparisons can better measure the impact of various adversarial attacks on different semantics that are transmitted over a wireless channel.      
\begin{figure*}[htp]
\centering
    \subfloat[]{
        \includegraphics[width=.33\textwidth]{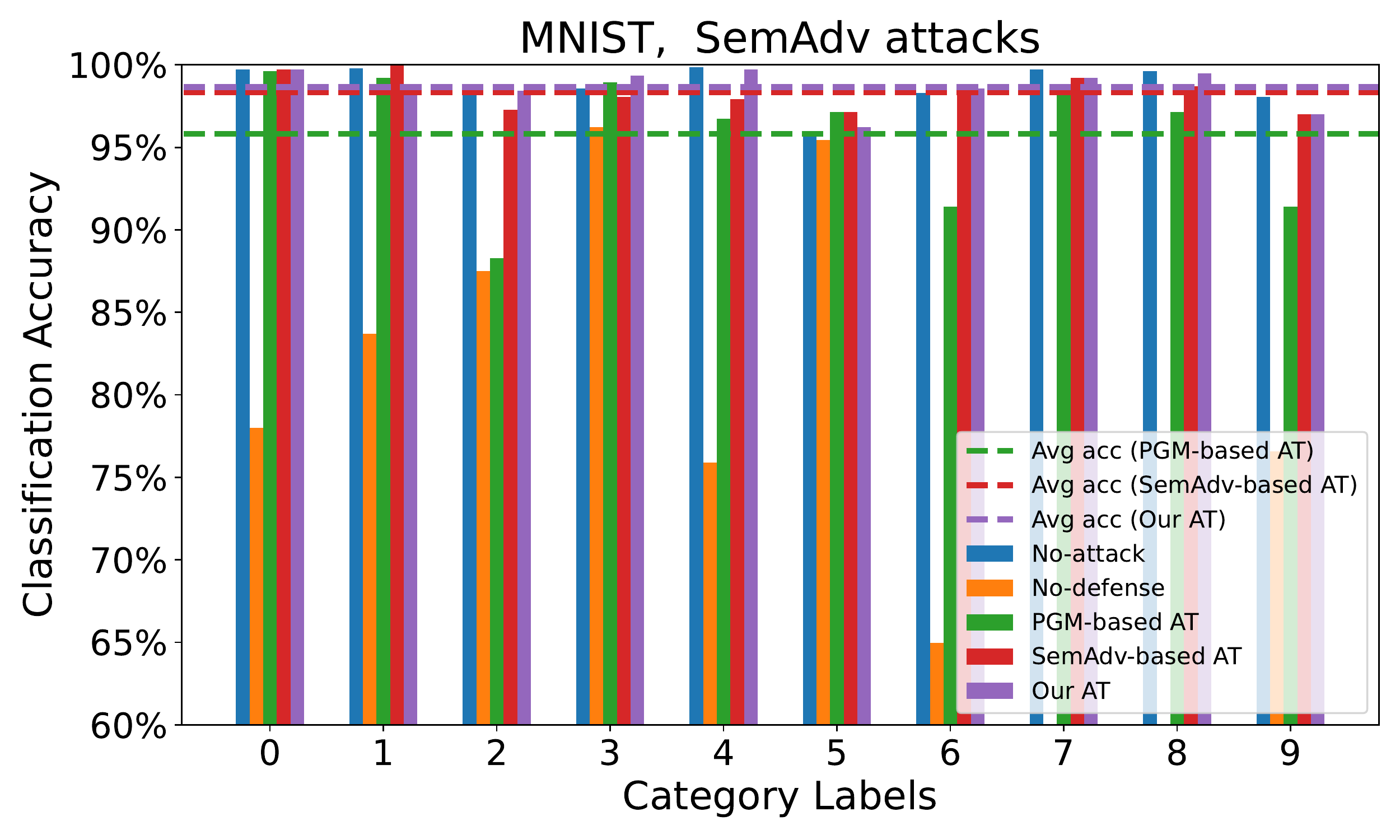}\hfill
    }
    \subfloat[]{
        \includegraphics[width=.33\textwidth]{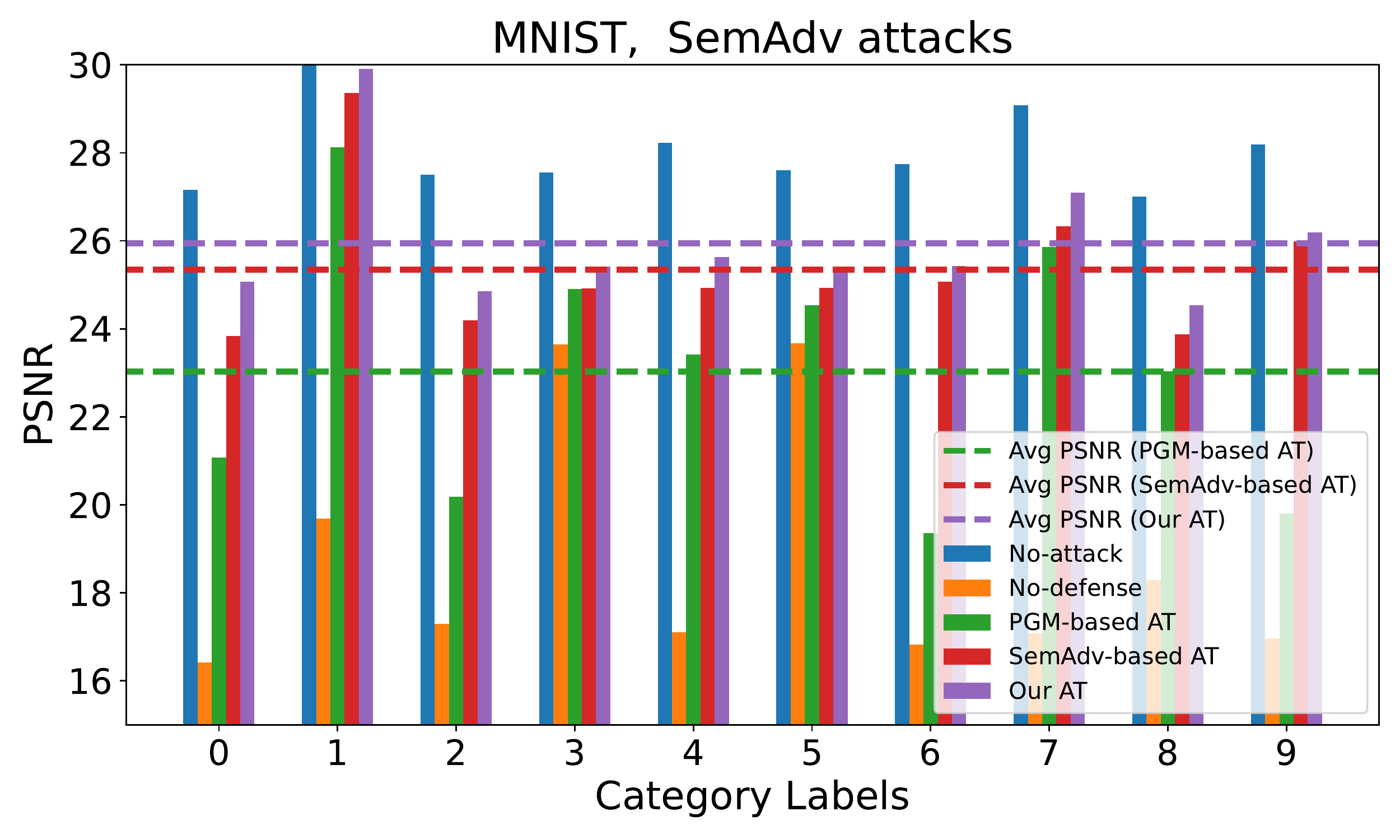}\hfill
    }
    \subfloat[]{
\includegraphics[width=.33\textwidth]{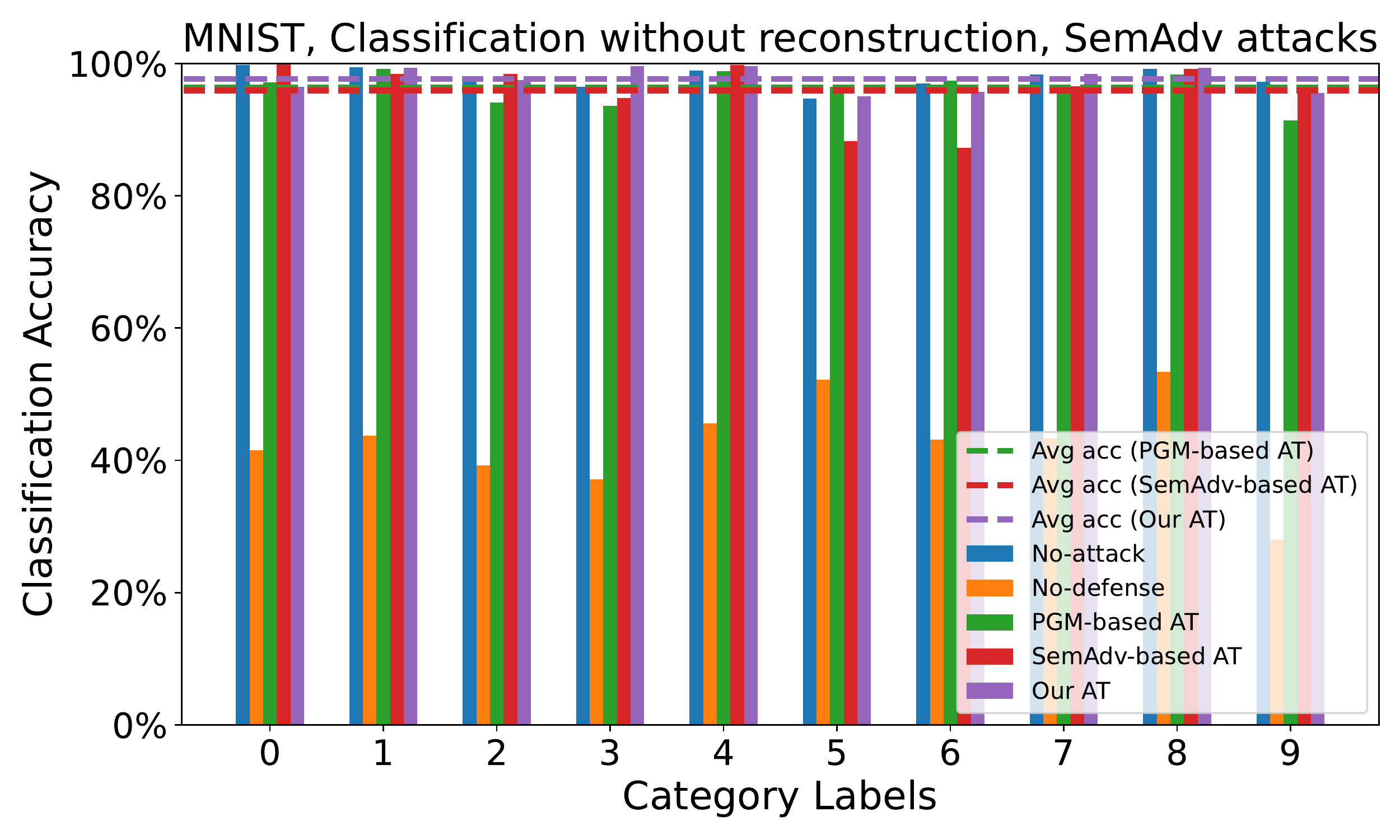}\hfill
    
}
    
    \subfloat[]{
        \includegraphics[width=.33\textwidth]{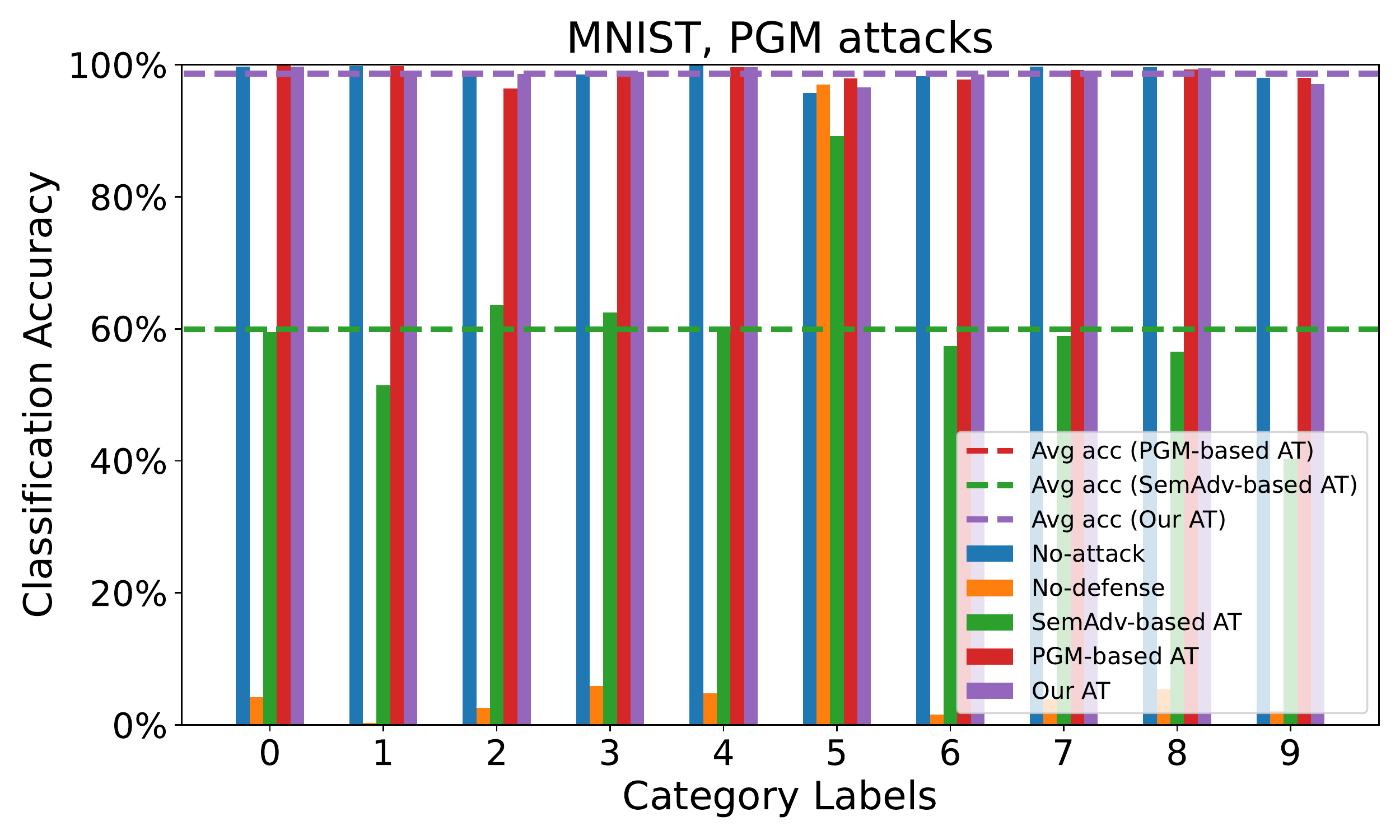}\hfill
    }
    \subfloat[]{
        \includegraphics[width=.33\textwidth]{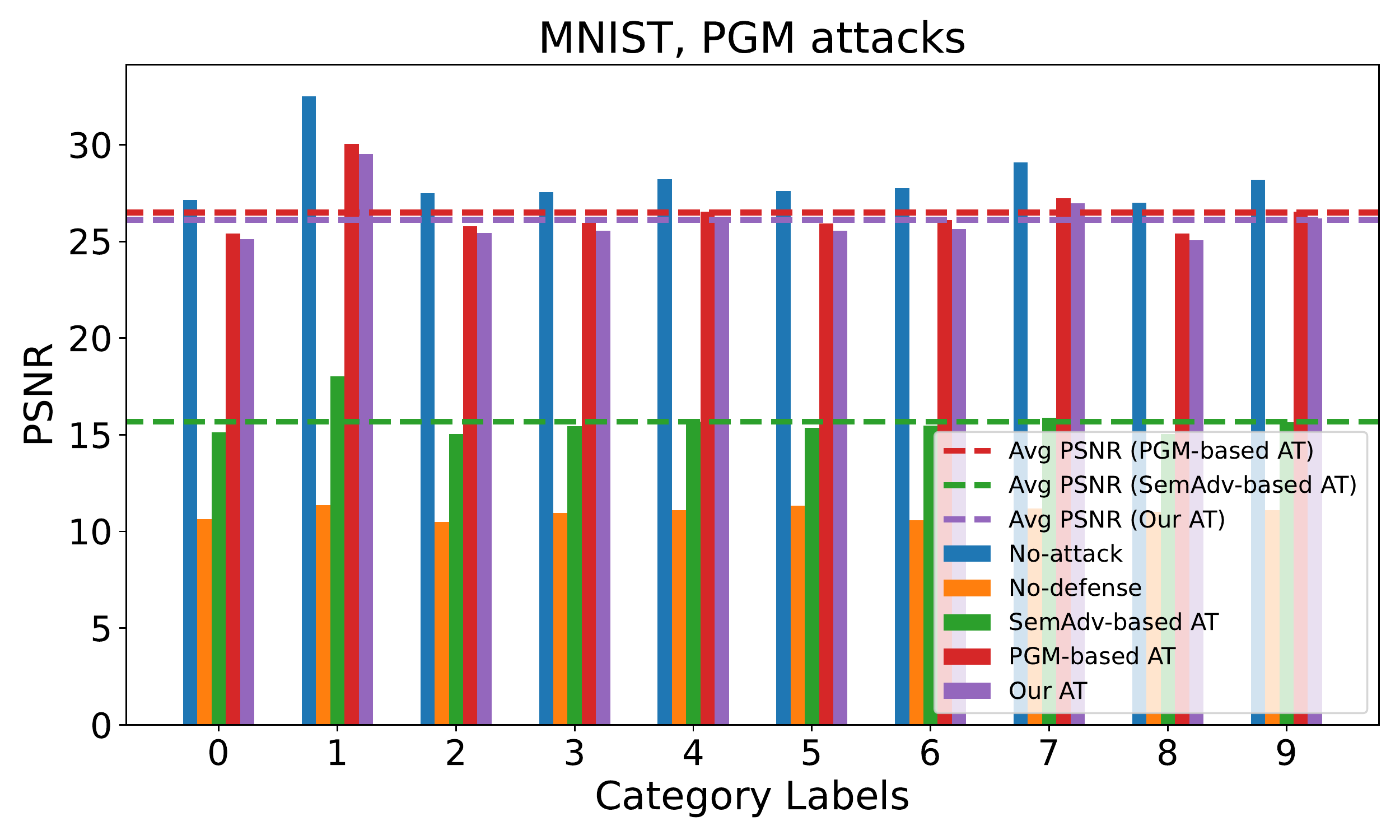}\hfill
    }
    \subfloat[]{
\includegraphics[width=.33\textwidth]{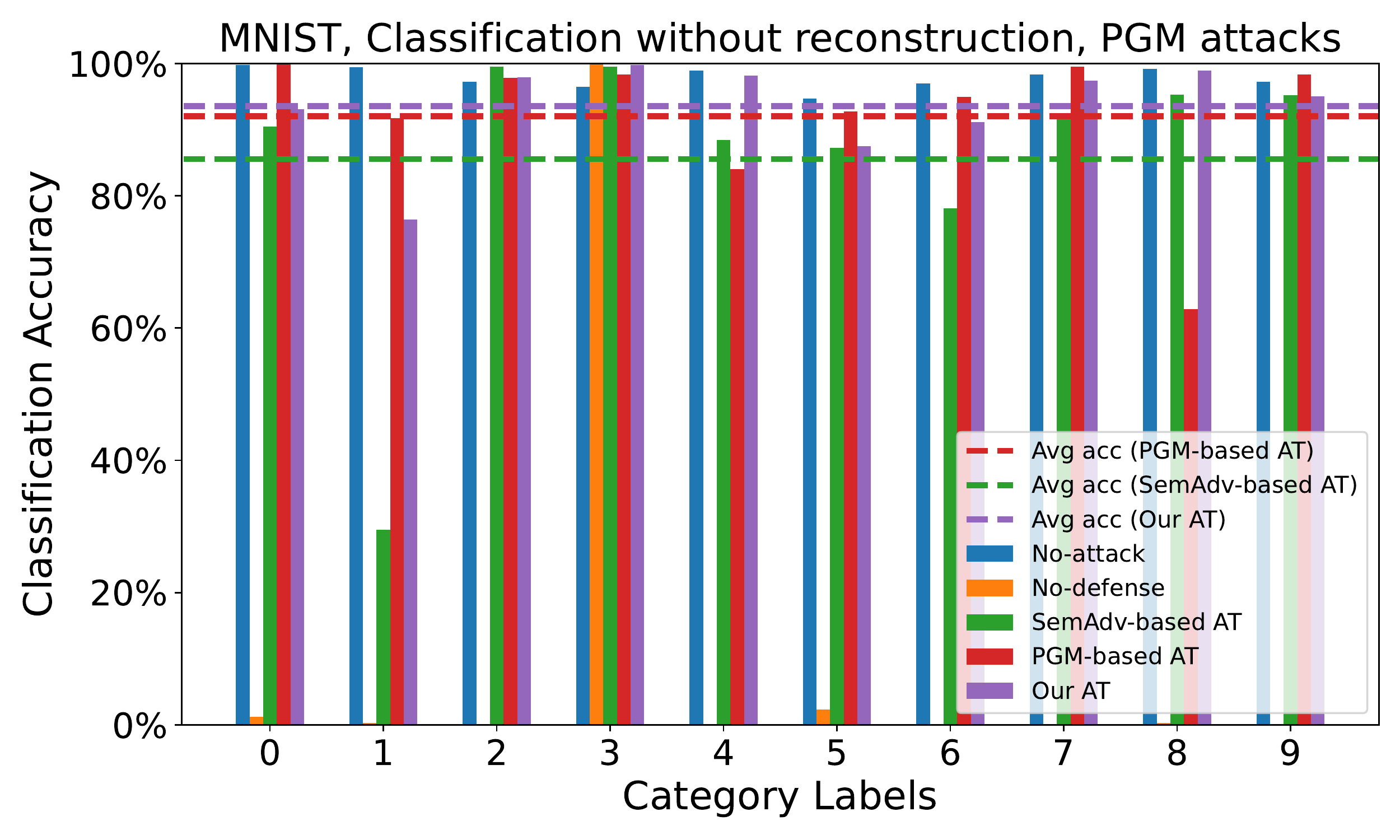}\hfill
}
    \caption{The comparisons of adversarial robustness for various defense strategies on the MNIST dataset, including ``PGM based AT'', ``SemAdv based AT'', and ``Our AT''. Here the abbreviation AT indicates adversarial training. SNR and PSR are configured as $10$dB and -$4$dB, respectively. Under SemAdv attacks, (a), (b), and (c) demonstrate the robustness of different defense mechanisms measured by classification accuracy and PSNR, respectively. Under PGM attacks, (d), (e), and (f) demonstrate the comparisons of adversarial robustness of different defenses with the two metrics. (c) and (f ) are comparisons without reconstructing images. A vertical comparison of these pictures shows our algorithm's ability to defend against both attack methods.}
    \label{fig:defense_mnist}
\end{figure*}

\subsubsection{Classification Accuracy} Figures \ref{fig:attacks} (a), (d) and (g) show the comparisons of classification accuracy under various attacks, and we configure SNR and Perturbation-to-signal ratio (PSR) \cite{sadeghi2019physical} as $10$dB and -$4$dB, respectively. We set the $2$-nd class type as the targeting semantics, which is digital number $2$ in MNIST, \textit{bird} in CIFAR10, and \textit{frogs} in ImageNet, respectively. Figures 6 (c) and (e) show the comparisons of classification accuracy without reconstructing images under various attacks, and we configure SNR and Perturbation-to-signal ratio (PSR) \cite{sadeghi2019physical} as $10$dB and -$4$dB, respectively. We set the $9$-th class type as the targeting semantics to indicate the universality of the attack effect. In general, we observe that the classification accuracy is significantly degraded by adversarial attacks on the three datasets, and adversarial attacks are more destructive than conventional jamming attacks. This observation aligns with the findings that are based on end-to-end autoencoder-based communication systems \cite{sadeghi2019physical}. We also have several interesting observations beyond the prior efforts:
\begin{itemize}
    \item Our proposed \texttt{SemAdv} is able to attack the targeting semantics and decrease the classification performance by $17$, $18$, and $60$ points, respectively, while it only causes less than $10$ points decrease on other categories with different semantics. Such an observation confirms our hypothesis that our attack is semantic-oriented and the interpretation of targeting semantics will be significantly degraded by the \texttt{SemAdv} attack.  
    \item Our \texttt{MobileSC} system tends to predict all images reconstructed by the receiver as the $6$-th class type under PGM adversarial attacks on CIFAR10. We posit this is possible because the neural models almost collapse under PGM attacks, due to the high sensitivity of the parameters. Although more destructive, PGM will be easily detected by the receiver to develop a defense strategy as all data transmitted in the system is affected.
\end{itemize}
   
\subsubsection{PSNR} Figures \ref{fig:attacks} {(b) reports} the performance comparisons in terms of PSNR under various attacks on MNIST, with SNR and PSR configured as $10$dB and -$4$dB, respectively. Interestingly, our proposed \texttt{SemAdv} causes slight PSNR performance degradation to all classes for the reconstructed images, although the classification accuracy is significantly degraded. Conversely, PSNR performance significantly drops by $50\%$, $25\%$ , and $20\%$ under PGM, PGD, and the Jamming attacks, respectively. The sharp performance decrease will be visually identified by detectors or human beings. These  results suggest that our proposed \texttt{SemAdv} will not significantly degrade the reconstruction quality of images and hence are more imperceptible than PGD and PGM attacks in semantic communication systems.  

\subsubsection{SSIM} Figures \ref{fig:attacks} (f) shows the performance comparisons in terms of SSIM against jamming attack, PGD, PGM, and \texttt{SemAdv} respectively on CIFAR10. Here the metric SSIM measures the similarity between an original image and its reconstructed one. We find that SSIM slightly drops with our proposed \texttt{SemAdv}, implying that minor distortions caused by attacks may not be perceived by human beings or detectors. However, SSIM is very sensitive under jamming, PGD, and PGM attacks. For example, the average SSIM is degraded by more than $60\%$ under PGM attacks. The significant SSIM fluctuation can be noticed by the system. These comparison results further confirm the superiority of our proposed \texttt{SemAdv} in generating imperceptible physical-layer adversaries for semantic communications.

\subsubsection{Misleading Rate} We also train our attacker to deceive the receiver to predict all data to a specific class type, i.e., the $6$-th class type, which is the dog category in CIFAR10 and digital image $6$ in the MNIST dataset. Then in the testing stage, we attack our ESC system under various PSR ranging from -$10$dB to -$2$dB. From Figures \ref{fig:attacks} (h) and (i), we observe that the misleading rate, which incorrectly predicts the reconstructed image to the $6$-th type, can reach $100\%$ on both two datasets. The results show that the attacks can be controllable in deep learning-based semantic communication systems. It should be noted that the perturbations generated by our \texttt{SemAdv} can also mislead the system to multiple class types if we train our generator in such a setting.    

In summary, the experimental results confirm that our proposed physical-layer adversarial attacks crafted by \texttt{SemAdv} are able to meet the four criteria mentioned at the beginning, which are semantic-oriented, imperceptible, input-agnostic, and controllable. 

\begin{figure*}[htp]
\centering
        \subfloat[]{
        \includegraphics[width=.33\textwidth]{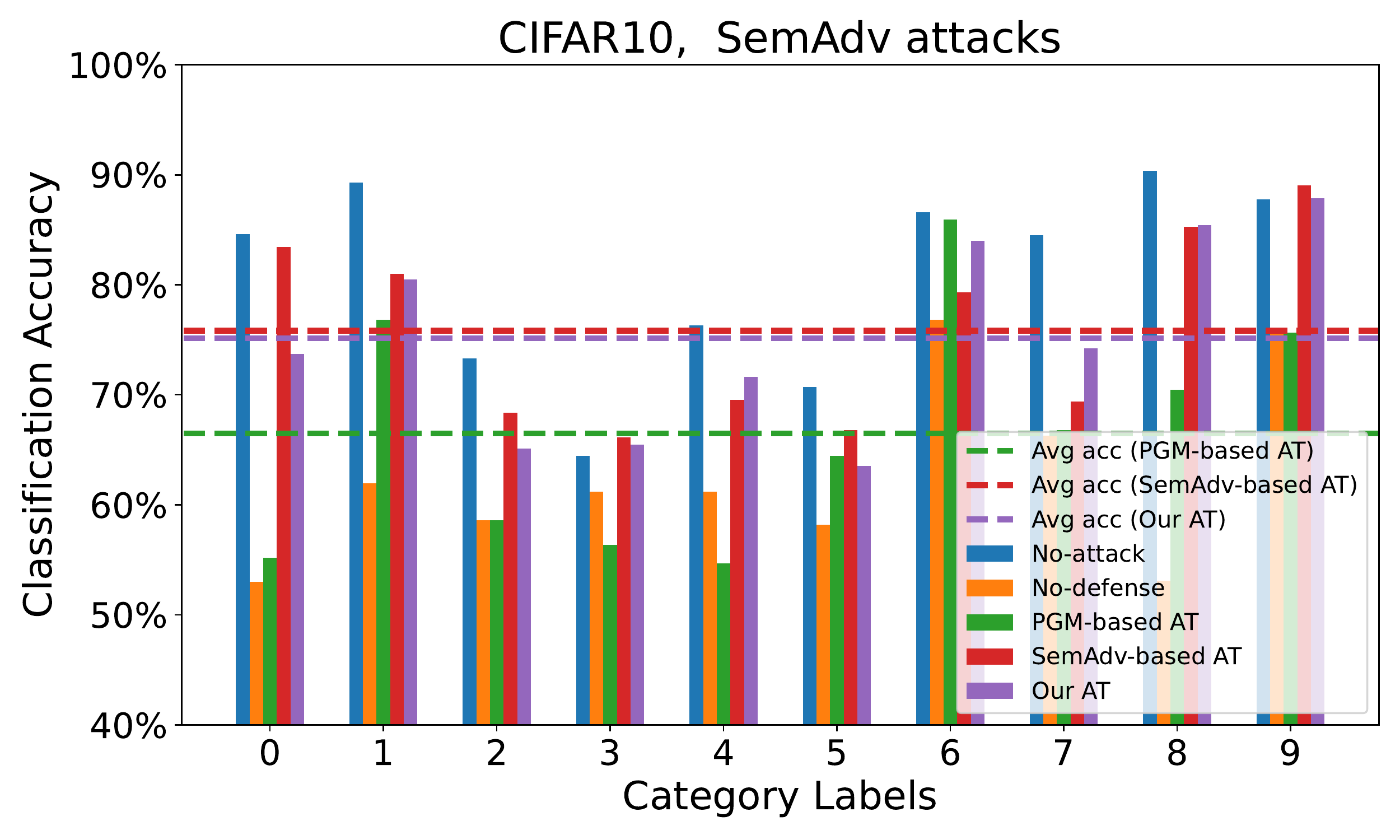}\hfill
    }
    \subfloat[]{
        \includegraphics[width=.33\textwidth]{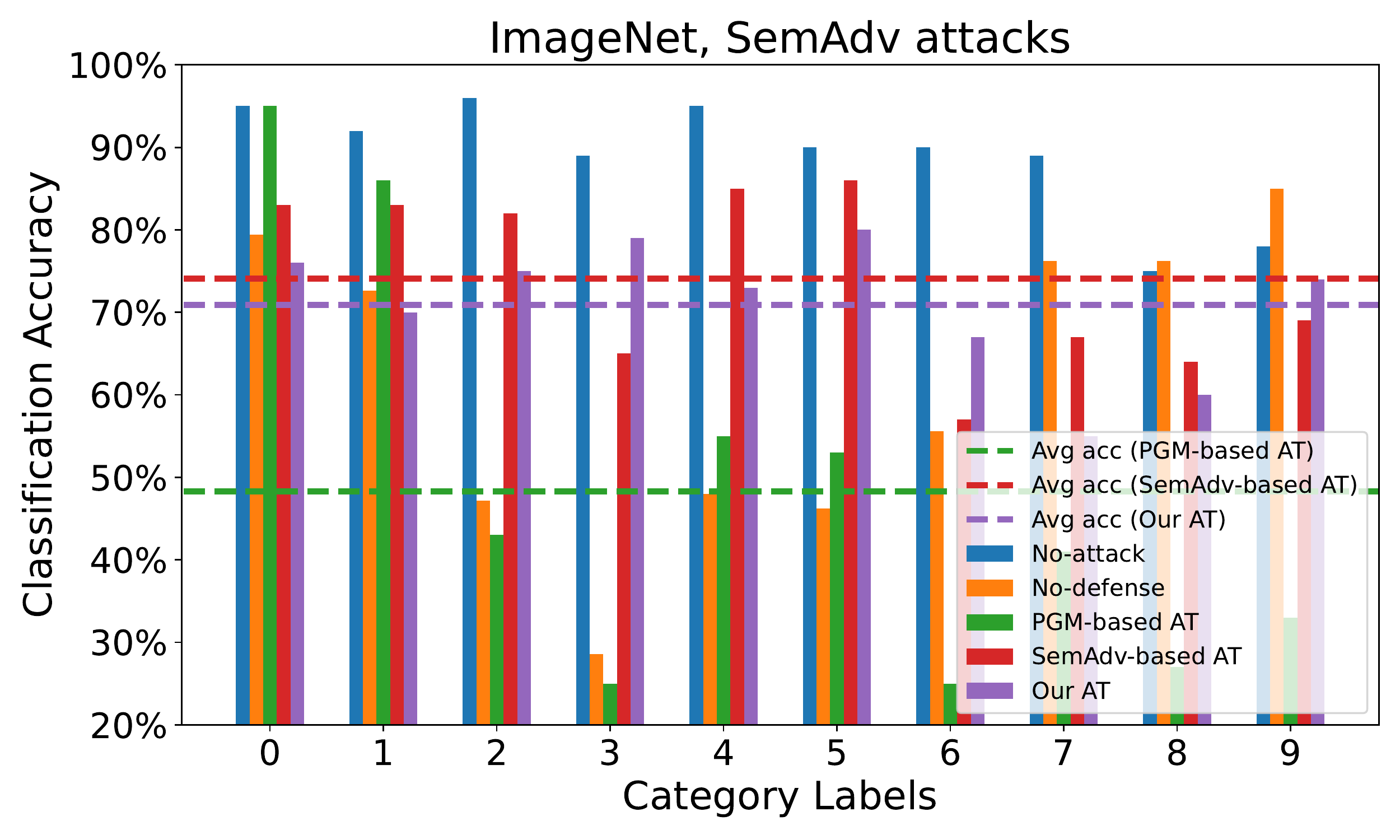}\hfill
    }
    \subfloat[]{
        \includegraphics[width=.33\textwidth]{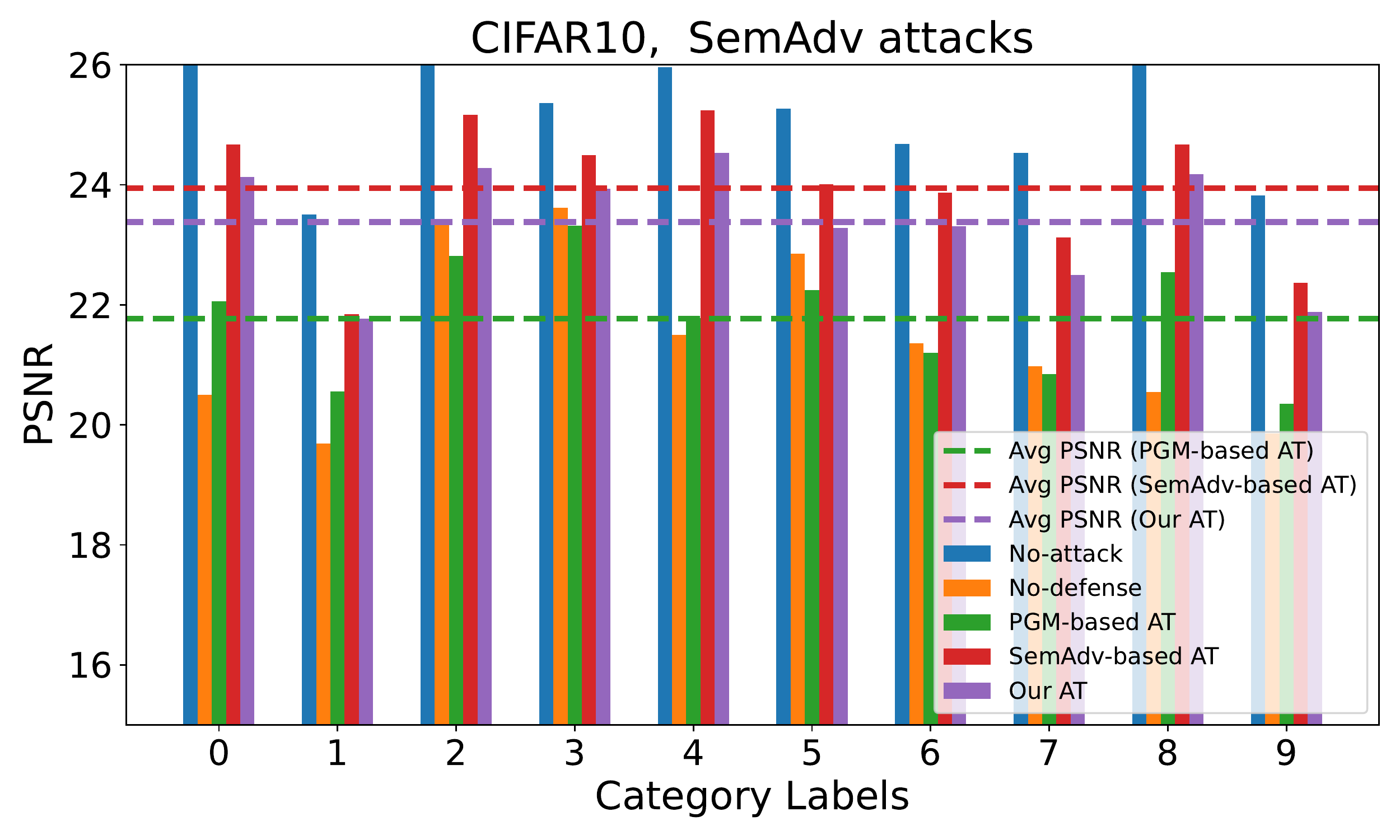}\hfill
    }

        \subfloat[]{
        \includegraphics[width=.33\textwidth]{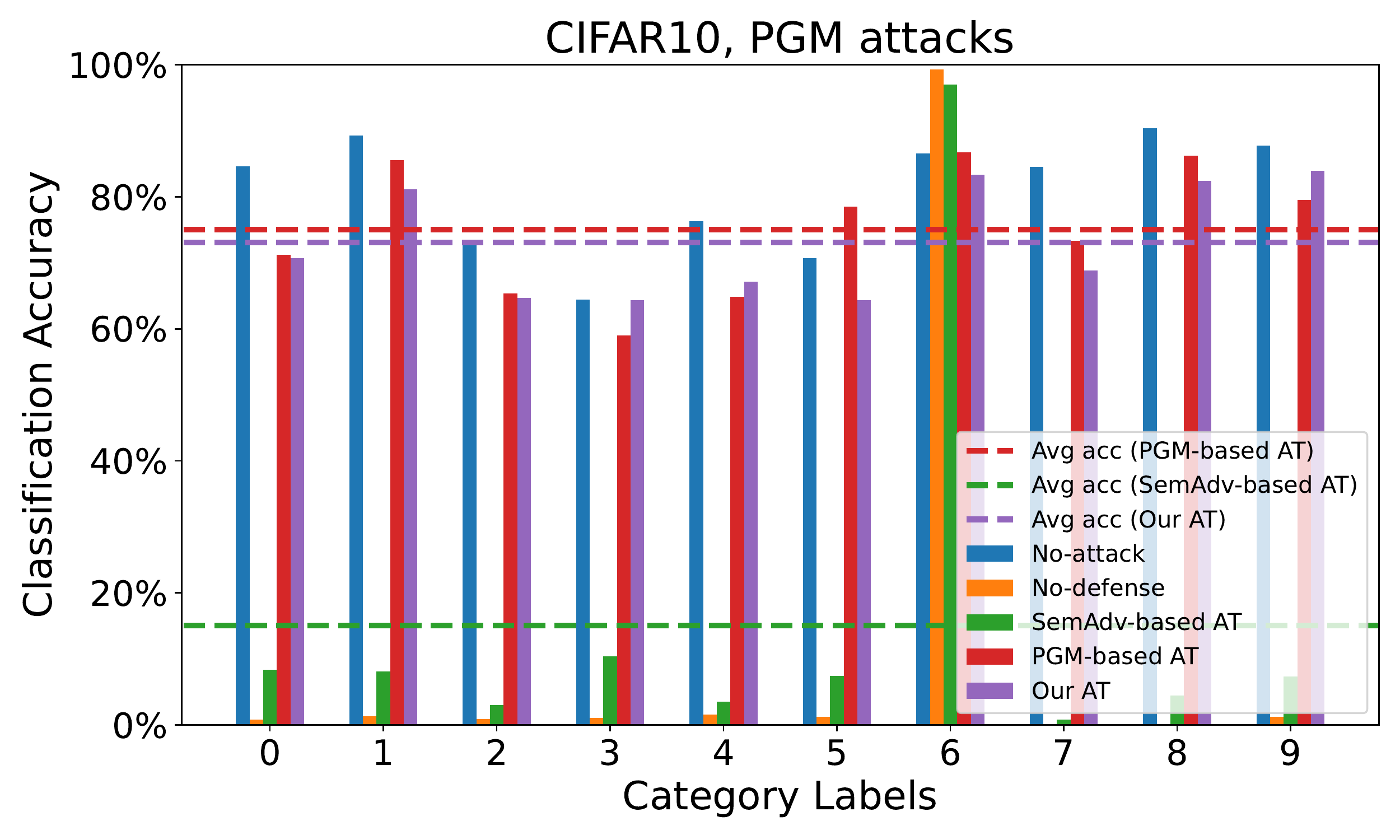}\hfill
    }
    \subfloat[]{
        \includegraphics[width=.33\textwidth]{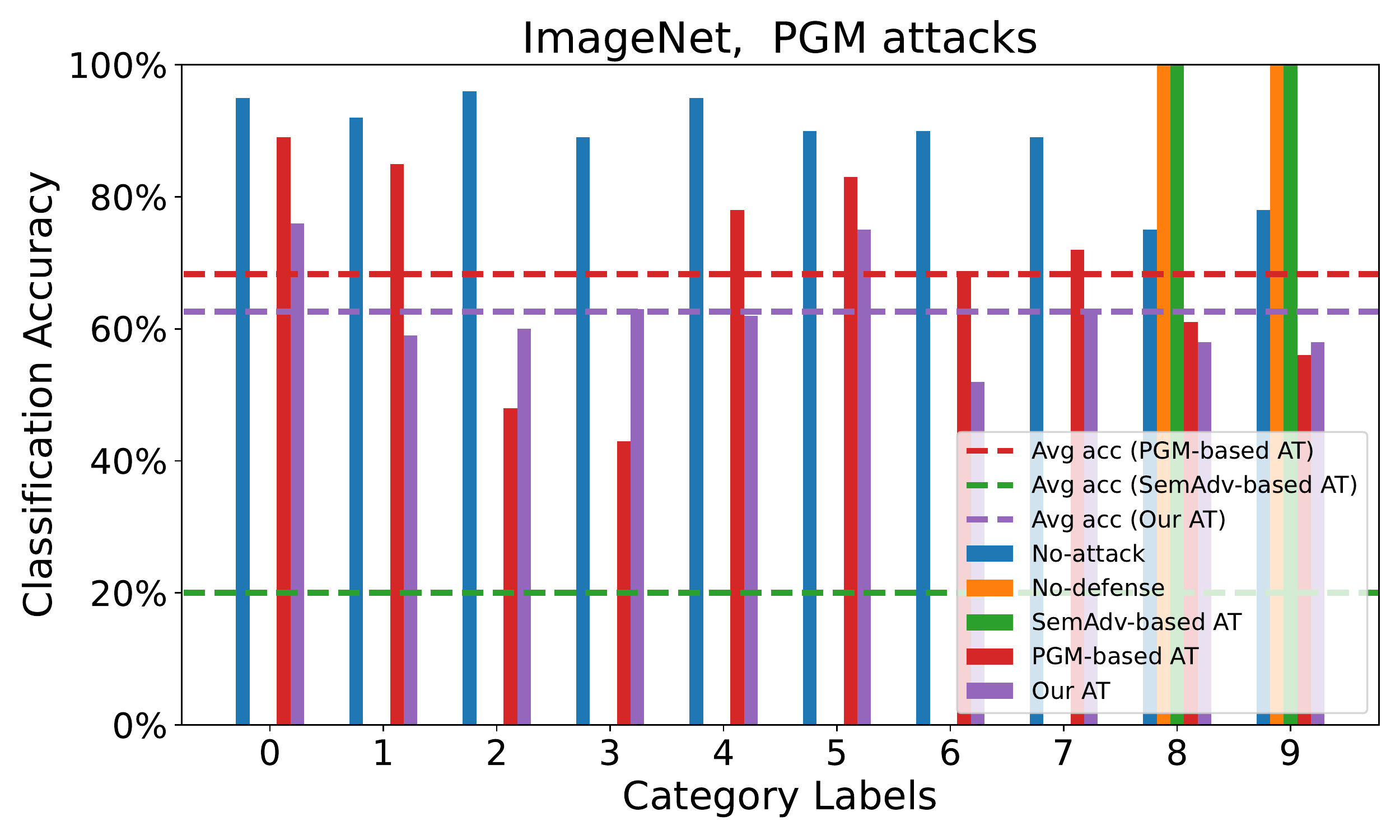}\hfill
    }
    \subfloat[]{
    
\includegraphics[width=.33\textwidth]{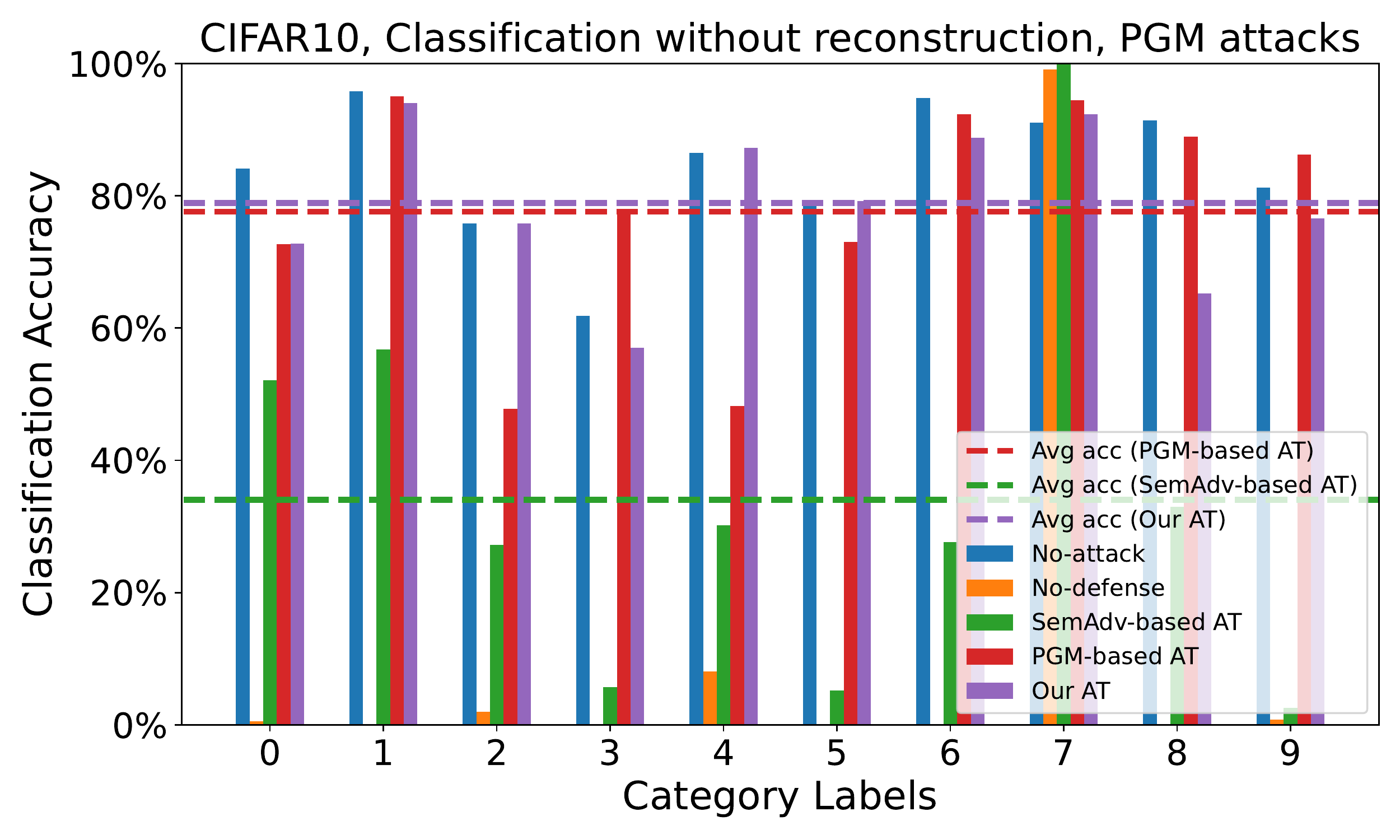}\hfill    
}
    \caption{The comparisons of adversarial robustness for various defenses include ``PGM based AT'', ``SemAdv based AT'', and ``Our AT'' on the CIFAR10 and ImageNet dataset, where the abbreviation AT refers to adversarial training. SNR and PSR are configured as $10$dB and -$4$dB, respectively. Under SemAdv attacks, (a), (b), and (c) show the comparisons of robustness in terms of classification accuracy and PSNR. (d), (e) and (f) demonstrate the comparisons under PGM attacks. (f) indicates classification without reconstructing images.}
    \label{fig:defense_cifar}
\end{figure*}

\subsection{Adversarial Robustness}
In this part, we begin to investigate to what extent different defense methods secure semantic interpretation and image reconstructions under various physical adversarial attacks. We conduct adversarial training (AT) with various defense strategies including ``AT with PGM'', ``AT with SemAdv'', and ``AT with ours'', which indicate that learnings are augmented with PGM perturbations, \texttt{SemAdv} and \texttt{SemAdv}+PGM perturbations, respectively. Then, we attack the semantic communication system hardened by the above ATs using \texttt{SemAdv} attacks and PGM attacks, respectively. We set the SNR and PSR as $10$dB and -$4$dB, respectively. Our proposed defense strategy based on ``\texttt{SemAdv}+PGM'' is termed as \texttt{SemMixed}. 

It is worth noting that in Section \ref{sec:perforance_under_attacks} the $\texttt{SemAdv}$ attacker aims to distort certain targeting semantics without any knowledge of input data. Differently, in this part, our \texttt{SemMixed} considers semantic adversaries for all class types in two AT experiments including ``AT with SemAdv'' and ``AT with ours'', such that the targeting semantics will be protected 
no matter which type of semantics is attacked. Therefore, we report the results based on the label of each semantic class, where the labels range from $0$ to $9$. The intuition is that, in practice, the semantics that  attackers are going to distort may also be agnostic, and therefore each semantic type is expected to be hardened against potential adversaries during AT. Figures \ref{fig:defense_mnist} and  \ref{fig:defense_cifar} show the comparison results of different defense mechanisms. We report the performance scores on each class label for the three datasets, and also give the average values (the dotted line in the same color) for a more clear demonstration.

\subsubsection{Classification Accuracy}
 Figures \ref{fig:defense_mnist} (a) and (d), Figures \ref{fig:defense_mnist} (c) and (f), Figures \ref{fig:defense_cifar} (a) and (d),  Figures \ref{fig:defense_cifar} (b) and (e), and Figures \ref{fig:defense_cifar} (f) show the comparisons in terms of classification accuracy under \texttt{SemAdv} and PGM attacks. We observe that both \texttt{SemAdv} and PGM attackers can severely threaten to unprotected semantic communications. Augmenting either PGM or \texttt{SemAdv} separately to training instances can benefit the system's robustness. Among the three defenses under the two attacks, our proposed \texttt{SemMixed}, which randomly combines \texttt{SemAdv} and PGM during the training procedure, consistently achieves comparable robustness to the best one that is specifically trained for a single attack. We also find that the classification accuracy of \texttt{MixedAT} remains stable on each semantic class ranging from 0 to 9. Conversely, the variance of a single defense scheme is much larger on different class types. The results confirm the effectiveness of our proposed defense approach in securing the interpretation of semantics, the central to the success of a deep learning-based semantic communication system. Equipped with \texttt{SemMixed}, our $\texttt{MobileSC}$ is able to provide more reliable transmissions for the safety-critical applications to make inferences based on semantics.  

We see in Figures \ref{fig:defense_mnist} (a) and (d), and Figures \ref{fig:defense_mnist} (c) and (f) that the average classification accuracy with AT on MNIST is able to reach near 100\%, which is comparable to the one without attacks. While from Figures \ref{fig:defense_cifar} (a) and (d) on CIFAR10, we find that system hardened by various ATs still has a 5\% gap to catch up. This is probably because the semantics involved in CIFAR are much more complex than the ones in MNIST, and further understanding such a phenomenon would be an interesting research direction in future work.   

\subsubsection{PSNR and SSIM}
Figures \ref{fig:defense_mnist} (b) and (e), and Figures \ref{fig:defense_cifar} (c) demonstrate the comparisons of robustness in terms of PSNR using the three different defenses. We observe that each AT benefits the robustness of PSNR, improving the image reconstruction quality under various adversarial attacks. However, ``AT with only PGM or AT'' with \texttt{SemAdv} cannot effectively defend the \texttt{SemAdv} attacks or PGM attacks, as \texttt{SemAdv} attacks haven't been seen by the ``AT with PGM'' defense, and PGM attacks are unseen to the ``AT with \texttt{SemAdv}''. Our $\texttt{SemMixed}$, which is indicated as ``AT with Ours'', is able to defend against the two attacks with competitive PSNR results. We find that our $\texttt{SemMixed}$ consistently performs best under two attacks simultaneously on MNIST. These observations further confirm the merits of our proposed $\texttt{SemMixed}$ in mitigating the blind spots of the semantic communication systems, facilitating reliable semantics understanding at the receiver side.

\begin{table}
\caption{Comparisons of model size and inference time on CIFAR10.}
\scalebox{0.95}{
\begin{tabular}{lcc}
\toprule
\textbf{Model}   & \textbf{Number of Parameters} & \textbf{Inference time} \\ 
\midrule
Our Semantic Encoder & 805,464                                   & 5.6s (10,000 images)         \\
Our Semantic Decoder & 804,955                                   & 5.6s (10,000 images)         \\
Mobilenet V2 Encoder & 1,476,032                                   & 6.2s (10,000 images)         \\
Mobilenet V2 Decoder & 1,475,521                                    & 6.2s (10,000 images)         \\
JSCC Encoder \cite{yang2022ofdm}     & 2,760,128                                   & 6.7s (10,000 images)       \\
JSCC Decoder \cite{yang2022ofdm}     & 2,759,617                                   & 6.7s (10,000 images)         \\ 
\bottomrule
\end{tabular}}
\label{tab:speed}
\end{table}

\section{Discussions and Case Study}
\label{sec:discussion}
So far, we have shown the superiority of our proposed physical-layer adversarial attacks $\texttt{SemAdv}$ and the corresponding defense strategy $\texttt{SemMixed}$ for the robustness of semantic communications on the three benchmarks. In this section, we take a further step to show some more interesting observations based on comparisons with previous deep learning-based approaches for content-oriented communications and conventional block-structure communication systems. We also visualize each method with a case selected from CIFAR10.

\subsection{Memory and Computation Efficiency} We compare our semantic encoder and semantic decoder with ones of the prior work JSCC \cite{yang2022ofdm} and MobileNet V2 \cite{sandler2018mobilenetv2}. Table \ref{tab:speed} reports that our proposed semantic encoder and decoder only need $64\%$ and $40\%$ fewer parameters and can speed up by $10\%$ compared with the existing encoders and decoders. We attribute such a significant gain to our proposed SemBlock. The size of the encoder or decoder is only about 3.2MB as it only involves about $805,464$ parameters with an int 32 data type. We believe the mobile device can afford such a small neural model for deployment. These results further confirm our claims that the proposed ESC framework \texttt{MobileSC} will be more practical for deployment in mobile devices with limited battery and memory capacity. 

\subsection{Comparisons with Conventional Wireless Communications}

\subsubsection{Performance Comparisons}
We refer to previous works \cite{xie2020deep,yang2022ofdm} to compare our proposed $\texttt{MobileSC}$ system with a series of conventional block-structure communication systems, e.g., ``1/2LDPC+QPSK'' and ``1/2LDPC+16QAM'', in terms of transmission efficiency, PSNR, and SSIM, respectively. We use JPG2000 \cite{rabbani2002book} for source encoding/decoding and configure the SNR as $10$dB for both two systems for fair comparisons. 

\begin{table}
\centering
\caption{comparisons of transmission efficiency and reconstruction performance}
\begin{tabular}{@{}lcll@{}}
\toprule
\textbf{System} & \textbf{Number of OFDM symbols} & \textbf{PSNR}        & \textbf{SSIM}        \\ \midrule
1/2LDPC+BPSK       & 76                    & 19.77 & 0.77 \\
1/2LDPC+QPSK       & 38                     & 18.31 & 0.71 \\
1/2LDPC+16QAM      & 19                     & 13.75 & 0.48 \\
1/2LDPC+64QAM      & 9                     & 11.03 & 0.28 \\ 
\midrule
MobileSC       & 6                      & 25.13  & 0.85 \\
\bottomrule
\end{tabular}
\label{tab:efficiency}
\end{table}

Table \ref{tab:efficiency} reports the comparison results in terms of the number of OFDM symbols transmitted for each image, PSNR and SSIM. Under the same settings, our proposed \texttt{MobileSC} needs to transmit only $6$ OFDM symbols for each image in CIFAR10. While the number for a classical communication system can be as high as $76$ (``1/2LDPC+QPSK''), which is nearly $12$ times larger than the data transmitted by the semantic communication system. Meanwhile, the image quality recovered by the \texttt{MobileSC} receiver is much better than the ones on the conventional communication systems, e.g., PSNR of \texttt{MobileSC} can be $25.13$, which is much higher than the results of ``1/2LDPC+QPSK'' ($19.77$). Although the conventional system ``1/2LDPC+64QAQ'' needs to transmit $9$ OFDM symbols for an image, which is much smaller than $76$, the corresponding PSNR and SSIM scores ($11.03$) are much lower than the ones of ``1/2LDPC+BPSK'' ($19.77$). We observe that the classical system requires a careful trade-off between the transmission efficiency and the recovered image quality. However, our \texttt{MobileSC} doesn't need such a complex trade-off and is able to achieve much higher PSNR/SSIM scores with a much smaller number of symbols to transmit. We can draw a conclusion that our \texttt{MobileSC} is much more efficient (e.g., 153 times) for image transmissions, at the same time, more effective for image reconstruction.    

\begin{figure}[!t]
\centering
\includegraphics[width=3.5in]{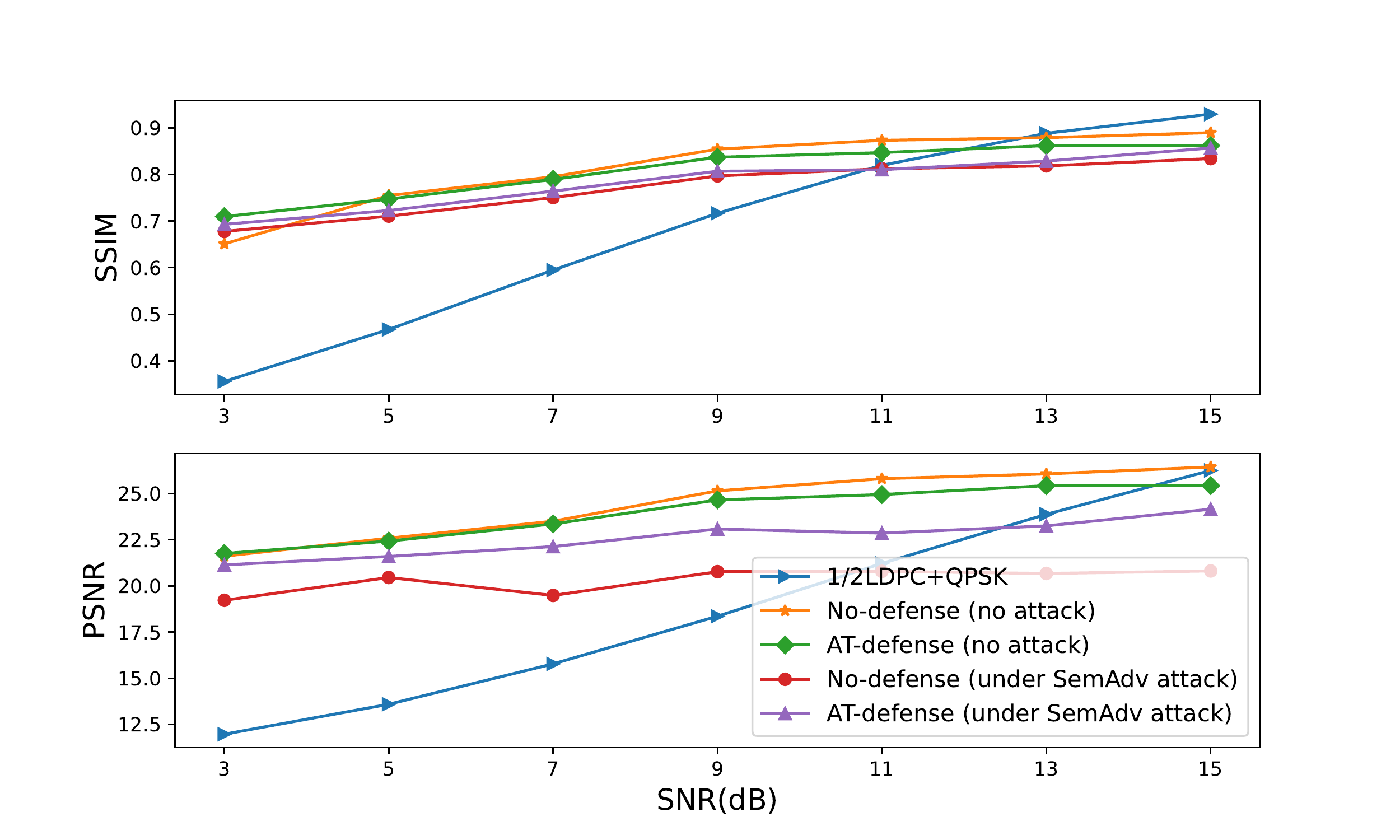}
\caption{Comparisons between $\texttt{MobileSC}$ and a classical communication system.}
\label{fig:classical}
\end{figure}

\subsubsection{Robustness Comparisons}
Figure \ref{fig:classical} compares our proposed $\texttt{MobileSC}$ with a classical wireless communication system ``1/2LDPC+QPSK'' in terms of PSNR and SSIM. We select these two metrics and don't use the classification accuracy, as it is not feasible to add a classifier to a conventional block-wise system in such a setting. We show the robustness of the two systems under various settings, e.g., ``no attack'' and ``under SemAdv attack''. Some interesting observations are given as follows. 
\begin{itemize}
    \item Without physical adversarial attacks, the PSNR and SSIM scores of our \texttt{MobileSC} (the color orange, ``No-defense'') are much higher in the relatively low SRN regime ($\le$ $10$dB) and comparable in the relatively high SNR regime compared with the classical communication system (color blue, ``1/2LDPC+QPSK''). The rationale behind this is that \texttt{MobileSC} is able to reconstruct high-quality images with powerful DNN. The comparisons demonstrate the superiority of semantic communications in image reconstruction over noised wireless channels.
    \item Our $\texttt{MobileSC}$ can even perform better under adversarial attacks, compared with ``1/2LDPC+QPSK'' with the natural noise. We only consider the natural noise measured by SNR for ``1/2LDPC+QPSK'' as the concept of adversaries is not applicable in conventional block-structure wireless communications. These results show that our semantic communication system can be more robust under attacks. 
    \item We use ``No-defense (under SemAdv attacks)'' and ``AT-defense (under SemAdv attacks)'' to denote the original $\texttt{MobileSC}$ system and the reinforced system based on our proposed adversarial training method \texttt{SemMixed}. We find that adversarial training benefits the robustness of $\texttt{MobileSC}$ under physical semantic attacks in various SNR regimes, although AT is unable to provide further image quality improvement measured by PSNR and SSIM.
\end{itemize}
From the above observations, we are able to conclude that our $\texttt{MobileSC}$ can be more robust than the conventional block-wise communication systems, showing great potential for the practical deployment of semantic communications in the future.



\begin{figure}[!t]
\centering
\includegraphics[width=3.2in]{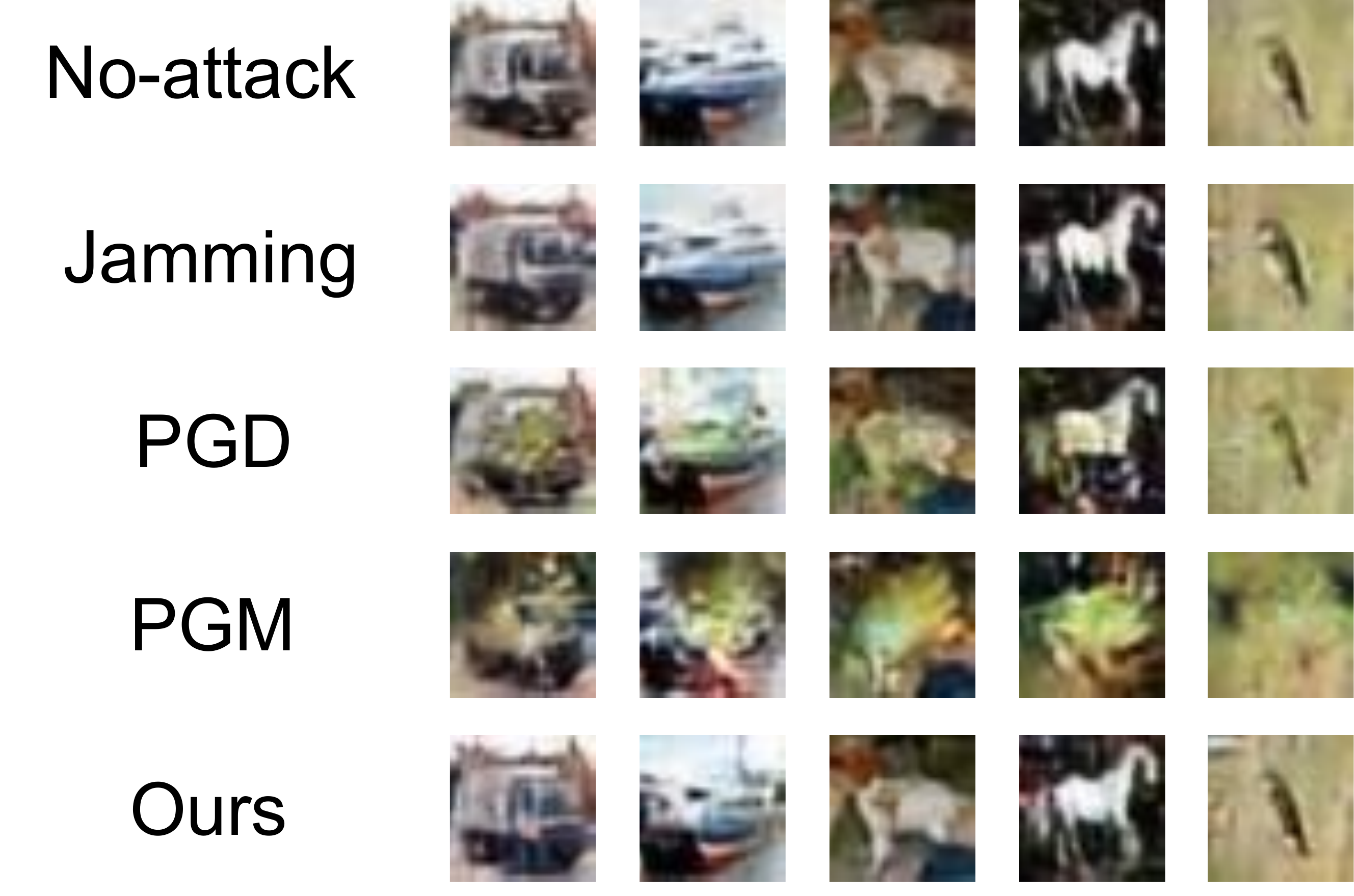}
\caption{Case study on CIFAR10 under various attacks.}
\label{fig:case_cifar}
\end{figure}

\subsection{Case Study}
We visualize a case study on CIFAR10 to demonstrate the effect of each attack on image reconstruction on our proposed semantic communication systems. Figure \ref{fig:case_cifar} shows the visualizations on CIFAR10. The reconstruction quality we are able to perceive from the image is closely related to two metrics PSNR and SSIM as discussed in Figure \ref{fig:attacks}. We observe that the distortions caused by the Jamming attack and our method are comparable. This aligns with our findings based on quantitative results in Figures \ref{fig:attacks} (f). Under \texttt{SemAdv} attacks, we are still able to identify a bird after reconstruction, even if the targeting semantics of the attacks belong to the "bird" category. Conversely, PGD and PGM attacks are able to bring obviously distortions, and we can hardly recognize the content. The visualizations illustrate that our attacks are much more imperceptible compared with PGD and PGM attacks. 
We can draw a conclusion that our \texttt{SemAdv} attacks can be treated as the Gaussian noise if we measure the reconstruction quality based on PSNR and SSIM. 

\section{Conclusion}
\label{sec:conclusion}
This paper studies the physical-layer adversarial robustness of deep learning-based semantic communication systems. We first present a semantic communication framework \texttt{MobileSC}, which considers computation constraints and energy efficiency over the wireless environment, showing great potential for practical deployment in real-world scenarios. Equipped with such a framework, we then propose a novel physical adversarial attacker \texttt{SemAdv} that can craft semantic adversarial perturbations over wireless channels. Unlike the previous works, the distortions generated by \texttt{SemAdv} are semantic-oriented, imperceptible, input-agnostic, and controllable, better characterizing the real-world effects. Based on such perturbations, we then conduct adversarial training to harden the system's robustness against semantic attacks. Experiments on three popular datasets show the effectiveness of our proposed methods in terms of classification accuracy, PSNR, and SSIM, as well as a novel metric misleading rate that is proposed for the robustness evaluation of the ESC system. In the discussion section, we also show that our \texttt{MobileSC} can even be more robust than classical block-wise communication systems in the low SNR regime.


\section{Acknowledgement}
\label{sec:acknowledgement}
This work was supported in part by the National Key R\&D Program of China (No. 2022YFB2902200), in part by the Joint funds for Regional Innovation and Development of the National Natural Science Foundation of China (No. U21A20449), and in part by the National Research Foundation, Singapore and Infocomm Media Development Authority under its Future Communications Research \& Development Programme.

\bibliographystyle{IEEEtran}
\bibliography{IEEEabrv,robust}

\begin{thebibliography}{10}
\providecommand{\url}[1]{#1}
\csname url@samestyle\endcsname
\providecommand{\newblock}{\relax}
\providecommand{\bibinfo}[2]{#2}
\providecommand{\BIBentrySTDinterwordspacing}{\spaceskip=0pt\relax}
\providecommand{\BIBentryALTinterwordstretchfactor}{4}
\providecommand{\BIBentryALTinterwordspacing}{\spaceskip=\fontdimen2\font plus
\BIBentryALTinterwordstretchfactor\fontdimen3\font minus
  \fontdimen4\font\relax}
\providecommand{\BIBforeignlanguage}[2]{{%
\expandafter\ifx\csname l@#1\endcsname\relax
\typeout{** WARNING: IEEEtran.bst: No hyphenation pattern has been}%
\typeout{** loaded for the language `#1'. Using the pattern for}%
\typeout{** the default language instead.}%
\else
\language=\csname l@#1\endcsname
\fi
#2}}
\providecommand{\BIBdecl}{\relax}
\BIBdecl

\bibitem{ITU}
ITU-R, ``Imt traffic estimates for the years 2020 to 2030,''
  \url{https://www.itu.int/dms_pub/itu-r/opb/rep/R-REP-M.2370-2015-PDF-E.pdf},
  July 2015, [Online].

\bibitem{Lin2023TransmissiveMA}
J.~Lin, G.~Wang, S.~Atapattu, R.~He, G.~Yang, and C.~Tellambura, ``Transmissive
  metasurfaces assisted wireless communications on railways: Channel strength
  evaluation and performance analysis,'' \emph{IEEE Transactions on
  Communications}, vol.~71, pp. 1827--1841, 2023.

\bibitem{Yang2020ReconfigurableIS}
G.~Yang, Y.~Liao, Y.-C. Liang, O.~Tirkkonen, G.~Wang, and X.~Zhu,
  ``Reconfigurable intelligent surface empowered device-to-device communication
  underlaying cellular networks,'' \emph{IEEE Transactions on Communications},
  vol.~69, pp. 7790--7805, 2020.

\bibitem{Li2010ANF}
J.~Li, H.~Zhang, X.~Xu, X.~Tao, T.~Svensson, C.~Botella, and B.~Liu, ``A novel
  frequency reuse scheme for coordinated multi-point transmission,'' \emph{2010
  IEEE 71st Vehicular Technology Conference}, pp. 1--5, 2010.

\bibitem{Xu2016EnhancingSC}
M.~Xu, X.~Tao, F.~Yang, and H.~Wu, ``Enhancing secured coverage with comp
  transmission in heterogeneous cellular networks,'' \emph{IEEE Communications
  Letters}, vol.~20, pp. 2272--2275, 2016.

\bibitem{Liu2014JointBA}
X.~Liu, F.~Gao, G.~Wang, and X.~Wang, ``Joint beamforming and user selection in
  multicast downlink channel under secrecy-outage constraint,'' \emph{IEEE
  Communications Letters}, vol.~18, pp. 82--85, 2014.

\bibitem{Li2021PrivacyPreservedFL}
Y.~Li, X.~Tao, X.~Zhang, J.~Liu, and J.~Xu, ``Privacy-preserved federated
  learning for autonomous driving,'' \emph{IEEE Transactions on Intelligent
  Transportation Systems}, vol.~23, pp. 8423--8434, 2021.

\bibitem{you2021towards}
X.~You, C.-X. Wang, J.~Huang, X.~Gao, Z.~Zhang, M.~Wang, Y.~Huang, C.~Zhang,
  Y.~Jiang, J.~Wang \emph{et~al.}, ``Towards 6g wireless communication
  networks: Vision, enabling technologies, and new paradigm shifts,''
  \emph{Science China Information Sciences}, vol.~64, no.~1, pp. 1--74,
  November 2021.

\bibitem{xiao2020toward}
Y.~Xiao, G.~Shi, Y.~Li, W.~Saad, and H.~V. Poor, ``Toward self-learning edge
  intelligence in 6g,'' \emph{IEEE Communications Magazine}, vol.~58, no.~12,
  pp. 34--40, January 2020.

\bibitem{strinati20216g}
E.~C. Strinati and S.~Barbarossa, ``6g networks: Beyond shannon towards
  semantic and goal-oriented communications,'' \emph{Computer Networks}, vol.
  190, p. 107930, May 2021.

\bibitem{xie2020deep}
H.~Xie, Z.~Qin, G.~Y. Li, and B.-H. Juang, ``Deep learning based semantic
  communications: An initial investigation,'' in \emph{IEEE Global
  Communications Conference (GLOBECOM)}, December 2020, pp. 1--6.

\bibitem{luo2022semantic}
X.~Luo, H.-H. Chen, and Q.~Guo, ``Semantic communications: Overview, open
  issues, and future research directions,'' \emph{IEEE Wireless
  Communications}, vol.~29, no.~1, pp. 210--219, January 2022.

\bibitem{lecun2015deep}
Y.~LeCun, Y.~Bengio, and G.~Hinton, ``Deep learning,'' \emph{Nature}, vol. 521,
  no. 7553, pp. 436--444, May 2015.

\bibitem{weaver1953recent}
W.~Weaver, ``Recent contributions to the mathematical theory of
  communication,'' \emph{ETC: A Review of General Semantics}, pp. 261--281,
  Autumn 1953.

\bibitem{bao2011towards}
J.~Bao, P.~Basu, M.~Dean, C.~Partridge, A.~Swami, W.~Leland, and J.~A. Hendler,
  ``Towards a theory of semantic communication,'' in \emph{IEEE Network Science
  Workshop (NSW)}, June 2011, pp. 110--117.

\bibitem{o2017introduction}
T.~O’shea and J.~Hoydis, ``An introduction to deep learning for the physical
  layer,'' \emph{IEEE Transactions on Cognitive Communications and Networking},
  vol.~3, no.~4, pp. 563--575, October 2017.

\bibitem{farsad2018deep}
N.~Farsad, M.~Rao, and A.~Goldsmith, ``Deep learning for joint source-channel
  coding of text,'' in \emph{IEEE International Conference on Acoustics, Speech
  and Signal Processing (ICASSP)}, April 2018, pp. 2326--2330.

\bibitem{kramer1991nonlinear}
M.~A. Kramer, ``Nonlinear principal component analysis using autoassociative
  neural networks,'' \emph{AIChE Journal}, vol.~37, no.~2, pp. 233--243,
  February 1991.

\bibitem{xie2021deep}
H.~Xie, Z.~Qin, G.~Y. Li, and B.-H. Juang, ``Deep learning enabled semantic
  communication systems,'' \emph{IEEE Transactions on Signal Processing},
  vol.~69, pp. 2663--2675, April 2021.

\bibitem{sadeghi2019physical}
M.~Sadeghi and E.~G. Larsson, ``Physical adversarial attacks against end-to-end
  autoencoder communication systems,'' \emph{IEEE Communications Letters},
  vol.~23, no.~5, pp. 847--850, February 2019.

\bibitem{moosavi2017universal}
S.-M. Moosavi-Dezfooli, A.~Fawzi, O.~Fawzi, and P.~Frossard, ``Universal
  adversarial perturbations,'' in \emph{Proceedings of the IEEE conference on
  computer vision and pattern recognition}, July 2017, pp. 1765--1773.

\bibitem{wong2020learning}
E.~Wong and J.~Z. Kolter, ``Learning perturbation sets for robust machine
  learning,'' in \emph{International Conference on Learning Representations},
  July 2020.

\bibitem{maini2020adversarial}
P.~Maini, E.~Wong, and Z.~Kolter, ``Adversarial robustness against the union of
  multiple perturbation models,'' in \emph{International Conference on Machine
  Learning (ICML)}.\hskip 1em plus 0.5em minus 0.4em\relax (PMLR), July 2020,
  pp. 6640--6650.

\bibitem{madaan2021learning}
D.~Madaan, J.~Shin, and S.~J. Hwang, ``Learning to generate noise for
  multi-attack robustness,'' in \emph{International Conference on Machine
  Learning (ICML)}.\hskip 1em plus 0.5em minus 0.4em\relax (PMLR), July 2021,
  pp. 7279--7289.

\bibitem{leino2021globally}
K.~Leino, Z.~Wang, and M.~Fredrikson, ``Globally-robust neural networks,'' in
  \emph{International Conference on Machine Learning (ICML)}.\hskip 1em plus
  0.5em minus 0.4em\relax PMLR, July 2021, pp. 6212--6222.

\bibitem{bahramali2021robust}
A.~Bahramali, M.~Nasr, A.~Houmansadr, D.~Goeckel, and D.~Towsley, ``Robust
  adversarial attacks against dnn-based wireless communication systems,'' in
  \emph{Proceedings of the ACM SIGSAC Conference on Computer and Communications
  Security}, November 2021, pp. 126--140.

\bibitem{hu2022robust}
Q.~Hu, G.~Zhang, Z.~Qin, Y.~Cai, and G.~Yu, ``Robust semantic communications
  against semantic noise,'' \emph{arXiv preprint arXiv:2202.03338}, February
  2022.

\bibitem{sandler2018mobilenetv2}
M.~Sandler, A.~Howard, M.~Zhu, A.~Zhmoginov, and L.-C. Chen, ``Mobilenetv2:
  Inverted residuals and linear bottlenecks,'' in \emph{Proceedings of the IEEE
  Conference on Computer Vision and Pattern Recognition}, June 2018, pp.
  4510--4520.

\bibitem{yang2022ofdm}
M.~Yang, C.~Bian, and H.-S. Kim, ``Ofdm-guided deep joint source channel coding
  for wireless multipath fading channels,'' \emph{IEEE Transactions on
  Cognitive Communications and Networking}, vol.~8, no.~2, pp. 584--599,
  February 2022.

\bibitem{dorner2017deep}
S.~D{\"o}rner, S.~Cammerer, J.~Hoydis, and S.~Ten~Brink, ``Deep learning based
  communication over the air,'' \emph{IEEE Journal of Selected Topics in Signal
  Processing}, vol.~12, no.~1, pp. 132--143, December 2017.

\bibitem{aoudia2018end}
F.~A. Aoudia and J.~Hoydis, ``End-to-end learning of communications systems
  without a channel model,'' in \emph{52nd Asilomar Conference on Signals,
  Systems, and Computers}, October 2018, pp. 298--303.

\bibitem{ye2018channel}
H.~Ye, G.~Y. Li, B.-H.~F. Juang, and K.~Sivanesan, ``Channel agnostic
  end-to-end learning based communication systems with conditional gan,'' in
  \emph{IEEE Globecom Workshops (GC Wkshps)}, February 2018, pp. 1--5.

\bibitem{bourtsoulatze2019deep}
E.~Bourtsoulatze, D.~B. Kurka, and D.~G{\"u}nd{\"u}z, ``Deep joint
  source-channel coding for wireless image transmission,'' \emph{IEEE
  Transactions on Cognitive Communications and Networking}, vol.~5, no.~3, pp.
  567--579, May 2019.

\bibitem{aoudia2019model}
F.~A. Aoudia and J.~Hoydis, ``Model-free training of end-to-end communication
  systems,'' \emph{IEEE Journal on Selected Areas in Communications}, vol.~37,
  no.~11, pp. 2503--2516, August 2019.

\bibitem{kurka2020deepjscc}
D.~B. Kurka and D.~G{\"u}nd{\"u}z, ``Deepjscc-f: Deep joint source-channel
  coding of images with feedback,'' \emph{IEEE Journal on Selected Areas in
  Information Theory}, vol.~1, no.~1, pp. 178--193, April 2020.

\bibitem{ye2020deep}
H.~Ye, L.~Liang, G.~Y. Li, and B.-H. Juang, ``Deep learning-based end-to-end
  wireless communication systems with conditional gans as unknown channels,''
  \emph{IEEE Transactions on Wireless Communications}, vol.~19, no.~5, pp.
  3133--3143, February 2020.

\bibitem{nan2023udsem}
G.~Nan, X.~Liu, X.~Lyu, Q.~Cui, X.~Xu, and P.~Zhang, ``Udsem: A unified
  distributed learning framework for semantic communications over wireless
  networks,'' \emph{IEEE Network}, 2023.

\bibitem{wang2022wireless}
S.~Wang, J.~Dai, Z.~Liang, K.~Niu, Z.~Si, C.~Dong, X.~Qin, and P.~Zhang,
  ``Wireless deep video semantic transmission,'' \emph{IEEE Journal on Selected
  Areas in Communications}, vol.~41, no.~1, pp. 214--229, 2022.

\bibitem{weng2021semantic-jsac}
Z.~Weng and Z.~Qin, ``Semantic communication systems for speech transmission,''
  \emph{IEEE Journal on Selected Areas in Communications}, vol.~39, no.~8, pp.
  2434--2444, June 2021.

\bibitem{Han2022SemanticPreservedCS}
T.~Han, Q.~Yang, Z.~Shi, S.~He, and Z.~Zhang, ``Semantic-preserved
  communication system for highly efficient speech transmission,'' \emph{IEEE
  Journal on Selected Areas in Communications}, vol.~41, pp. 245--259, 2022.

\bibitem{xie2021task}
H.~Xie, Z.~Qin, and G.~Y. Li, ``Task-oriented multi-user semantic
  communications for vqa task,'' \emph{IEEE Wireless Communications Letters},
  vol.~11, no.~3, pp. 553--557, December 2021.

\bibitem{xie2020lite}
H.~Xie and Z.~Qin, ``A lite distributed semantic communication system for
  internet of things,'' \emph{IEEE Journal on Selected Areas in
  Communications}, vol.~39, no.~1, pp. 142--153, November 2020.

\bibitem{singh2019beyond}
G.~Singh, R.~Ganvir, M.~P{\"u}schel, and M.~Vechev, ``Beyond the single neuron
  convex barrier for neural network certification,'' \emph{Advances in Neural
  Information Processing Systems}, vol.~32, December 2019.

\bibitem{Li2023BoostingPL}
Z.~Li, X.~Liu, G.~Nan, J.~Zhou, X.~Lyu, Q.~Cui, and X.~Tao, ``Boosting physical
  layer black-box attacks with semantic adversaries in semantic
  communications,'' 2023.

\bibitem{szegedy2013intriguing}
C.~Szegedy, W.~Zaremba, I.~Sutskever, J.~Bruna, D.~Erhan, I.~Goodfellow, and
  R.~Fergus, ``Intriguing properties of neural networks,'' \emph{arXiv preprint
  arXiv:1312.6199}, February 2013.

\bibitem{tramer2019adversarial}
F.~Tramer and D.~Boneh, ``Adversarial training and robustness for multiple
  perturbations,'' \emph{Advances in Neural Information Processing Systems},
  vol.~32, December 2019.

\bibitem{Qin2023SecuringSC}
Q.~Qin, Y.~Rong, G.~Nan, S.~Wu, X.~Zhang, Q.~Cui, and X.~Tao, ``Securing
  semantic communications with physical-layer semantic encryption and
  obfuscation,'' \emph{ArXiv}, vol. abs/2304.10147, 2023.

\bibitem{luo2022encrypted}
X.~Luo, Z.~Chen, M.~Tao, and F.~Yang, ``Encrypted semantic communication using
  adversarial training for privacy preserving,'' \emph{arXiv preprint
  arXiv:2209.09008}, September 2022.

\bibitem{tung2022deep}
T.-Y. Tung and D.~Gunduz, ``Deep joint source-channel and encryption coding:
  Secure semantic communications,'' \emph{arXiv preprint arXiv:2208.09245},
  August 2022.

\bibitem{chen2022Neuromorphic}
J.~Chen, N.~Skatchkovsky, and O.~Simeone, ``Neuromorphic wireless cognition:
  Event-driven semantic communications for remote inference,'' \emph{arXiv
  e-prints}, 2022.

\bibitem{yang2023energy}
Z.~Yang, M.~Chen, Z.~Zhang, and C.~Huang, ``Energy efficient semantic
  communication over wireless networks with rate splitting,'' \emph{arXiv
  preprint arXiv:2301.01987}, 2023.

\bibitem{huang2023semantic}
K.~Huang, Q.~Lan, Z.~Liu, and L.~Yang, ``Semantic data sourcing for 6g edge
  intelligence,'' \emph{arXiv preprint arXiv:2301.00403}, 2023.

\bibitem{Zhang2022SemanticCA}
Z.~Zhang, Q.~Yang, S.~He, and Z.~Shi, ``Semantic communication approach for
  multi-task image transmission,'' \emph{2022 IEEE 96th Vehicular Technology
  Conference (VTC2022-Fall)}, pp. 1--2, 2022.

\bibitem{Tang2023ContrastiveLB}
S.~Tang, Q.~Yang, L.~Fan, X.~Lei, Y.~Deng, and A.~Nallanathan, ``Contrastive
  learning based semantic communication for wireless image transmission,''
  \emph{ArXiv}, vol. abs/2304.09438, 2023.

\bibitem{shi2023excess}
Y.~Shi, S.~Shao, Y.~Wu, W.~Zhang, X.-G. Xia, and C.~Xiao, ``Excess distortion
  exponent analysis for semantic-aware mimo communication systems,''
  \emph{arXiv preprint arXiv:2301.04357}, 2023.

\bibitem{vaswani2017attention}
A.~Vaswani, N.~Shazeer, N.~Parmar, J.~Uszkoreit, L.~Jones, A.~N. Gomez,
  {\L}.~Kaiser, and I.~Polosukhin, ``Attention is all you need,''
  \emph{Advances in neural information processing systems}, vol.~30, December
  2017.

\bibitem{he2016identity}
K.~He, X.~Zhang, S.~Ren, and J.~Sun, ``Identity mappings in deep residual
  networks,'' in \emph{European Conference on Computer Vision}.\hskip 1em plus
  0.5em minus 0.4em\relax Springer, September 2016, pp. 630--645.

\bibitem{weinstein2009history}
S.~B. Weinstein, ``The history of orthogonal frequency-division multiplexing
  [history of communications],'' \emph{IEEE Communications Magazine}, vol.~47,
  no.~11, pp. 26--35, November 2009.

\bibitem{ochiai2002performance}
H.~Ochiai and H.~Imai, ``Performance analysis of deliberately clipped ofdm
  signals,'' \emph{IEEE Transactions on communications}, vol.~50, no.~1, pp.
  89--101, August 2002.

\bibitem{fan2017sanet}
H.~Fan and H.~Ling, ``Sanet: Structure-aware network for visual tracking,'' in
  \emph{Proceedings of the IEEE conference on computer vision and pattern
  recognition workshops}, July 2017, pp. 42--49.

\bibitem{xu2015empirical}
B.~Xu, N.~Wang, T.~Chen, and M.~Li, ``Empirical evaluation of rectified
  activations in convolutional network,'' \emph{arXiv preprint
  arXiv:1505.00853}, November 2015.

\bibitem{tsipras2018robustness}
D.~Tsipras, S.~Santurkar, L.~Engstrom, A.~Turner, and A.~Madry, ``Robustness
  may be at odds with accuracy,'' in \emph{International Conference on Learning
  Representations}, April 2018.

\bibitem{hendrycks2019benchmarking}
D.~Hendrycks and T.~Dietterich, ``Benchmarking neural network robustness to
  common corruptions and perturbations,'' in \emph{7th International Conference
  on Learning Representations (ICLR)}, May 2019.

\bibitem{pytorch}
``Pytorch,'' \url{https://pytorch.org/}, April 2022, [Online].

\bibitem{rabbani2002book}
M.~Rabbani, ``Book review: Jpeg2000: Image compression fundamentals, standards
  and practice,'' 2002.

\end{thebibliography}


\begin{IEEEbiography}[{\includegraphics[width=1in,height=1.25in,clip,keepaspectratio]{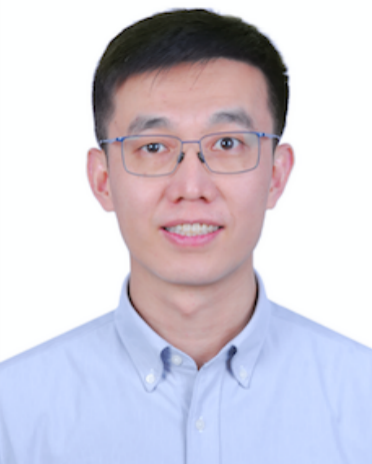}}]{Guoshun Nan}
is a tenure-track professor at Beijing University of Posts and Telecommunications (BUPT). He is a member of the National Engineering Research Center for Mobile Network Technologies and a member of ACL. He has broad interest in natural language processing, computer vision, machine learning and wireless communications, such as information extraction, model robustness, multimodal retrieval and next generation wireless networks. He has published papers in top-tier conferences and journals including ACL, CVPR, EMNLP, SIGIR, IJCAI, CKIM, SIGCOMM, IEEE Network, Computer Networks, Journal of Network and Computer Applications. He served as a reviewer for ACL, EMNLP, AAAI, IJCAI, Neurocomputing and IEEE Transactions on Image Processing.
\end{IEEEbiography}


\begin{IEEEbiography}[{\includegraphics[width=1in,height=1.25in,clip,keepaspectratio]{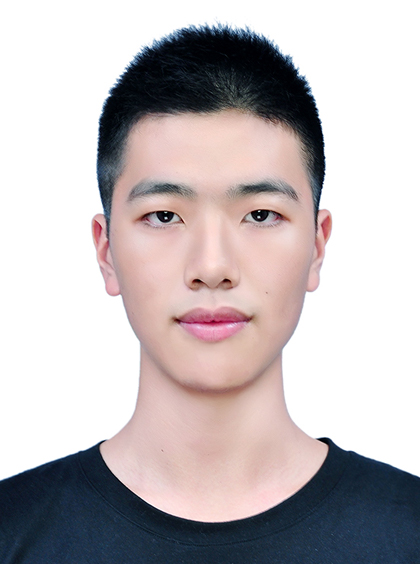}}]{Zhichun Li}
is an undergraduate student at Beijing University of Posts and Telecommunications (BUPT), Beijing, China. He is currently working as a research intern at the National Engineering Research Center of Mobile Network Security in BUPT. His research interests include wireless communications security and semantic communication.
\end{IEEEbiography}


\begin{IEEEbiography}[{\includegraphics[width=1in,height=1.25in,clip,keepaspectratio]{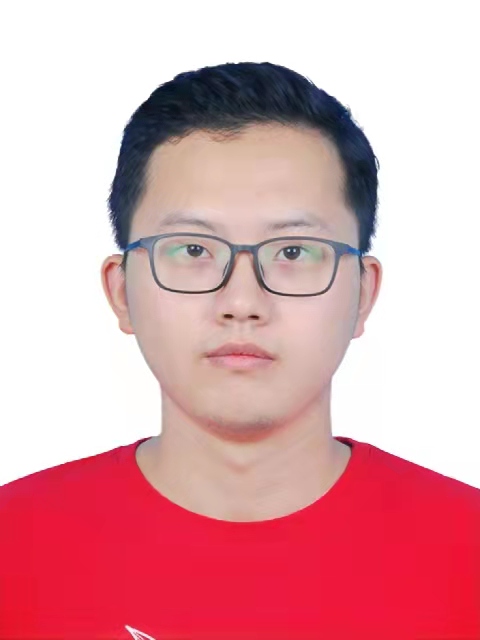}}]{Jinli Zhai}
received the B.E.degree from Beijing University of Posts and Telecommunications(BUPT), Beijing, China, in 2020. He is currently working toward a Ph.D. degree with the National Engineering Laboratory of Mobile Network Technology in BUPT. His research interests include radio resource management, wireless communications security, federated learning via 6G, and semantic communication.
\end{IEEEbiography}


\begin{IEEEbiography}[{\includegraphics[width=1in,height=1.25in,clip,keepaspectratio]{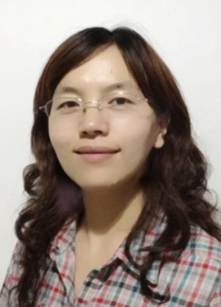}}]{Qimei Cui}
(M’09–SM’15)received the B.E. and M.S. degrees in electronic engineering from Hunan University, Changsha, China, in 2000 and 2003, respectively, and the Ph.D. degree in information and communications engineering from the Beijing University of Posts and Telecommunications (BUPT), Beijing, China, in 2006. She has been a Full Professor with the School of Information and Communication Engineering, BUPT, since 2014. She was a visiting Professor with the Department of Electronic Engineering, University of Notre Dame, IN, USA, in 2016. Her research interests include B5G/6G wireless communications, mobile computing and IoT. She serves as a Technical Program Chair of the APCC 2018, and a Track Chair of IEEE/CIC ICCC 2018, and a Workshop Chair of WPMC 2016. She also serves as a Technical Program Committee Member of several international conferences, such as the IEEE ICC, the IEEE WCNC, the IEEE PIMRC, the IEEE ICCC, the WCSP 2013, and the IEEE ISCIT 2012. She won the Best Paper Award at the IEEE ISCIT 2012, the IEEE WCNC 2014, and the WCSP 2019, and the Honorable Mention Demo Award at the ACM MobiCom 2009, and the Young Scientist Award at the URSI GASS 2014. She serves as Editor of SCIENCE CHINA Information Science , and Guest Editor of the EURASIP Journal on Wireless Communications and Networking and International Journal of Distributed Sensor Networks and Journal of Computer Networks and Comm.
\end{IEEEbiography}


\begin{IEEEbiography}[{\includegraphics[width=1in,height=1.25in,clip,keepaspectratio]{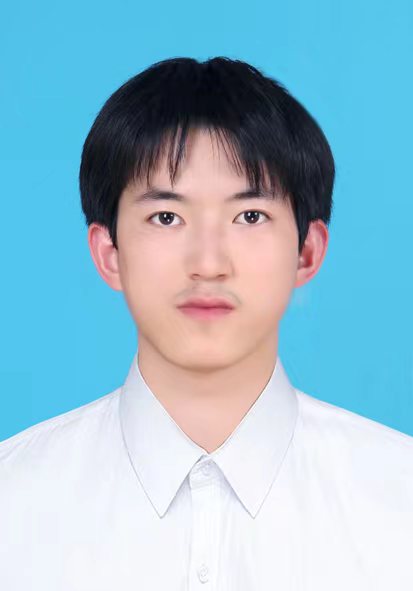}}]{Gong Chen}
 is a postgraduate student at Beijing University of Posts and Telecommunications (BUPT), Beijing, China. He is currently working at the National Engineering Laboratory of Mobile Network Technology in BUPT. His research interests include deep learning, wireless communications and semantic communications.
\end{IEEEbiography}


\begin{IEEEbiography}[{\includegraphics[width=1in,height=1.25in,clip,keepaspectratio]{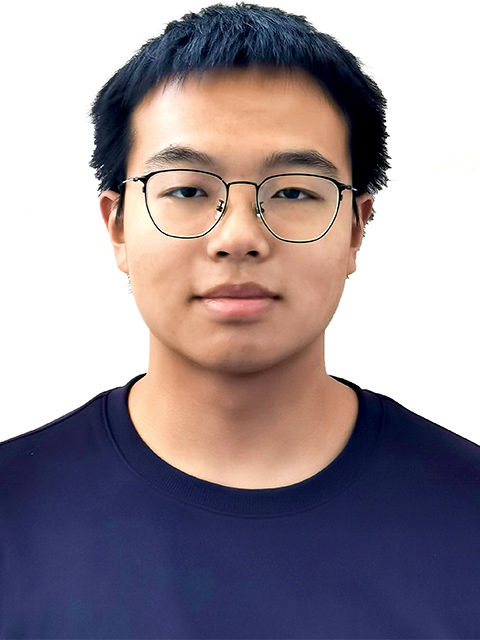}}]{Xin Du}
is an undergraduate student at Beijing University of Posts and Telecommunications(BUPT), Beijing, China. He is recently working on a college student entrepreneurship project of adversarial attack and defense on neural network. His research interests include model robustness and communication security.
\end{IEEEbiography}


\begin{IEEEbiography}[{\includegraphics[width=1in,height=1.25in,clip,keepaspectratio]{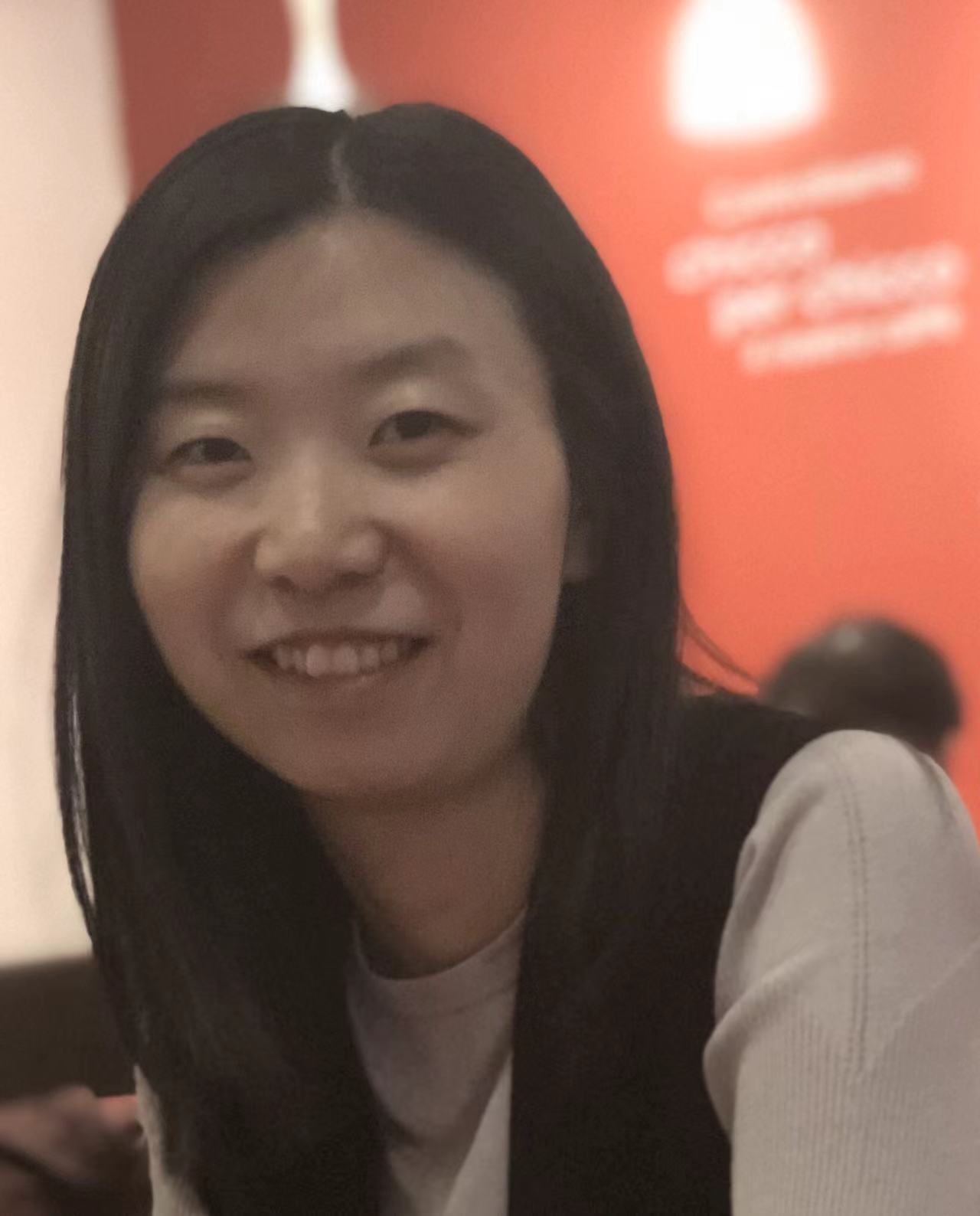}}]{Xuefei Zhang}
(Member, IEEE) received the B.S. and Ph.D. degrees in telecommunications engineering from the Beijing University of Posts and Telecommunications (BUPT) in 2010 and 2015, respectively. From September 2013 to August 2014, she was visiting the School of Electrical and Information Engineering, University of Sydney, Australia. She is currently an Associate Professor with the School of Information and Communication Engineering, BUPT. Her research area includes semantic communication and satellite-terrestrial integrated network.
\end{IEEEbiography}


\begin{IEEEbiography}[{\includegraphics[width=1in,height=1.25in,clip,keepaspectratio]{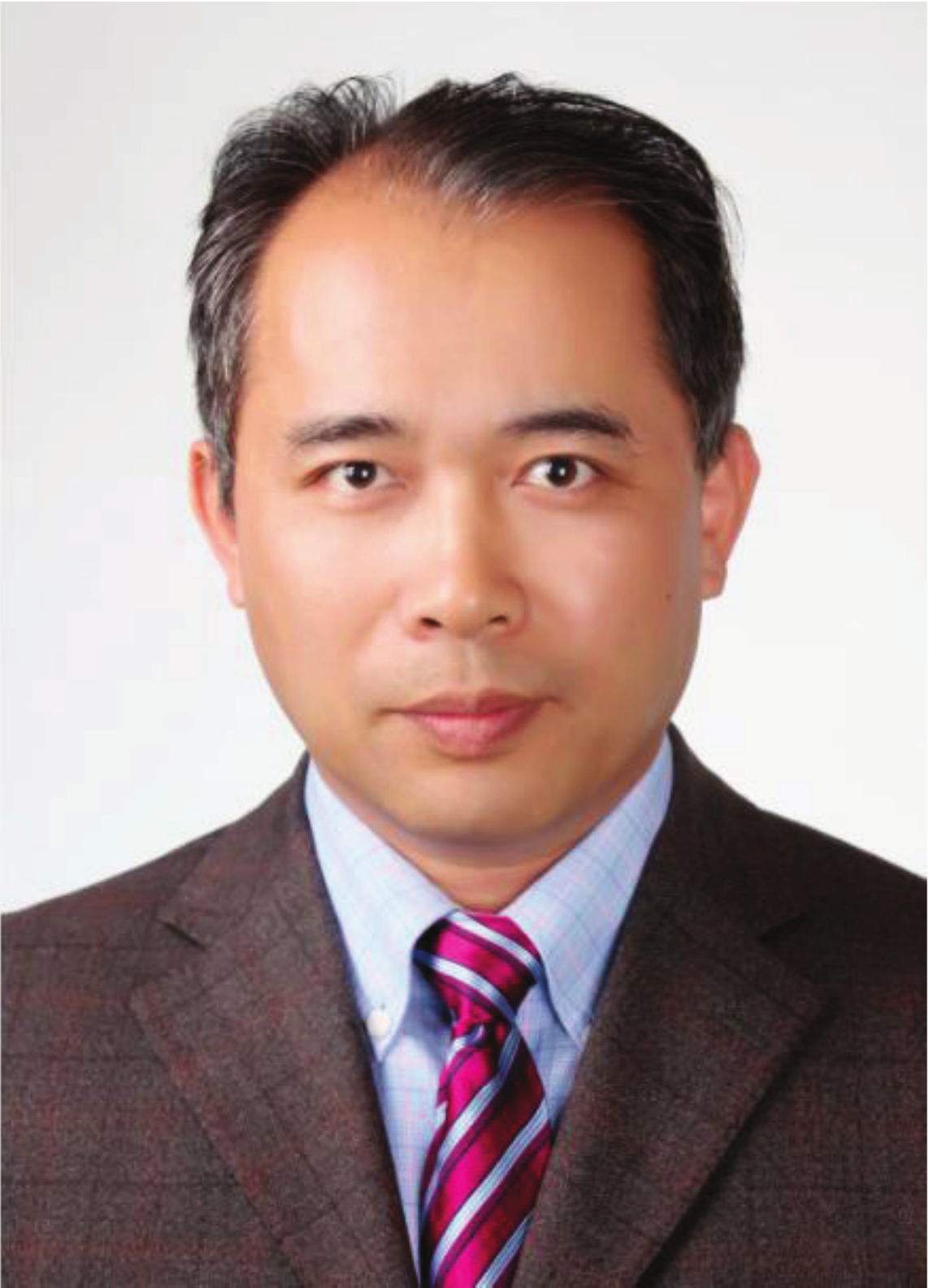}}]{Xiaofeng Tao}
received the B.S. degree in electrical engineering from Xi’an Jiaotong University, Xi’an, China, in 1993, and the M.S. and Ph.D. degrees in telecommunication engineering from Beijing University of Posts and Telecommunications(BUPT), Beijing, China, in 1999 and 2002, respectively. He is a Professor in BUPT, a Fellow of the Institution of Engineering and Technology, and Chair of the IEEE ComSoc Beijing Chapter. He has authored or co-authored over 200 papers and three books in wireless communication areas. He focuses on 5G/B5G research.
\end{IEEEbiography}

\vspace{-5mm}

\begin{IEEEbiography}[{\includegraphics[width=1in,height=1.25in,clip,keepaspectratio]{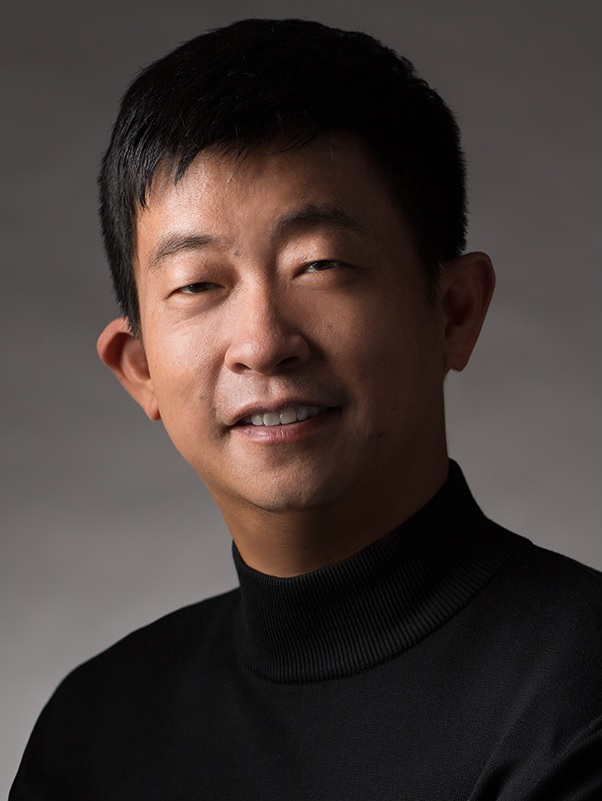}}]{Zhu Han}
(S’01–M’04-SM’09-F’14) received the B.S. degree in electronic engineering from Tsinghua University, in 1997, and the M.S. and Ph.D. degrees in electrical and computer engineering from the University of Maryland, College Park, in 1999 and 2003, respectively. 

From 2000 to 2002, he was an R\&D Engineer of JDSU, Germantown, Maryland. From 2003 to 2006, he was a Research Associate at the University of Maryland. From 2006 to 2008, he was an assistant professor at Boise State University, Idaho. Currently, he is a John and Rebecca Moores Professor in the Electrical and Computer Engineering Department as well as in the Computer Science Department at the University of Houston, Texas. Dr. Han’s main research targets on the novel game-theory related concepts critical to enabling efficient and distributive use of wireless networks with limited resources. His other research interests include wireless resource allocation and management, wireless communications and networking, quantum computing, data science, smart grid, security and privacy.  Dr. Han received an NSF Career Award in 2010, the Fred W. Ellersick Prize of the IEEE Communication Society in 2011, the EURASIP Best Paper Award for the Journal on Advances in Signal Processing in 2015, IEEE Leonard G. Abraham Prize in the field of Communications Systems (best paper award in IEEE JSAC) in 2016, and several best paper awards in IEEE conferences. Dr. Han was an IEEE Communications Society Distinguished Lecturer from 2015-2018, AAAS fellow since 2019, and ACM distinguished Member since 2019. Dr. Han is a 1\% highly cited researcher since 2017 according to Web of Science. Dr. Han is also the winner of the 2021 IEEE Kiyo Tomiyasu Award, for outstanding early to mid-career contributions to technologies holding the promise of innovative applications, with the following citation: ``for contributions to game theory and distributed management of autonomous communication networks."
\end{IEEEbiography}

\vspace{-5mm}

\begin{IEEEbiography}[{\includegraphics[width=1in,height=1.25in,clip,keepaspectratio]{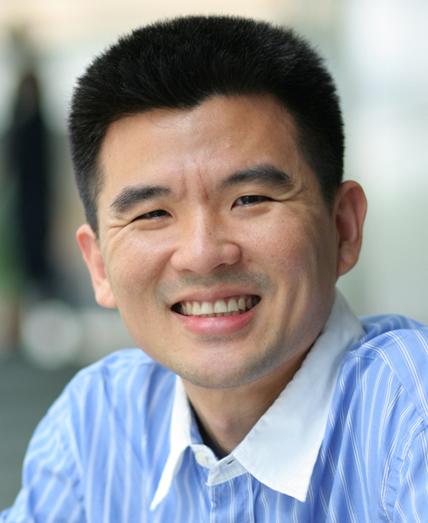}}]{Tony Q.S. Quek}
(S'98-M'08-SM'12-F'18) received the B.E.\ and M.E.\ degrees in electrical and electronics engineering from the Tokyo Institute of Technology in 1998 and 2000, respectively, and the Ph.D.\ degree in electrical engineering and computer science from the Massachusetts Institute of Technology in 2008. Currently, he is the Cheng Tsang Man Chair Professor with Singapore University of Technology and Design (SUTD) and ST Engineering Distinguished Professor. He also serves as the Director of the Future Communications R\&D Programme, the Head of ISTD Pillar, and the Deputy Director of the SUTD-ZJU IDEA. His current research topics include wireless communications and networking, network intelligence, non-terrestrial networks, open radio access network, and 6G.

Dr.\ Quek has been actively involved in organizing and chairing sessions, and has served as a member of the Technical Program Committee as well as symposium chairs in a number of international conferences. He is currently serving as an Area Editor for the {\scshape IEEE Transactions on Wireless Communications}. 

Dr.\ Quek was honored with the 2008 Philip Yeo Prize for Outstanding Achievement in Research, the 2012 IEEE William R. Bennett Prize, the 2015 SUTD Outstanding Education Awards -- Excellence in Research, the 2016 IEEE Signal Processing Society Young Author Best Paper Award, the 2017 CTTC Early Achievement Award, the 2017 IEEE ComSoc AP Outstanding Paper Award, the 2020 IEEE Communications Society Young Author Best Paper Award, the 2020 IEEE Stephen O. Rice Prize, the 2020 Nokia Visiting Professor, and the 2022 IEEE Signal Processing Society Best Paper Award. He is a Fellow of IEEE and a Fellow of the Academy of Engineering Singapore.\end{IEEEbiography}

\vfill

\end{document}